\documentclass[review]{elsarticle}

\usepackage[colorlinks,citecolor=blue,linktoc=all,linkcolor=cyan]{hyperref}
\usepackage{graphicx}

\usepackage[T1]{fontenc}
\usepackage{dsfont}               
\usepackage{mathrsfs}             
\usepackage{slashed}              
\usepackage{amsmath}
\usepackage{amssymb}
\usepackage{amsbsy}
\usepackage{amsfonts}
\usepackage{physics}
\usepackage{bm}
\usepackage{bbm}
\usepackage{xcolor}
\usepackage{braket}
\usepackage{bbold}

\newcommand{\inner}[2]{\langle #1 | #2 \rangle}
\newcommand{\avg}[1]{\langle #1 \rangle}
\newcommand{\nc}{\newcommand}

\newcommand{\beq}{\begin{equation}}
\newcommand{\eeq}{\end{equation}}
\nc{\bfx}{{\bf x}}
\nc{\bfy}{{\bf y}}
\nc{\bfz}{{\bf z}}
\nc{\bfxh}{{\bf \hat{x}}}
\nc{\bfyh}{{\bf \hat{y}}}
\nc{\bfzh}{{\bf \hat{z}}}
\nc{\bfj}{{\bf j}}
\nc{\bfr}{{\bf r}}
\nc{\bfrh}{{\bf \hat{r}}}
\nc{\bfR}{{\bf R}}
\nc{\bfk}{{\bf k}}
\nc{\bfK}{{\bf K}}
\nc{\bfq}{{\bf q}}
\nc{\bfp}{{\bf p}}
\nc{\bfv}{{\bf v}}
\nc{\bfs}{{\bf s}}
\nc{\bfA}{{\bf A}}
\nc{\bfB}{{\bf B}}
\nc{\bfJ}{{\bf J}}
\nc{\bfL}{{\bf L}}
\nc{\bfS}{{\bf S}}
\nc{\bfT}{{\bf T}}
\nc{\bfY}{{\bf Y}}
\nc{\bfl}{{\bf l}}
\nc{\bfP}{{\bf P}}
\nc{\bfjex}{{\bf j}^{\rm ext}}
\nc{\jex}{j^{\rm ext}}
\nc{\bfSg}{{\bf \Sigma}}
\nc{\bfsg}{{\bm \sigma}}
\nc{\bfta}{{\bm \tau}}
\nc{\bfvh}{{\bf \hat{v}}}
\nc{\bfqh}{{\bf \hat{q}}}
\nc{\bfeh}{{\bf \hat{e}}}
\nc{\bfkh}{{\bf \hat{k}}}
\nc{\reduce}[3]{\braket{ #1 \Vert #2 \Vert #3 }}

\newcommand{\SU}[1]{\ensuremath{\mathrm{SU}( #1 )}}

\newcommand{\SpR}[1]{\ensuremath{\mathrm{Sp}( #1,\mathbb{R} )}}
\newcommand{\braketop}[3]{\ensuremath{\left\langle #1 \right| #2 \left| #3 \right\rangle}}
\newcommand{\Nmax}{$N_{\rm max}$}
\newcommand{\hw}{\ensuremath{\hbar\Omega}}
\newcommand{\Be}{$^8\mathrm{Be}$}
\newcommand{\Li}{$^8\mathrm{Li}$}
\newcommand{\NNLOopt}{NNLO$_\mathrm{opt}$}
\newcommand{\bt}{$\beta$ }

\newcommand{\half}{\frac{1}{2}}

\newcommand{\eg}{\emph{e.g.}}
\newcommand{\etal}{\emph{et al.}}

\newcommand {\red} [1]{\textcolor{black}{#1}}

\numberwithin{equation}{section}
\numberwithin{table}{section}
\numberwithin{figure}{section}

\journal{Progress in Particle and Nuclear Physics}

\usepackage{titlesec}
\usepackage{sectsty}
\titleformat{\section}{\normalfont\Large\bfseries}{\thesection}{1em}{}
\titleformat{\subsection}{\normalfont\large\bfseries}{\thesubsection}{1em}{}
\titleformat{\subsubsection}{\normalfont\normalsize\bfseries}{\thesubsubsection}{1em}{}

\begin{document}
	
	\begin{frontmatter}
		\title{The Role of \textit{Ab Initio} Beta-Decay Calculations in Light Nuclei \\ for Probes of  Physics Beyond the Standard Model}

		
		\author[frib]{Grigor H. Sargsyan\corref{mycorrespondingauthor}}
		\cortext[mycorrespondingauthor]{Corresponding author}
		\ead{sargsyan at frib}
		
\author[lanl]{Garrett B. King}
\author[tau]{Ayala Glick-Magid}
\author[utk]{Chien-Yeah Seng}
        
		\address[frib]{Facility for Rare Isotope Beams, Michigan State University, East Lansing, Michigan 48824, USA.}
		\address[lanl]{Theoretical Division, Los Alamos National Laboratory, Los Alamos, NM 87545, USA}
		\address[tau]{School of Physics and Astronomy, Tel-Aviv University, Tel-Aviv 69978, Israel}
        \address[utk]{Department of Physics and Astronomy, University of Tennessee, Knoxville, TN 37996, USA}
        
		\begin{abstract}

Precision beta decay experiments serve as powerful probes of physics beyond the Standard Model, enabling stringent tests of fundamental symmetries of nature. In particular, these experiments primarily focus on precise determinations of the Cabibbo-Kobayashi-Maskawa matrix element $V_{ud}$ and the search for exotic weak currents, both of which depend critically on theoretical calculations of radiative, recoil-order, and isospin-breaking corrections with quantified uncertainties. In recent years, \emph{ab initio} nuclear many-body methods—grounded in realistic nucleon-nucleon interactions and systematically improvable approximations—have advanced considerably in their ability to compute these higher-order corrections for various nuclei. This review provides a comprehensive overview of state-of-the-art \emph{ab initio} calculations of beta-decay corrections, encompassing both radiative corrections and recoil-order terms, and examines their significance for precision tests of the Standard Model. We discuss the theoretical formalisms employed, including the integration of effective field theory frameworks with many-body approaches.  Particular attention is given to recent results for superallowed Fermi decays (e.g., $^{10}$C $\rightarrow$ $^{10}$B and $^{14}$O $\rightarrow$ $^{14}$C) and allowed Gamow-Teller transitions (e.g., $^{6}$He$\rightarrow$ $^{6}$Li, $^{8}$Li$\rightarrow$ $^{8}$Be, $^{8}$B$\rightarrow$ $^{8}$Be), where \emph{ab initio} calculations have achieved unprecedented precision. We also highlight emerging calculations for unique forbidden decays, which offer complementary sensitivity to BSM physics. Finally, we outline future directions aimed at extending the reach of \emph{ab initio} calculations to heavier nuclei and additional decay modes, thereby strengthening the synergy between theory and experiment in the ongoing search for new physics.

		\end{abstract}
		
		
	\end{frontmatter}
	
	\newpage
	
	\thispagestyle{empty}
	\tableofcontents
	

	\newpage
	\section{Introduction}\label{intro}
The search for beyond the Standard Model (BSM) physics is one of the most pressing open questions in modern particle physics. This area of research is vital for our understanding of the universe, from the origin of matter and the universe's matter-antimatter asymmetry to the properties of dark matter and the behavior of fundamental forces at high energies. Searches of BSM physics have been central topics for some of the largest high-energy particle accelerators in the world, such as the Large Hadron Collider (LHC). On the other hand, low-energy nuclear beta decays have emerged as a powerful and complementary probe of BSM physics that can rival the sensitivity of large-scale particle accelerator experiments. The precise measurements of beta decay spectra and transition rates, often obtained with high accuracy in experiments involving radioactive nuclei, have already produced some of the most stringent constraints on BSM physics, and hold promise for further advances in our understanding of the universe's fundamental laws.

Within the broader landscape of precision physics programs \cite{Cirigliano:2013xha, naviliat2013prospects, VosWT2015, Falkowski2021,Brodeur2023nuclear}, the study of beta decay processes stands out as a particularly vibrant and significant area. In the Standard Model (SM), these decays are fundamentally driven by the exchange of a W boson, connecting light quark and lepton currents. Crucially, the strength of the interaction between the W boson and quarks within the SM is governed by the Cabibbo-Kobayashi-Maskawa (CKM) matrix \cite{Cabibbo:1963yz,Kobayashi:1973fv}. Consequently, beta decays offer a unique opportunity to precisely extract $V_{ud}$, the upper left element of this fundamental matrix. Indeed, this method currently represents the most precise means of determining $V_{ud}$, by a substantial margin \cite{10.1093/ptep/ptaa104}. Deviations from the unitarity of the CKM matrix would indicate the existence of BSM physics, and a renewed interest in this topic has followed the recent suggestion of a CKM unitarity violation at $\sim3 \sigma$ \cite{cirigliano2023scrutinizing}, which favors non-zero right-handed quark couplings (see, e.g. \cite{Cirigliano:2023nol} and references therein). 

 Furthermore, searches for BSM physics in beta decay test the left-handed vector minus axial-vector ($V - A$) structure of the weak interaction by looking for minuscule admixtures of non-SM currents. In an effective low-energy description, possible contributions from right-handed vector, scalar, and tensor currents--potentially arising from new heavy mediators--would modify decay observables relative to SM expectations. Precision measurements of beta-decay correlation coefficients (for example, electron–antineutrino correlations and angular/momentum correlations) and the beta-spectrum shape provide sensitivity to these exotic couplings \cite{VosWT2015,Falkowski2021,severijns2006tests, gonzalez2019new}. Together with radiative and recoil-order corrections, the results place stringent constraints on new physics scenarios such as exotic scalar or tensor interactions.

The extraction of fundamental parameters from precision beta-decay data depends on theoretical inputs for radiative corrections, isospin-breaking effects, and recoil-order terms. Given the precision that current measurements achieve, it is vital to have an accurate description of these higher-order corrections with quantified uncertainties to be able to distinguish between new physics and conventional nuclear physics effects.  
 Nuclear \emph{ab initio} methods--based on realistic nucleon-nucleon interactions--have made substantial progress in computing corrections to the beta decay for an expanding set of nuclei \cite{glick2022nuclear, Sargsyan_A8, King:2022zkz, LongfellowGSB2024, Gennari:2024sbn,Cirigliano:2024rfk,Cirigliano:2024msg,King:2025fph}. 
\emph{Ab initio} models are able to provide first-principles descriptions of nuclei and beta decay corrections with quantified uncertainties. The calculations from \emph{ab initio} methods are systematically improvable as the nucleon-nucleon interactions and many-body truncations are refined \cite{Hergert:2020bxy, Ekstrom:2022yea}. These advances strengthen the reliability of beta-decay observables used to test the SM and constrain BSM physics.

This review gives an overview of recent \emph{ab initio} calculations of beta decay radiative and recoil-order corrections and discusses their significance to the precision measurements that probe physics beyond the Standard Model. In the following sections, we detail the many-body methods that have been employed to perform \emph{ab initio} calculations of these corrections and briefly discuss their application span for future studies.  Moreover, we outline the formalism essential for deriving radiative corrections necessary for the precise extraction of the $V_{ud}$ matrix element of the CKM matrix, highlighting some of the most recent results. Furthermore, we examine advancements in computing nuclear recoil-order corrections to beta decays that need to be accounted for in high-precision experiments that probe BSM physics. We conclude with a summary and an outlook to future studies hence aiming to illuminate the path toward more accurate tests of fundamental theories.
 
	%
	%

\section{Many-body methods}\label{sec:many.body}

The landscape of {\it ab initio} nuclear theory~\cite{Hergert:2020bxy}, which has increasingly come to define many-body calculations that describe nuclei at the finest resolution scale possible and are systematically improvable~\cite{Ekstrom:2022yea}, has grown steadily over the last decade. While the fundamental theory of the strong interaction is QCD, and there has been progress in describing nuclei with lattice approaches~\cite{Beane:2010em}, the finest degrees of freedom to describe most few-body nuclei are protons and neutrons (collectively, nucleons). While there are a number of {\it ab initio} approaches in the literature, we will limit the scope of this
section to
methods based on the no-core configuration interaction (NCCI) and quantum Monte Carlo (QMC) approaches as they are the ones relevant for our subsequent discussion. 

Generically, one would like to solve the $A$-body Schr\"{o}dinger equation with a nuclear Hamiltonian that can be schematically represented as,
\begin{equation}
    H = \sum_i T_i + V_\mathrm{Coul} + \sum_{i<j} v_{ij} + \sum_{i<j<k} V_{ijk} + \ldots \, , \label{Eq:intH}
\end{equation}
where $T_i$ are the kinetic energies of the nucleons, $V_\mathrm{Coul}$ is the Coulomb interaction between the protons, and $v_{ij}$ and $V_{ijk}$ are two- and three-nucleon interactions, respectively. In principle, there could be four-nucleon and other many-nucleon forces; however, the studies reviewed in this work retain at most three-nucleon contributions. \red{There are a number of approaches to modeling the nuclear interaction, and we will describe these more in detail in Section~\ref{sec:hamiltonian}. For now, once a model for $H$ is adopted, one needs solve the Schr\"{o}dinger Equation,}
\begin{equation}
    H\ket{\Psi(J^{\pi};T,T_z)} = E\ket{\Psi(J^{\pi};T,T_z)}\, , \label{Eq:ShrEqn}
\end{equation}
to obtain the nuclear wave function $\Psi$ with the appropriate spin-parity $J^{\pi}$, isospin $T$, and isospin projection $T_z$ corresponding to an energy $E$. Using these wave functions, one is interested in evaluating transitions mediated by weak charge ($\rho$) and current ($\bfj$) operators. \red{We elaborate further on these operators in Section~\ref{sec:currents}.}

\subsection{No-core configuration interaction approaches}

\subsubsection{No-core shell model}
{\it Ab initio} configuration-interaction models, such as the no-core shell model (NCSM) and the symmetry-adapted no-sore shell model (SA-NCSM) adopt a complete orthonormal basis $\psi_i$, such that the expansion of $\Psi(J^{\pi};T,T_z)$ in terms of  unknown coefficients $d_k$,
$\Psi(J^{\pi};T,T_z) = \sum_{k} d_k \psi_k(\vec r_1, \vec r_2, \ldots, \vec r_A)$,
 renders Eq. (\ref{Eq:ShrEqn}) into a matrix eigenvalue equation,
\begin{equation}
\sum_{k'} H_{k k'} d_{k'} = E d_k, \label{eq:Eig}
\end{equation}
where the many-body Hamiltonian matrix elements are
$H_{k k'} =\braketop{\psi_k}{H}{\psi_{k'}} $ 
and are calculated for the given interaction (\ref{Eq:intH}). The many-body basis is a finite set of antisymmetrized products of  single-particle states (Slater determinants), referred to as a ``model space''. 
Typically, the single-particle states  of a three-dimensional spherical HO are used:
 $\phi_{\eta (\ell \half) j}(\mathbf{r};b)=R_{\eta \ell}(r;b)\mathcal{Y}_{(\ell \half)j}(\hat{r})=\sum_{m\sigma} C_{\ell m \half \sigma}^{j m_j} R_{\eta \ell}(r;b)Y_{\ell m}(\hat r)\chi_{\half \sigma}$, with radial wavefunctions $R_{\eta \ell}(r;b)$ and spin functions $\chi_{\half \sigma}$,
where $\eta=2n_r+\ell$ is the HO shell number, $\ell$ is coupled with spin-{1/2}~ 
to $j$, and  the oscillator length $b=\sqrt{ \frac{\hbar }{m\Omega} }$ with oscillator frequency $\hw$ (an additional quantum number $t_z$ is added to distinguish between protons and neutrons). 

\begin{figure}
	\centering
    \includegraphics[width=0.5\columnwidth]{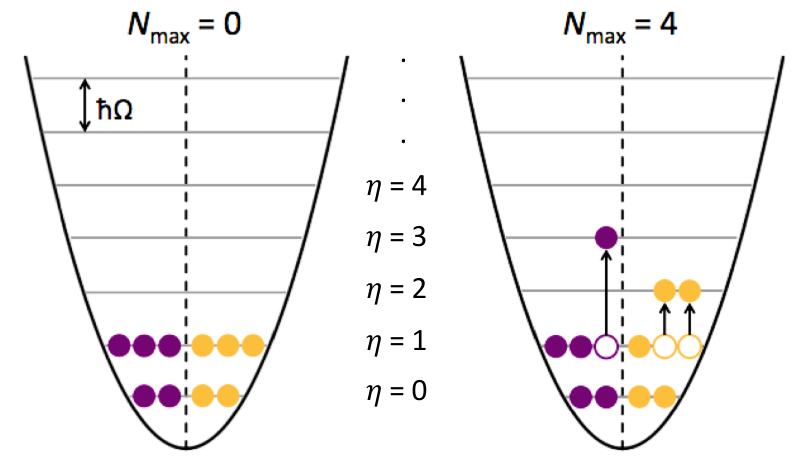}
	\caption{\Nmax = 0 and
\Nmax = 4 configurations in $^{10}$B in a standard particle-based NCSM (a single configuration is
shown in each case).\label{fig:Nmax}} 
\end{figure} 

In calculations, the model space is finite and is provided for given $N_{\max}$, 
which is the total HO excitations above the nuclear configuration of the lowest HO energy allowed by the Pauli principle (Fig. \ref{fig:Nmax}). The minimal model space for each nucleus corresponds to $N_{\max}$= 0, and it increases in steps of 2 for states sharing the same parity. Odd $N_{\max}$ values encompass states with the opposite parity. 
It is important that in the conventional NCSM with complete model spaces truncated by $N_{\rm max}$ (see the review \cite{BarrettNV13}) the center-of-mass (c.m.) wavefunction can be factored out exactly since the c.m. operator ($\hat N_{\rm c.m.}$) does not mix c.m. states with different HO excitations \cite{VERHAAR1960508}.
Such a model space allows for preservation of translational invariance of the nuclear self-bound system and provides solutions in terms of single-particle HO wavefunctions that are analytically known. 

\begin{figure}
	\centering
    \includegraphics[width=0.95\columnwidth]{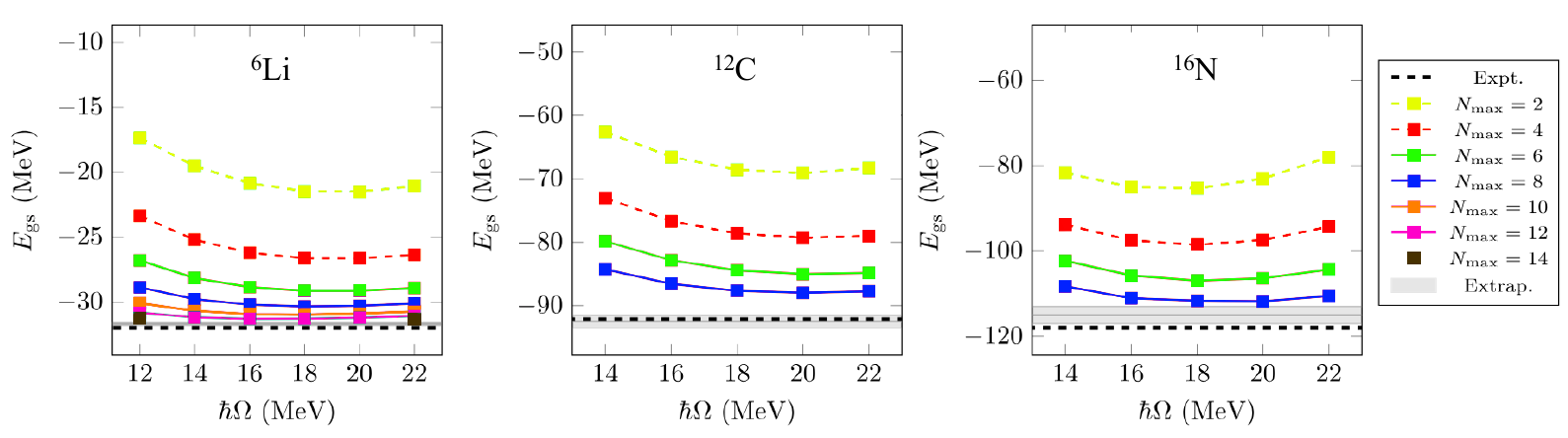}
	\caption{Ground-state energies of $^6$Li, $^{12}$C, and $^{16}$N obtained with the NN-N4LO + 3N$_\mathrm{lnl}^*$ interaction \cite{Kravvaris:2022eyf} with different HO frequencies. Data for the figure is taken from Ref. \cite{Jokiniemi:2024zdl}.\label{fig:NCSM_en}} 
\end{figure} 

With larger model spaces utilized in the  no-core shell-model theory, the eigensolutions  converge to the exact values, and the corresponding observables become independent from the \hw~basis parameter. Examples of ground state energy (eigenvalue) convergence with increasing \Nmax~are presented in Fig. \ref{fig:NCSM_en}. The eigenvectors obtained by solving Eq. (\ref{eq:Eig}) are then used to calculate various nuclear observables, such as radii, electromagnetic transition strengths and beta decay strengths.  Several methods have been used for extrapolating observables calculated in finite model spaces and specific \hw~values to the infinite model space value \cite{FurnstahlHP12, CoonAKVMV2012,WendtFPS15, MarisVS2009, Shanks55}. As \Nmax~increases, the number of configurations in the model space grows combinatorially. Consequently, the computational cost rises rapidly, typically limiting NCSM calculations to light nuclei or to relatively small model spaces. 

\red{A key strength of the NCSM is its controlled, variational convergence with increasing \Nmax~and the exact removal of center-of-mass contamination. When large model spaces are computationally available, and convergence is achieved, NCSM provides practically exact predictions of eigenvalues and eigenvectors of the many-body problem. This makes it well suited to obtain \emph{ab initio} matrix elements for light nuclei where complete model spaces are computationally attainable. Its principal limitation is the factorial growth of the configuration space with $A$ and \Nmax, which restricts practical calculations to light (p‑shell and lighter sd‑shell) systems or requires aggressive truncations for heavier nuclei. Another challenge is that  the spherical HO basis used in NCSM is poorly suited to describe exterior exponential wavefunction tails. Hence, long-range observables typically require larger model spaces to converge. Such a challenge usually does not arise in coordinate based methods such as QMC approaches described below. For beta‑decay observables NCSM calculations tend to deliver high‑quality Gamow–Teller and Fermi matrix elements (and consistent transition densities) with controlled extrapolation error for $A\leq16 - 18$, but systematic uncertainty grows when model-space truncation or \hw~dependence becomes significant. In addition, certain beta decay higher-order correction operators act on the long-range part of the wavefunction and may require larger \Nmax~calculations. However, correlations with other observables can help obtain robust predictions even in comparatively small model spaces \cite{Sargsyan_A8, CaprioMF2025I, CaprioFM2025II}. }

\subsubsection{Symmetry-adapted no-core shell model}

The NCSM many-body basis can be arranged into  \SU{3} subspaces of given deformation, and these can further be grouped into \SpR{3}$\supset$\SU{3} subspaces of given nuclear shape, referred to as symmetry-adapted (SA) bases. The symmetries of interest are \SU{3} and \SpR{3}.
The \SU{3} group is the exact symmetry  of the three-dimensional spherical HO, while the \SpR{3} (symplectic) symmetry represents the dynamical symmetry  of the three-dimensional spherical HO. Such near symmetries were first recognized by Bohr \& Mottelson (1975 Physics Nobel Prize) \cite{BohrMottelson69}, followed by Elliott's  seminal work   \cite{Elliott58,Elliott58b,ElliottH62}  and the microscopic no-core formulation by Rowe \& Rosensteel \cite{RosensteelR77,Rowe85}.

The SA-NCSM basis states are labeled schematically as
\begin{equation}
    \ket{\vec{\gamma},N(\lambda \; \mu) \kappa L; (S_pS_n)S; J},
\end{equation}
where $S_p$, $S_n$, and $S$ denote the proton, neutron, and total intrinsic spins, respectively, and $(\lambda~\mu)$ indicates a set of quantum numbers that label an SU(3) irreducible representation (or "irrep"). This means that each basis state is labeled according to  \SU{3}$_{(\lambda\,\mu)}\times$\SU{2}$_S$ by $S$ and $(\lambda\,\mu)$ quantum numbers with $\lambda=N_z-N_x$ and $\mu=N_x-N_y$, where $N_x+N_y+N_z=N$ is the total number of the HO quanta distributed in the $x$, $y$, and $z$ direction (in addition to other quantum numbers that are needed for complete labeling). The label $\kappa$ distinguishes multiple instances of the same orbital angular momentum $L$ within the parent irrep $(\lambda~ \mu)$. The angular momentum $L$ is coupled with the intrinsic spin $S$ to form the total angular momentum $J$. The symbol $\vec{\gamma}$ schematically denotes the additional quantum numbers required to specify the distribution of nucleons across the major HO shells, including their single-shell and inter-shell quantum numbers. The SU(3)  labels $(\lambda~ \mu)$ bring forward important information about nuclear shapes and deformation, according to an established mapping \cite{CastanosDL88, RosensteelR77, LeschberD87}. For example, (0 0), $(0 ~\mu)$ and $(\lambda~ 0)$
describe spherical, oblate and prolate deformation, respectively.  This basis organization allows traditional and complete model spaces to be augmented for large $N$ by a subset of the SA basis states (referred to ``SA model spaces''). Since small complete model spaces (up to a small $N$ value) are often sufficient to describe less deformed shapes, the additional SA basis states in high $N$ subspaces are exactly those needed for the full description of spatially expanded (very deformed or clustered) configurations \cite{LauneyDD16,DytrychLDRWRBB20}.

Similarly to NCSM, in SA-NCSM (see the review \cite{LauneyDD16}), the c.m. wavefunction can be factored out exactly since c.m. states are SU(3) scalar, and do not mix SU(3) subspaces of the SA-NCSM \cite{VERHAAR1960508,Hecht:1971xlg,Millener92}. It should be noted that while the SA basis is complete and related to the NCSM basis by a unitary transformation, the SA basis states are not constructed by this transformation; rather, they are generated via an efficient group-theoretical algorithm \cite{LangrDDLT19}.

\begin{figure}
	\centering
    \includegraphics[width=0.99\columnwidth]{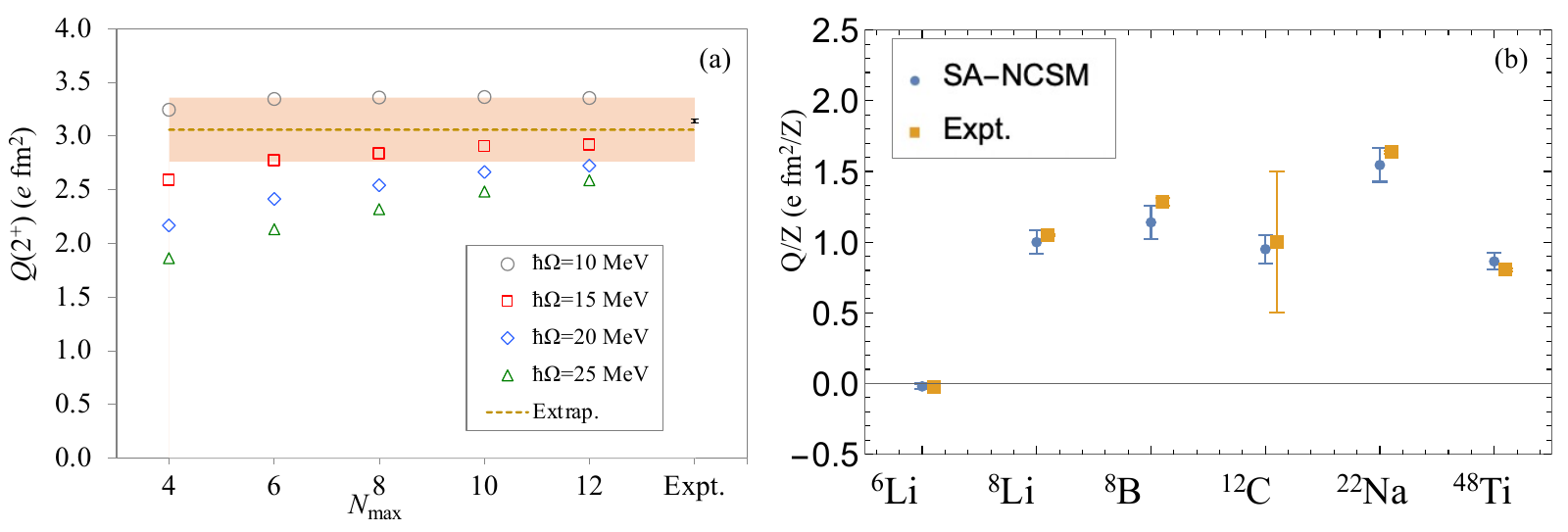}
	\caption{(a) Convergence of the $^8$Li ground state quadrupole moment vs. \Nmax~ calculated using SA-NCSM with different HO parameters ($\hbar\Omega$)  and compared to the experimental value from \cite{BorremansBBG2005Q2}. The dashed horizontal line with a band shows the infinite-space extrapolated value. Figure adapted from Ref. \cite{Sargsyan_A8} with permission. (b) Quadrupole moments of various nuclei divided by their proton numbers calculated using SA-NCSM and compared to the experimental values from Ref. \cite{STONE2016}. Uncertainties of some of the experimental values are smaller than the size of the marker. 
    \label{fig:SANCSM_Q}} 
\end{figure} 

The outstanding feature of SA-NCSM is its ability to express solutions in a physically meaningful basis that captures the collective behavior and symmetries underlying nuclear dynamics.  Large model spaces are needed to represent essential collective modes and to connect to continuum degrees of freedom within the interaction range. Beyond this range, SA-NCSM uses exact Coulomb eigenfunctions. The aim is to attain unprecedented accuracy across systems heavier than the lightest nuclei, while supplying nuclear-structure and reaction information for theory and experiment. Hence, SA-NCSM enables \emph{ab initio} calculations for spherical and deformed nuclei up to the Titanium region \cite{DytrychLDRWRBB20,DreyfussLESBDD20, Ruotsalainen19, PhysRevC.100.014322, LauneySOTANCP42018,LauneyMD_ARNPS21, baker24_physrevc.110.034605, burrows_2025} (Fig. \ref{fig:SANCSM_Q}), incorporating crucial correlations in the wavefunction, such as collective and clustering effects that are notoriously difficult to tackle in polynomial-scaling methods \cite{LauneyDSBD2020,LauneyMD_ARNPS21,launey2025ab}.

\red{The SA‑NCSM preserves the formal advantages of the NCSM (\emph{ab initio}, translational invariance, and convergence in HO spaces) while unitarily transforming the basis to the collective SU(3)/\SpR{3} basis. This transformation allows one to reduce the effective dimension needed to capture collective and clustering physics, extending practical calculations to heavier and more deformed systems. For beta‑decay applications, SA‑NCSM often yields improved descriptions of transition strengths that are sensitive to collective spatial structure and allows for expansion of the NCSM model space to even larger \Nmax~values. Challenges include selection of the most optimal model space for each nucleus and transformation of uncommon operators into the the SA basis, such as the ones needed for calculations of higher-order corrections to beta decays. Also, additional uncertainties in predictions may arise due to the selection of the model space.}

\subsection{Quantum Monte Carlo approaches}

\begin{figure}
	\centering
    \includegraphics[width=0.5\columnwidth]{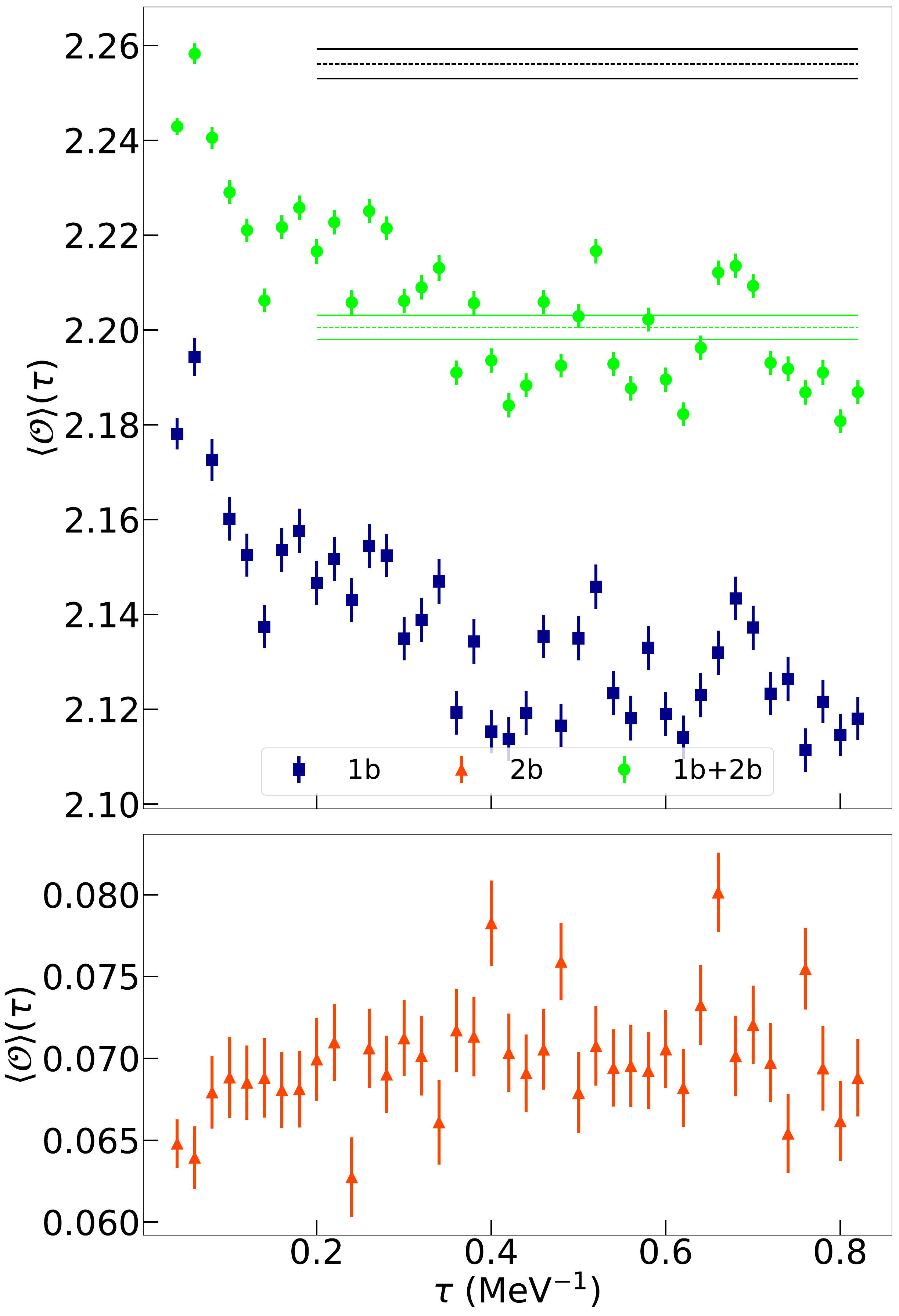}
	\caption{GFMC propagation of the $^6$He $\to$ $^6$Li Gamow-Teller $\beta$ decay transition matrix element for the NV2+3-Ia nuclear Hamiltonian. The one-body matrix element is shown in blue in the top panel, and is compared with the matrix element retaining one- and two-body transition operators shown in green. The two-body contribution is isolated in the bottom panel and plotted in orange-red color. The average GFMC estimate is shown by the green dotted line, with statistical uncertainty represented by the solid green lines. The average is compared to a variational Monte Carlo calculation with central value represented by the dashed black line, and statistical uncertainties shown with the solid black lines. Figure reprint with permission from Ref.~\cite{King:2020wmp}.} 
    \label{fig:gfmc.prop}
\end{figure} 

In this section, we outline QMC approaches to solving the many-body Schr\"{o}dinger Equation for strongly-correlated nucleons. These methods comprise a suite of stochastic approaches to solve the many-body problem non-perturbatively, retaining the full complexity of interparticle correlations generated by the the nuclear interaction. \red{These approaches are formulated in coordinate space, allowing for wave functions that can handle both short-range correlations and long-range effects like clustering on the same footing. The coordinate space formulation also allows for the use of interactions with a hard-core in momentum space, which are challenging for the NCCI approaches. However, in order to limit variance in the stochastic process, these methods typically employ local interactions (though there have been efforts to incorporate non-local terms in the potential~\cite{Lynn:2012fq,Curry:2023mkm}).} In the literature, there are reviews of both QMC approaches to study nuclei~\cite{Carlson:2014vla,Gandolfi:2020pbj}, and more recently their application to the study of electroweak processes in light nuclei~\cite{King:2024zbv}. Here, we aim to cover the salient features of the methods so that QMC results may be understood in the broader context of {\it ab initio} modeling of $\beta$ decays, and direct interested readers to the aforementioned reviews for detailed discussion. 

\subsubsection{Variational Monte Carlo}

The QMC approach in nuclei begins with variational Monte Carlo (VMC)~\cite{Wiringa:1991kp}, which requires an ansatz for the form of the trial wave function $\ket{\Psi_T}$ that will be optimized. The most generic form of a trial wave function factorizes the system into a model wave function $\ket{\Phi}$ that encodes the quantum numbers and long-range structure of the system, and a correlation operator $\hat{F}$ that reflects the impact of the nuclear interaction at short distances; {\it i.e.}, 
\begin{equation}
    \ket{\Psi_T} = \hat{F}\ket{\Phi}\, .
\end{equation}
The precise form of $\hat{F}$ and $\ket{\Phi}$ depend on which diffusion Monte Carlo approach is used to improve upon the VMC calculations, and will be discussed in the following two sections. Here, it is worth noting simply that the variational wave function is optimized by minimizing the energy expectation value,
\begin{equation}
    E_T = \frac{\mel{\Psi_T}{H}{\Psi_T}}{\inner{\Psi_T}{\Psi_T}} \leq E_0 \, ,
\end{equation}
where $E_0$ is the true energy of the ground state. This integral can be evaluated with Metropolis Monte Carlo techniques. In the following sections, we will refer to the optimized variational state as $\ket{\Psi_V}$. 

\subsubsection{Green's function Monte Carlo}

\red{While the VMC approach optimizes the energy expectation value and encodes correlations through the ansatz, it is not an exact calculation.} Green's function Monte Carlo is a diffusion Monte Carlo technique that improves upon the result from VMC. Generally, diffusion approaches work by solving the time-dependent Schr\"{o}dinger Equation,
\begin{equation}
i\frac{\partial}{\partial t}\ket{\Psi(t)} = (H-E_T)\ket{\Psi(t)}\, , \label{eq:tdse}
\end{equation}
making a change of variable to imaginary time $\tau=it$, where $E_T$ is an arbitrary energy offset. Eq.~\ref{eq:tdse} is now a diffusion equation in $\tau$, the solution of which is an exponential,
\begin{equation}
\ket{\Psi(\tau)} = e^{-(H-E_T)\tau}\ket{\Psi(0)}\, .    
\label{eq:diffusion}
\end{equation}
Taking our initial state to be $\ket{\Psi(0)}=\ket{\Psi_V}$, and noting that any state can be expanded in the exact eigenstates $\ket{\psi_i}$ of H as,
\begin{equation}
\ket{\Psi_V} = \sum_{i=0}^{\infty} c_i\ket{\psi_i}\, ,
\end{equation}
with complex coefficients $c_i$, we can obtain the true ground state $\ket{\psi_0}$ by choosing $E_T = E_0$ taking the limit,
\begin{equation}
\lim_{\tau\to\infty} e^{-(H-E_0)\tau}\ket{\Psi_T} \propto c_0\psi_0\, . \label{eq:tau.infty}
\end{equation}
In practice, this is achieved by acting on the state in several small imaginary time steps $\Delta\tau$,
\begin{equation}
\ket{\Psi(\tau)} = \left[ e^{-(H-E_0)\Delta\tau}\right]^n\ket{\Psi_V}\, .
\label{eq:short.time}
\end{equation}
When performing a GFMC calculation, one must consider the evolution of several sets of configurations $\bfR$, where $\bfR$ is used as short-hand to denote the set of $3A$ spatial coordinates and $2^AA!/(N!Z!)$ spin-isospin degrees of freedom. Projecting Eq.~\ref{eq:diffusion} for a short imaginary time onto configuration space, and using a completeness of the states $\ket{\bfR'}$, we obtain,
\begin{align}
\Psi(\bfR,\tau+\Delta\tau) &= \int d\bfR' \mel{\bfR}{e^{-(H-E_0)\Delta\tau}}{\bfR'}\Psi(\bfR',\tau)\\
&= \int d\bfR' G(\bfR,\bfR',\Delta\tau)\Psi(\bfR',\tau)\, , \label{eq:path.int}
\end{align}
where we define the short-time propagator $G(\bfR,\bfR',\Delta\tau)=\mel{\bfR}{e^{-(H-E_0)\Delta\tau}}{\bfR'}$, and denote the matrix element $\inner{\bfR}{\Psi(\tau)}$ as $\Psi(\bfR,\tau)$. The propagation in imaginary time can thus be performed by sampling $G(\bfR,\bfR',\Delta\tau)$ with Monte Carlo techniques to evolve the configurations to the next time step. Expectation values of operators must also be computed as a function of imaginary time, and averaged once convergence is reached. Figure~\ref{fig:gfmc.prop} shows this for the $^6$He Gamow-Teller $\beta$ decay matrix element computed in Ref.~\cite{King:2020wmp} with GFMC using the NV2+3-Ia chiral interaction~\cite{Piarulli:2016vel,Piarulli:2017dwd}. 

The form of the VMC trial wave function used in nuclear structure calculations has the form~\cite{Wiringa:2000gb} 
\begin{equation}
\ket{\Psi_T} = \mathcal{S}\prod_{i<j}^A \left[ 1 + U_{ij} + \sum_{k \neq i,j}^A U_{ijk} \right] \ket{\Psi_J}\, ,
\label{eq:trial}
\end{equation}
with $U_{ij}$ and $U_{ijk}$ being two- and three-body correlation operators designed to reflect the impact of the nuclear interaction at short distances, and $\mathcal{S}$ being symmetrization operator. The component $\Psi_J$ serves the role of the unperturbed model wave function and contains information about the symmetries and quantum numbers of the system. Details of the specific form of $\Psi_J$ are discussed in the review of Ref.~\cite{Carlson:2014vla}. \red{For $p$-shell nuclei, nucleons are placed into the $s$- and $p$-shell orbitals with additional correlation functions that encode the cluster structure of the system ({\it e.g}, $^6$Li has an $\alpha$-deuteron cluster structure).} The correlation operators $U_{ij}$ are spin and isospin dependent, and have a radial dependence that comes from solutions to a set of coupled screened two-body Schr\"{o}dinger equations for different spin-isospin ($ST$) channels of the $N\!N$ system. Instead, $U_{ijk}$ comes directly from the three-body force operators, and has variational parameters governing the strength of each term, as well as re-scaling the interparticle separations. Specific forms of the correlation operators can be found in Ref.~\cite{Pudliner:1997ck}.

\subsubsection{Auxiliary field diffusion Monte Carlo}

The spin-isospin basis of configuration space for the GFMC wave function grows exponentially with $A$, thus limiting its application to light nuclei. The auxiliary field diffusion Monte Carlo (AFDMC) approach \cite{Carlson:2014vla, Schmidt:1999lik, Lonardoni:2018nob} have been developed to circumvent this limitation. Like GFMC, one solves a diffusion equation in imaginary time. Now, however, one assumes a wave function with single particle representation, whose projection onto a spin state $S$ is given by~\cite{Lonardoni:2018nob},
\begin{equation}
    \inner{S}{\Psi} \propto \xi_{\alpha_1}(s_1)\xi_{\alpha_2}(s_2) \ldots \xi_{\alpha_A}(s_A)\, ,
\end{equation}
This wave function has the advantage that it scales polynomially in $A$, allowing for the computation of nuclei that are too large to access with GFMC. An important consideration is how operators quadratic in spin act on the wave function. Consider the example,
\begin{eqnarray} \nonumber
       \mel{S}{\bfsg_1\cdot\bfsg_2}{\Psi} &\propto& 2\xi_{\alpha_1}(s_2)\xi_{\alpha_2}(s_1) \ldots \xi_{\alpha_A}(s_A) - \xi_{\alpha_1}(s_1)\xi_{\alpha_2}(s_2) \ldots \xi_{\alpha_A}(s_A) \\ 
    &=& 2\inner{S'}{\Psi} - \inner{S}{\Psi}\, ,
\end{eqnarray}
which demonstrates that the basis is not closed to the action of quadratic operations. Linear operations only involve rotations of spins or isospins of single nucleons. Thus, to maintain the computational advantage of the single particle basis, one can linearize quadratic spin and isospin operators \red{in the propagator} with the Hubbard-Stratonovich transformation,
\begin{equation}
    e^{-\frac{\lambda}{2}\mathcal{O}^2} = \frac{1}{\sqrt{2\pi}}\int dx e^{-\frac{x^2}{2}}e^{x\sqrt{-\lambda}\mathcal{O}}\, ,
\end{equation}
where the $x$ are called auxiliary fields, and the integral can be performed via Monte Carlo sampling drawing from the probability distribution $P(x) = \exp[-x^2/2]$. Details of the linearization of the operators can be found in Ref.~\cite{Gandolfi:2020pbj}, and this transformation allows for the imaginary time propagation of up to quadratic spin-isospin operators in imaginary time when using the AFDMC basis. 

\red{The form of the VMC trial wave function used in the AFDMC calculation is~\cite{Lonardoni:2018nob,Gandolfi:2020pbj}, 
\begin{equation}
\ket{\Psi_T} = \left(1 + \sum_{i<j} U_{ij} + \sum_{i<j<k} U_{ijk}\right) \ket{\Phi_J}\, ,
\end{equation}
where $\Phi_J$ is the Jastrow wave function. For the AFDMC calculation, $\Phi_J$ has a simpler structure than $\Psi_J$ for the GFMC; namely, it is composed of two- and three-body scalar correlation operators acting on a sum of Slater determinants coupled to the appropriate quantum numbers of the system.}

\red{Both the GFMC and AFDMC wave functions add correlations to the best VMC wave function through the removal of spurious excited state contamination; however, due to the fermion sign problem, the variational wave functions can impact the results and quality of calculations. This is because one typically mitigates the sign problem through the use of constrained path approaches~\cite{Wiringa:2000gb} which use the trial wave function to guide the sampling by removing configurations that will contribute noise. This is done by discarding configurations which have zero overlap with $\Psi_T$. The GFMC wave function is well-suited for studying light nuclei $A\leq 12$ and the ansatz for the trial wave function allows it make predictions for nuclei where long-range cluster structures are important. The AFDMC method is able to address heavier nuclei than the GFMC method, but does not explicitly include clustering effects. Studying open-shell nuclei is also made more complicated by the need to include a large number of slater determinations in the trial wave function~\cite{Gandolfi:2020pbj}.}

\subsubsection{Mixed Estimates for off-diagonal transitions}

Ideally, for off-diagonal matrix elements of some operator $\mathcal{O}$ between some initial state $i$ and final state $f$, one would compute,
\begin{equation}
\avg{\mathcal{O}(\tau)} = \frac{\mel{\Psi^f(\tau)}{\mathcal{O}}{\Psi^i(\tau)}}{\sqrt{\inner{\Psi^f(\tau)}{\Psi^f(\tau)}}\sqrt{\inner{\Psi^i(\tau)}{\Psi^i(\tau)}}}\, ;    
\end{equation}
however, such a calculation would make analyzing a variety of matrix elements numericall quite costly, as each expectation value would require its own GFMC or AFDMC propagation. Because of this, the standard approach for off-diagonal matrix elements in diffusion Monte Carlo is to compute a mixed estimate, which is a suitable approximation given the quality of the VMC wave functions used in nuclear calculations. One assumes the GFMC state at each $\tau$ is a perturbation on top of the VMC wave function; {\it i.e.}, $\Psi(\tau) = \Psi_V + \delta \Psi$. Working to first order in $\delta \Psi$, we have that~\cite{Pervin:2007sc}, 
\begin{eqnarray}\nonumber
\avg{\mathcal{O}(\tau)} &\approx& \sqrt{\frac{\inner{\Psi^f_V}{\Psi^f(\tau)}}{\inner{\Psi^i_V}{\Psi^i_V}}}\frac{\mel{\Psi^f(\tau)}{\mathcal{O}}{\Psi^i_V}}{\inner{\Psi^f(\tau)}{\Psi^f_V}} + \sqrt{\frac{\inner{\Psi^i_V}{\Psi^i(\tau)}}{\inner{\Psi^f_V}{\Psi^f_V}}}\frac{\mel{\Psi^i(\tau)}{\mathcal{O}^{\dagger}}{\Psi^f_V}}{\inner{\Psi^i(\tau)}{\Psi^i_V}}\\ 
&&- \frac{1}{2} \left[ \sqrt{\frac{\inner{\Psi^f_V}{\Psi^f_V}}{\inner{\Psi^i_V}{\Psi^i_V}}}\frac{\mel{\Psi^f_V}{\mathcal{O}}{\Psi^i_V}}{\inner{\Psi^f_V}{\Psi^f_V}} + \sqrt{\frac{\inner{\Psi^i_V}{\Psi^i_V}}{\inner{\Psi^f_V}{\Psi^f_V}}}\frac{\mel{\Psi^i}{\mathcal{O}^{\dagger}}{\Psi^f}}{\inner{\Psi^i_V}{\Psi^i_V}}\right]\, . 
\label{eq:mixed.exact}
\end{eqnarray}
The quantities $\mel{\Psi_f(\tau)}{\mathcal{O}}{\Psi^i_V}/\inner{\Psi_f(\tau)}{\Psi^f_V}$ and $\mel{\Psi_i(\tau)}{\mathcal{O}^{\dagger}}{\Psi^f_V}/\inner{\Psi_i(\tau)}{\Psi^i_V}$ are obtained from the GFMC walk, and the term in brackets is the VMC matrix element for the off diagonal transition averaged over the VMC walks for both the initial and final state to improve the statistical uncertainty of the calculation. A further approximation typically made in the literature is as follows,
\begin{eqnarray}\nonumber
\avg{\mathcal{O}(\tau)} &\approx& \sqrt{\frac{\inner{\Psi^f_V}{\Psi^f_V}}{\inner{\Psi^i_V}{\Psi^i_V}}}\frac{\mel{\Psi^f(\tau)}{\mathcal{O}}{\Psi^i_V}}{\inner{\Psi^f(\tau)}{\Psi^f_V}} + \sqrt{\frac{\inner{\Psi^i_V}{\Psi^i_V}}{\inner{\Psi^f_V}{\Psi^f_V}}}\frac{\mel{\Psi^i(\tau)}{\mathcal{O}^{\dagger}}{\Psi^f_V}}{\inner{\Psi^i(\tau)}{\Psi^i_V}}\\ 
&&- \frac{1}{2} \left[ \sqrt{\frac{\inner{\Psi^f_V}{\Psi^f_V}}{\inner{\Psi^i_V}{\Psi^i_V}}}\frac{\mel{\Psi^f_V}{\mathcal{O}}{\Psi^i_V}}{\inner{\Psi^f_V}{\Psi^f_V}} + \sqrt{\frac{\inner{\Psi^i_V}{\Psi^i_V}}{\inner{\Psi^f_V}{\Psi^f_V}}}\frac{\mel{\Psi^i}{\mathcal{O}^{\dagger}}{\Psi^f}}{\inner{\Psi^i_V}{\Psi^i_V}}\right]\, . 
\label{eq:mixed.exact}
\end{eqnarray}

\subsection{{Nuclear Hamiltonians and current operators}} \label{sec:hamiltonian}

\red{So far, we have reviewed the approaches to solve the many-body Schr\"{o}dinger Equation for light nuclei; however, another important ingredient in these calculations are the models of the two- and three-nucleon interactions $v_{ij}$ and $V_{ijk}$, respectively that enter the Hamiltonian in Eq.~\ref{Eq:intH}.  The interaction between nucleons is not fundamental, and thus one must model these potentials. Historically, the community adopted phenomenological interactions to model the nuclear interaction-- such as, for instance, the Argonne $v_{18}$ (AV18)~\cite{WiringaSS95} and CD-Bonn potentials~\cite{Machleidt01}-- which proved reasonably successful to understand a variety of phenomena. These approaches, despite their strong predictive capabilities, are not systematically improvable. In more recent years, approaches based on chiral effective field theory ($\chi$EFT) have come to the forefront. It is beyond the scope of this review to cover the full details of this approach, and interested readers are directed to the reviews of Refs.~\cite{BedaqueVKolck02,EntemM03,Epelbaum:2014sza,RevModPhys.92.025004}. The salient feature of $\chi$EFT is that the separation of scales between low-energy QCD ($Q \approx m_{\pi}$) and the scale of heavy meson ($\Lambda_\chi \approx 1$ GeV) allows one to derive a nuclear interaction that can be expanded in powers of $\varepsilon_{\chi} = Q/\Lambda_{\chi}$. The underlying QCD is subsumed in short-range contact interactions whose strengths are parameterized by ``low-energy constants" (LECs) which must be fit to data in order to obtain a nuclear interaction that one can use in many-body calculations. Several models of the nuclear interaction have been obtained with various procedures to fit the LECs. 
For a good overview of the state of the field, we direct interested readers to the recent review in Refs.~\cite{Machleidt:2024bwl}. While similarly successful to the phenomenological approach, it is worth mentioning that one must adopt a particular power counting scheme for the $\chi$EFT expansion. Determining the appropriate power counting remains an open question in low-energy nuclear theory~\cite{Kaplan:1998tg,vanKolck:2020plz}. Thus, the $\chi$EFT approach is {\it in principle} systematically improvable and the works reviewed here comprise the best effort of the community toward {\it ab initio} modeling of $\beta$ decays. Ongoing research in low-energy nuclear interactions aims at advancing the field toward a truly systematically improvable approach that builds upon the efforts that we detail in this article. In the remainder of this section, we provide details on some of the nuclear Hamiltonians adopted for studies of $\beta$ decay in light nuclei, separated by  local and non-local potential models. We also briefly discuss the notion of electroweak transition operators.}

\subsubsection{{Local interactions}}

\red{Because it makes the sampling more efficient~\cite{Carlson:2014vla}, one typically adopts local interactions in QMC calculations that depend on interparticle spacings $\bfr = \bfr_i - \bfr_j$ between particles $i$ and $j$. Historically, the AV18~\cite{Wiringa:1994wb} supplemented by three-nucleon forces was the typical interaction used in QMC approaches. More recently, interactions developed in $\chi$EFT have become available. In this section, we will review the salient features of these potentials.}

\red{The AV18 is a phenomenological model of the nuclear interaction. Included in the potential are the dominant one-pion exchange (OPE) contributions, as well as intermediate-range contributions that approximate two-pion exchange (TPE) with radial dependencies that assume this exchange is dominated by the tensor Yukawa contribution. Finally, there is a short-range potential whose shape is given by a Woods-Saxon form~\cite{Wiringa:1994wb}. The potential can be subdivided into a charge-independent potential with 14 operators,
and additional 4 terms that break charge-independence and charge-symmetry.
The radial functions for the intermediate- and short-range potentials contain 42 parameters which were constrained with 4301 $np$ and $pp$ scattering data from the Nijmegen partial wave analysis~\cite{Stoks:1993tb} with a $\chi^2$/datum of ${\sim}1.1$. The AV18 model is considered a ``hard-core'' interaction, as the maximum value of the central potential takes a value ${\sim}2$ GeV~\cite{Carlson:2014vla}. It is typical to supplement the AV18 with either the Urbana X (UX)~\cite{Wiringa:2014} or Illinois-7 (IL7)~\cite{Carlson:2014vla} three-body interactions. The UX is a phenomenological model of the three-nucleon force that hybridizes two other models; namely, the IL7 and Urbana IX (UIX) models~\cite{Carlson:2014vla}. The UX supplements the long-range two-pion P-wave and central S-wave repulsion of the UIX with a two-pion S-wave term, taking the strengths of all three of these terms from the IL7 parametrization. The IL7 contains three-pion ring diagrams involving one or two intermediate $\Delta$-isobars, in addition to the terms of the UX model, and is fit to ground state and low-lying excitation energies of $A\leq 10$ nuclei. Because of this, the AV18+IL7 in combination with QMC methods makes robust predictions for the properties of light nuclei~\cite{Carlson:2014vla}. The hard-core of the interaction, however, makes it difficult to use in configuration interaction approaches due to the need for large model spaces in order to obtain convergence~\cite{Hergert:2020bxy}. Further, it is not systematically improvable in the same way as $\chi$EFT as there is no prescription to determine which terms are left out of the theory and what would be their estimated impact on observables.}

\red{With the advent of softer chiral interactions that made {\it ab initio} computations possible for heavier nuclei~\cite{Hergert:2020bxy}, an effort was made to develop local chiral interactions that would have a softer core than the AV18, be systematically improvable with the $\chi$EFT expansion, and still be efficient for QMC calculations. One such model that was developed with this goal in mind was the Norfolk model of the nuclear interaction. It is composed of a two-body interaction (NV2) and three-body interaction (NV3), and collectively it is denoted as NV2+3. The NV2 potential was derived by first developing a minimally nonlocal two-body force retaining nucleon and $\Delta$ isobar degrees of freedom by means of Fierz transformations up to $\mathcal{O}(\varepsilon_\chi^3)$~\cite{Piarulli:2014bda} in the power counting prescription of Weinberg~\cite{Weinberg:1991um}. The model phenomenologically includes terms at the next order in the expansion. While non-localities arise at $\mathcal{O}(\varepsilon_{\chi}^4)$, only local terms at this order necessary to provide a good description of nucleon-nucleon scattering were retained to form the NV2 model~\cite{Piarulli:2016vel}. This force is supplemented by a $\mathcal{O}(\varepsilon_\chi^3)$ $\Delta$-full three-body force, the NV3, based on terms first derived by van Kolck {\it et al.}~\cite{vanKolck:1994yi} and Epelbaum {\it et al.}~\cite{Epelbaum:2002vt}. In total, the full NV2+3 has 28 unknown low-energy constants governing the strength of 26 two-nucleon contacts and two three-nucleon contacts.}

\red{Various fitting schemes were used to obtain eight NV2+3 interactions; namely, choices for which two-nucleon data to fit, how to regularize singularities in the long-range terms of the potential, and how to fit the three-body force. There are two classes (I and II) of the NV2 that differ only in the range of energy over which they are fitted-- class I up to 125 MeV, and class II up to 200 MeV. For each class, two 
cutoffs 
were chosen to smear out the contact interactions and remove singularities in the pion-exchange terms. Two models correspond to ``soft'' cutoffs 
-- models NV2-Ia and NV2-IIa-- and two more to ``hard'' cutoffs 
-- models NV2-Ib and NV2-IIb. These correspond to cutoffs of roughly ${\sim} 500$ MeV and ${\sim} 550$ MeV in momentum space. Class I (II) models were fit to about 2700 (3700) $np$ and $pp$ scattering data with a $\chi^2$/datum $\lesssim 1.1$ ($\lesssim 1.4$)~\cite{Piarulli:2016vel,Piarulli:2014bda}. The NV3 LECs were fit with either strong interaction data only (the trinucleons energies and the $nd$ doublet scattering length)~\cite{Piarulli:2017dwd}, and no additional annotations are used to denote these model classes. Another choice for fitting the NV3 was to use strong and weak data simultaneously (the trinucleon energies and the tritium Gamow-Teller $\beta$-decay matrix element)~\cite{Baroni:2018fdn}, and these models are denoted with a star ($*$). These interactions are usually employed in GFMC calculations of light nuclei and have successfully reproduced static~\cite{Piarulli:2017dwd} and electroweak~\cite{King:2024zbv} properties of light nuclei.}

\red{Another model of local interaction was developed in Refs.~\cite{Gezerlis:2013ipa,Gezerlis:2014zia,Tews:2015ufa,Lynn:2015jua}. The two-nucleon interaction model was derived in $\chi$EFT with pion and nucleon degrees of freedom, and also followed the Weinberg power-counting. Terms are retained up to $\mathcal{O}(\varepsilon_{\chi}^3)$, including the leading momentum-independent charge-independence- and charge-symmetry-breaking contacts~\cite{Gezerlis:2013ipa,Gezerlis:2014zia}. Isospin-symmetry breaking effects are also included through modifications to pion-exchange terms accounting for the pion mass splitting. The two models of local interactions described in this section differ in their choices of regulator. For the NV2 potential, the regulator functions are given by~\cite{Piarulli:2016vel} and take both a long- and short-range cutoff, while for the interactions of Gezerlis {\it el al}, the regulator in Ref.~\cite{Gezerlis:2014zia} takes a single cutoff parameter.
With the recommended coordinate space cutoff parameter values, the models have cutoffs roughly corresponding to the range ${\sim}400 - {\sim}500$ MeV in momentum space~\cite{Lynn:2017fxg}. 
The models were fit to the Nijmegen partial wave analysis~\cite{Stoks:1993tb} at energies
up to 150 MeV.
This potential was supplemented with a three-nucleon force up to the same order $\mathcal{O}(\varepsilon_{\chi}^3)$ in Weingberg's power-counting~\cite{Tews:2015ufa}, introducing two additional LECs, similarly to the NV3. These LECs were fit using $n-\alpha$ scattering $P$-wave phase shifts and the $^4$He binding energy~\cite{Lynn:2015jua}. Because of Fierz-rearrangement ambiguities, different operator structures for the contact term corresponding to the LEC $c_E$ could be chosen, and three different structures were fit~\cite{Lynn:2015jua}. The two that commonly appear in the literature are the $E\mathbb{1}$ and $E\tau$ models, corresponding to the structures $\delta(r_{ij})\delta(r_{kj})$ and $\bfta_i\cdot\bfta_k\delta(r_{ij})\delta(r_{kj})$, respectively, for particles $i$, $j$, and $k$. The delta functions are of course smeared using the local regulator. This interaction model is typically employed in AFDMC calculations, reproducing the static properties of nuclei~\cite{Lynn:2017fxg,Lonardoni:2018nob} while also being used for studies of nuclear and neutron matter~\cite{Buraczynski:2015oaa,Buraczynski:2016jia,Buraczynski:2019jyr,Riz:2020ude,Tews:2018kmu,Lonardoni:2019ypg}.}

\subsubsection{{Non-local interactions}}
In configuration interaction models non-local interactions are typically used. However, local interactions constructed for QMC applications have also been employed, especially in earlier studies. The difficulty with using the local interactions is that they exhibit a slow model-space convergence in configuration-space calculations because they tend to require a significant repulsive core at short distance to describe nucleon-nucleon scattering data. Some of the most commonly used non-local NN potentials--also utilized by the studies in this work--are briefly reviewed below. 

JSP16 \cite{shirokov2004nucleon, Shirokov2010nn} is a nonlocal, phase-equivalent J-matrix inverse-scattering NN interaction constructed in an ``ab exitu'' approach by fitting NN data plus selected properties of light nuclei (via NCSM calculations). JISP16 is soft and engineered to reduce the need for explicit three-nucleon forces and it has been widely used in NCSM studies. It is expected to work only for nuclei with masses $A<16$ as systematic overbinding occurs for heavier masses. This interaction’s construction differs conceptually from EFT-based potentials (no systematic EFT expansion or truncation-error protocol). It has been shown to reproduce experimental binding energies and excitation spectra in light nuclei with quick convergence \cite{Shirokov2010nn}. 

NNLOopt \cite{Ekstrom13} is a chiral NN interaction optimized at next-to-next-to-leading order (NNLO) to NN data with an emphasis on a high-quality description of NN elastic scattering up to about 125 MeV and few-body observables. It uses a regulator cutoff of 500 MeV. The optimization produces a comparatively soft chiral NN force for which explicit three-nucleon contributions in $A=3,4$ systems are reduced relative to earlier chiral parametrizations. It is therefore often used in \emph{ab initio} configuration interaction studies where reduced 3N effects simplify applications. It has yielded energy spectra and structure and reactions observables in close agreement with experimental data (see \cite{launey2025ab} and references therein).

NNLOsat \cite{NNLOsat2015} is a chiral interaction in which two- and three-nucleon LECs were optimized simultaneously to NN scattering data and selected binding energies and charge radii (few-body systems and some medium-mass isotopes). It uses a regulator cutoff of 450 MeV and the joint fit yields improved reproduction of nuclear radii and saturation properties compared with NN-only fits, at the cost of a different balance of long- and short-range contributions. NNLOsat is therefore frequently chosen for studies that require realistic radii and bulk saturation (e.g., medium-mass structure and response functions). However, this interaction is only fitted
to low-energy scattering data, up to 35 MeV, and it has a poor reproduction of p-wave NN phase shifts.

N3LO-EM potential \cite{EntemM03} is a nonlocal, momentum-space chiral NN interaction constructed at fourth order in the chiral expansion that includes one- and multi-pion exchanges plus a full set of contact terms fit to NN phase shifts (yielding $\chi^2$/datum near unity comparable to phenomenological forces). It employs spectral-function regularization and a typical cutoff of around $500$ MeV, fitting NN data up to several hundred MeV. Moderate 3N contributions are needed to reproduce experimental binding energies. This potential is widely used as a baseline NN interaction in many \emph{ab initio} studies typically after being evolved via Similarity Renormalization Group (SRG) methods to achieve faster convergence of observables. We note that after the SRG evolution, the operators of the observables should also be evolved to be compatible with the obtained wavefunctions. This step is occasionally omitted in some studies due to its technical difficulty with a varying effect on different observables. 

The Entem–Machleidt–Nosyk N4LO potential \cite{Entem:2017gor} provides a family of nonlocal, momentum-space chiral NN interactions constructed consistently up to fifth order (N4LO) with spectral-function regularization and long-range $\pi$N LECs. The NN potentials are fit to the world NN data below pion-production
threshold of the year of 2016. The potential of the highest order (N4LO) reproduces the world
NN data with the $\chi^2$/datum of 1.15. Regulator cutoffs of 450, 500 and 550 MeV are used. It requires considerable 3N forces to reproduce binding energies and is typically SRG evolved to achieve faster convergence of observables (see, \eg~\cite{Gennari:2024sbn}). 

Because none of the NN interactions described above were fitted to beta-decay data (except for some 3N forces), their predictive performance for weak observables cannot be assumed \emph{a priori}. For this reason, it is desirable to perform calculations with multiple interactions (with different regulator choices) to assess interaction dependence and quantify systematic uncertainties.

\subsubsection{{Weak charge and current operators}}\label{sec:currents}

\red{The interaction of external probes with individual nucleons and pairs of correlated nucleons is decomposed into one- and  two-body charge and current operators,
\begin{eqnarray}
\rho    &=& \sum_i {\rho}_i({\bf q}) + \sum_{i<j} {\rho}_{ij}({\bf q}) + \dots\ , \\
\nonumber
{\bf j} &=&  \sum_i {\bf j}_i({\bf q}) + \sum_{i<j} {\bf j}_{ij}({\bf q}) +\dots  \ ,
\end{eqnarray}
where ${\bf q}$ is the momentum transferred to the nucleus. When studying weak processes in nuclei, such as $\beta$ decay, there are both vector and axial current contributions. The single nucleon charge and current operators are obtained from a non-relativistic reduction, and the leading-order vector charge $\rho_V$ and axial current $\bfj_A$ correspond to the familiar Fermi and Gamow-Teller operators, respectively; namely, 
\begin{eqnarray}
    \rho_V(\bfq) &=& \frac{g_V}{2}\sum_i \tau_{a,i}e^{i\bfq\cdot\bfr_i} \, , \\
    \bfj_{A,a}(\bfq) &=& -\frac{g_A}{2}\sum_i \bfsg_{i}\tau_{a,i}e^{i\bfq\cdot\bfr_i} \, ,
\end{eqnarray} 
where $g_V=1$, $g_A=1.2751$~\cite{ParticleDataGroup:2024cfk} is the nucleon axial coupling constant, and $a$ indicates the isospin index. The charge-changing weak current can be constructed by combining the $x$ and $y$ isospin components for an operator $\mathcal{O}_a$ as $\mathcal{O}_{\pm} = \mathcal{O}_x \pm i\mathcal{O}_y$, where $+$ ($-$) indicates an isospin raising (lowering) operators. }

\red{Studies based solely on single nucleon operators are not always sufficient to describe experimental data. For instance, two-body currents are necessary to describe electromagnetic data at both low- and high-energy regimes, including measured magnetic moments~\cite{Pastore:2012rp,Chambers-Wall:2024uhq,Miyagi:2023zvv,Martin:2023dhl} and electron-nucleus cross sections~\cite{Carlson:1997qn,Lovato:2016}. These contributions have also been shown to be important for weak processes, such as Gamow-Teller $\beta$ decay matrix elements~\cite{Gysbers:2019uyb,Pastore:2017uwc,King:2020wmp} and muon capture rates~\cite{Marcucci:2011jm,King:2021jdb,Jokiniemi:2021qqg,Jokiniemi:2024zdl,Gnech:2023mvb}. These corrections account for processes in which external probes couple to pairs of correlated nucleons, and while they have been modeled with phenomenological approaches, recent calculations typically focus on currents derived in $\chi$EFT. }

\red{The pioneering work on vector and axial transition operators in $\chi$EFT was that of Park {\it et al.}~\cite{Park:1993jf,Park:1995pn}. More recently, two derivations based on time-ordered perturbation theory have appeared in the literature. One derivation is from the so-called Bochum-Bonn group~\cite{Kolling:2009iq,Kolling:2011mt,Krebs:2016rqz,Krebs:2019}, while the other is from what has come to be called the JLab-Pisa group~\cite{Pastore:2009is,Pastore:2011ip,Baroni:2015uza}. The former derivation uses the unitary transformation approach, which aims to find a unitary operator that decouples a purely nucleonic subspace from the rest of the pion-nucleon Fock space~\cite{Okubo:1954}. Differences between the two approaches arise in the treatment of irreducible diagrams. While the two approaches led to agreement when calculating box diagrams in the vector charge and current~\cite{Pastore:2011ip}, the groups found differences in box diagrams entering loop contributions to the axial current operator~\cite{Baroni:2018fdn,Krebs:2020rms} and the numerical impact of this was explored in Ref.~\cite{Baroni:2018fdn}. Both formulations are used in the study of electroweak properties of light nuclei, and interested readers are directed to the relevant references for more details on the expressions of these operators.}

\section{Radiative Corrections to superallowed beta decays\label{sec:RCsuperallowed}}

\subsection{Background}

Beta decays provide a perfect avenue to test predictions of SM at low energies~\cite{Brodeur2023nuclear}. As a most straightforward application, by measuring the decay lifetime one can access the parameter $V_{ud}$, which is the upper-left element of the Cabibbo-Kobayashi-Maskawa (CKM) matrix~\cite{Cabibbo:1963yz,Kobayashi:1973fv}. This allows us to test a major SM prediction, the so-called ``first-row CKM unitarity'':
\begin{equation}
    |V_{ud}|^2+|V_{us}|^2+|V_{ub}|^2=1~.
\end{equation}
A precision level of $10^{-4}$ in such a test will constrain new physics at multi-TeV scale, which is competitive to high-energy experiments at colliders. 

$V_{ud}$ can be extracted from beta decays of pion, free neutron, and nuclei. For pion which is theoretically the cleanest with recent lattice calculations of the electromagnetic radiative corrections (RC)~\cite{Feng:2020zdc,Yoo:2023gln}, the bottleneck is the large experimental uncertainty in the $\pi_{e3}$ branching ratio~\cite{Pocanic:2003pf}, although improvements are anticipated through the future PIONEER experiment~\cite{PIONEER:2022yag}. For free neutron decay, recent years have seen tremendous improvements in the study of RC~\cite{Seng:2018yzq,Seng:2018qru,Seng:2020wjq,Czarnecki:2019mwq,Shiells:2020fqp,Hayen:2020cxh,Ma:2023kfr,Cirigliano:2023fnz,VanderGriend:2025mdc,Cao:2025lrw,Moretti:2025qxt} based on dispersion relation, lattice QCD and effective field theory, but again the bottlenecks have been from the experimental side, including discrepancies in the neutron lifetime (see, e.g. Ref.\cite{Fuwa:2024cdf} and references therein) and axial coupling~\cite{Beck:2019xye,Beck:2023hnt,Markisch:2018ndu} measurements. 

At present, superallowed $0^+\rightarrow 0^+$ beta decays of $T=1$ nuclei provide the best avenue to extract $V_{ud}$. There are two major benefits in this channel:
(1) it is a pure Fermi transition, so the decay amplitude is largely fixed by isospin symmetry modulo small corrections (which is the aim of our study); (2)
currently, there are 23 measured transitions, 15 of which have partial half-life precision better than $0.23\%$~\cite{Hardy:2020qwl}, namely,  \red{${}^{10}\text{C}\rightarrow{}^{10}\text{B}$, ${}^{14}\text{O}\rightarrow{}^{14}\text{N}$, ${}^{22}\text{Mg}\rightarrow{}^{22}\text{Na}$, ${}^{26}\text{Si}\rightarrow{}^{26m}\text{Al}$, ${}^{34}\text{Ar}\rightarrow{}^{34}\text{Cl}$, ${}^{38}\text{Ca}\rightarrow{}^{38m}\text{K}$, ${}^{26m}\text{Al}\rightarrow{}^{26}\text{Mg}$, ${}^{34}\text{Cl}\rightarrow{}^{34}\text{S}$, ${}^{38m}\text{K}\rightarrow{}^{38}\text{Ar}$, ${}^{42}\text{Sc}\rightarrow{}^{42}\text{Ca}$, ${}^{46}\text{V}\rightarrow{}^{46}\text{Ti}$, ${}^{50}\text{Mn}\rightarrow{}^{50}\text{Cr}$, ${}^{54}\text{Co}\rightarrow{}^{54}\text{Fe}$, ${}^{62}\text{Ga}\rightarrow{}^{62}\text{Zn}$, and ${}^{74}\text{Rb}\rightarrow{}^{74}\text{Kr}$.} Averaging over these transitions largely reduces the experimental uncertainty. As a consequence, the error of $V_{ud}$ determined from this decay mode is dominated by theory uncertainties instead of experiment, and \textit{ab initio} calculations play a key role in reducing such uncertainties. 

\red{In our opinion, three among the 15 transitions listed above play a special role and are worth extra investment of theoretical efforts: ${}^{26m}\text{Al}\rightarrow{}^{26}\text{Mg}$ which has the best measured $ft$-value and may alone determine $V_{ud}$ as precise as the combined analysis of all superallowed transitions~\cite{Gorchtein:2025wli}, as well as the two lightest superallowed transitions  ${}^{10}\text{C}\rightarrow{}^{10}\text{B}$,  ${}^{14}\text{O}\rightarrow{}^{14}\text{N}$ that are most sensitive to BSM-induced scalar interactions~\cite{Hardy:2020qwl}. These two light transitions fit well into the context of this review and will be discussed in detail.}

\subsection{Theory framework}

$V_{ud}$ can be extracted from the superallowed decay half life $t$ through the following master formula~\cite{Hardy:2020qwl}:
\begin{equation}\label{eq:Vud}
|V_{ud}|_{0^+}^2=\frac{\pi^3\ln 2}{G_F^2m_e^5\mathcal{F}t(1+\Delta_R^V)}= \frac{2984.431(3)\,\text{s}}{\mathcal{F}t(1+\Delta_R^V)}~,
\end{equation}
where $\Delta_R^V$ is a nucleus-independent RC which is also present in the free neutron decay. All the SM theory inputs that depend on the decaying nuclear system reside in the following quantity:
\begin{equation}
    \mathcal{F}\equiv f(1+\delta_r')(1+\delta_\text{NS}-\delta_\text{C})~,\label{eq:bigF}
\end{equation}
where $f$ is the ``statistical rate function'' that comes from the phase space integral of the tree-level squared amplitude corrected by various leading \red{nucleus-dependent SM effects such as Fermi function and the nuclear weak form factor~\cite{Seng:2023cgl}, as well as atomic effects such as screening corrections~\cite{Rose:1936zz} and atomic overlap corrections~\cite{Hardy:2008gy}}. $\delta_r'$ is an ``outer'' RC which is largely independent of the detailed nuclear structure, $\delta_\text{C}$ is the isospin-symmetry-breaking (ISB) correction to the Fermi matrix element, and $\delta_\text{NS}$ is the nuclear structure (NS)-dependent RC. It is interesting to note that, since $V_{ud}$ is a nucleus-independent quantity, the product $\mathcal{F}t$ should also be nucleus-independent if SM is correct, despite that $\mathcal{F}$ and $t$ are separately nucleus-dependent. 

\begin{figure}
	\centering
    \includegraphics[width=0.3\columnwidth]{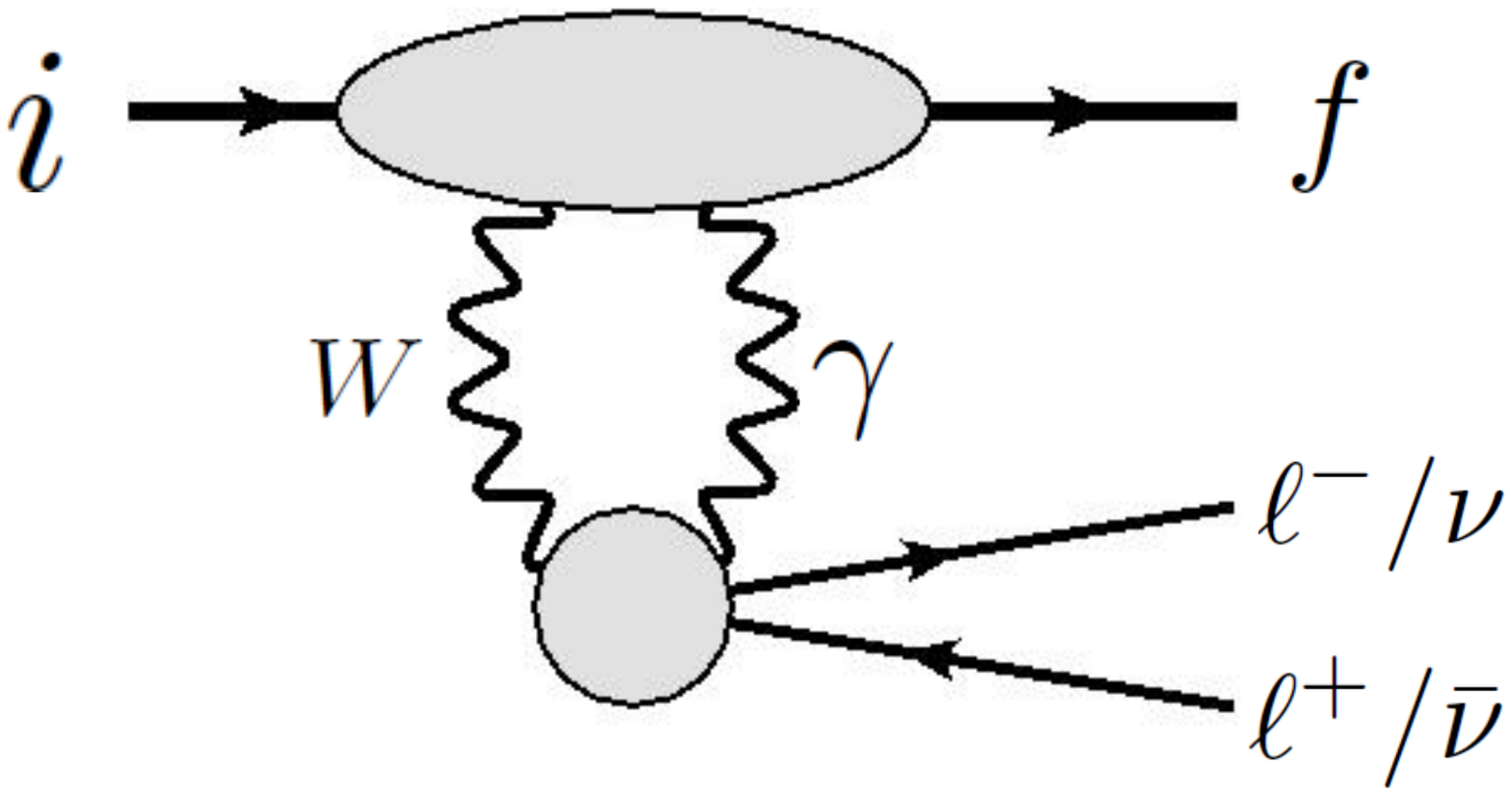}
   \includegraphics[width=0.5\columnwidth]{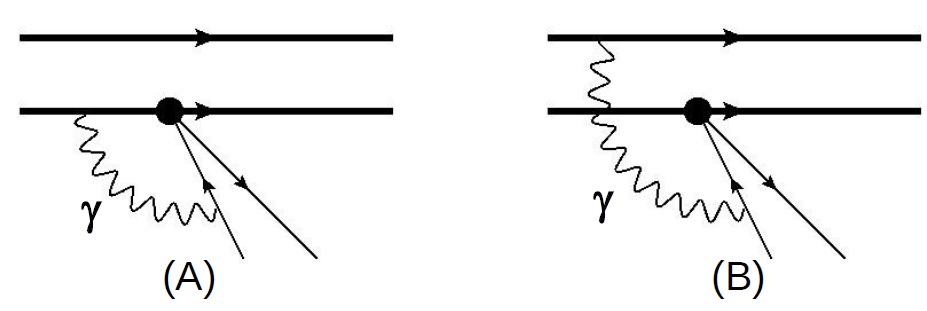}
	\caption{First diagram: A generic $\gamma W$-box diagram for $\beta^\pm$-decay. Second and third diagram: the ``traditional'' picture for $\delta_\text{NS}$.\label{fig:deltaNStrad}} 
\end{figure} 

$\delta_\text{NS}$ arises from the nucleus-dependent part of the so-called ``nuclear $\gamma W$-box diagram'', given by the first diagram in Fig. \ref{fig:deltaNStrad}. To be more specific, it is contributed by the ``axial'' box diagram, where the $W$-boson is coupled to the axial component of the nuclear weak current.  
The ``traditional'' picture to understand the origin of $\delta_\text{NS}$, is that it comes from the second and third diagrams in Fig. \ref{fig:deltaNStrad}:
\begin{itemize}
    \item Type A: Both the weak and the electromagnetic (EM) vertices are acting on the same nucleon, but it is a bound nucleon instead of a free nucleon, so the weak and EM couplings are shfited from the ``free'' values to the ``quenched'' values, that are numerically smaller.
    \item Type B: The weak and EM vertices can act on two different nucleons, so this is intrinsically a nucleus-dependent contribution.
\end{itemize}
Within this picture, one can write:
\begin{equation}
    \delta_\text{NS}=\delta_\text{NS,A}+\delta_\text{NS,B}.
\end{equation}
 Both of these terms have been calculated using shell model~\cite{Jaus:1989dh,Barker:1991tw,Towner:1992xm,Towner:1994mw}.

The recent dispersive representation of the nuclear $\gamma W$-box diagram~\cite{Seng:2018qru} showed that the traditional picture above is an incomplete description. In the dispersive approach, the integrand of the $\gamma W$ box diagram depends on the absorption spectrum of the target, and therefore $\delta_\text{NS}$ arises essentially from the difference between the nuclear and the nucleon absorption spectrum. Take the type-A contribution as an example: Since the quenched couplings are derived from low-energy nuclear excitations, the replacement of ``free'' coupling by ``quenched'' couplings essentially accounts for the contribution to $\delta_\text{NS}$ from low-lying nuclear excited states. It misses, however, more important nuclear effects such as the quasi-elastic absorption peak, which comes not from a highly off-shell nucleon, but from an almost on-shell nucleon loosely bounded to the nucleus. This means, the traditional shell model estimation of $\delta_\text{NS}$ contains a large theory systematics which was not included in their error budget. Ref. \cite{Seng:2018qru} estimated this systematic effect using a crude Fermi gas model, which results in an overall shift of $\delta_\text{NS}$ to more negative values, accompanied by an inflated total uncertainty. In the most recent critical survey of $V_{ud}$ from superallowed decays~\cite{Hardy:2020qwl},
\begin{equation}
    |V_{ud}|_{0^+}^2=0.94815(9)_\text{exp}(18)_{\Delta_R^V}(53)_{\delta_\text{NS}}(11)_{\delta_r'}(8)_{\delta_\text{C}}~,\label{eq:Vud2020}
\end{equation} 
we see that $\delta_\text{NS}$ is the dominant source of uncertainty in $V_{ud}$ at face value. Therefore, an urgent task in this field is to systematically reduce the theory uncertainty of $\delta_\text{NS}$ by a factor of $\sim 2$ or more for all important superallowed transitions.

\subsubsection{Current algebra formalism}

Refs. \cite{Seng:2018qru,Seng:2022cnq,Gorchtein:2023naa}
lay out the rigorous theory framework that allows an \textit{ab initio} calculation of $\delta_\text{NS}$, based on the current algebra approach formulated by Sirlin~\cite{Sirlin:1977sv}. It consists of writing the nuclear axial $\gamma W$-box diagram amplitude as follows:   
\begin{equation}
    \Box(E_e)=\frac{e^2}{M_F^{(0)}}\mathfrak{Re}\int\frac{d^4q}{(2\pi)^4}\frac{M_W^2}{M_W^2-q^2}\frac{\left[Q^2+M\nu\frac{p\cdot q m_e^2-p_e\cdot q p\cdot p_e}{M^2m_e^2-(p\cdot p_e)^2}\right]T_3(\nu,Q^2)}{[(p_e-q)^2-m_e^2+i\varepsilon](q^2+i\varepsilon)M\nu}~,
\label{eq:gWboxGen}
\end{equation}
where $M_F^{(0)}=\sqrt{2}$ is the Fermi matrix element in the isospin limit, $M$ is the mass of the decaying nucleus, and $E_e$ is the electron energy. The quantity
\begin{equation}
T_3(\nu,Q^2)=-\frac{2M\nu}{|\vec{q}|}\sum_X\Bigg[\frac{\langle \phi_f|J_\text{em}^x(\vec{q})|X\rangle\langle X|J_{W5}^{\dagger y}(-\vec{q})|\phi_i\rangle}{\nu_X-\nu-i\varepsilon}
+\frac{\langle \phi_f|J_{W5}^{\dagger y}(-\vec{q})|X\rangle\langle X|J_\text{em}^x(\vec{q})|\phi_i\rangle}{\nu_X+\nu-i\varepsilon}\Bigg]
\label{eq:T3GF}
\end{equation}
is the P-odd invariant amplitude of the so-called ``generalized Compton tensor'',  where $X$ represents all on-shell nuclear and hadronic intermediate states. The nuclear-structure-dependent RC is then just the difference between the nuclear and single-nucleon box diagram:
\begin{equation}
    \delta_\text{NS}=2(\langle\Box_\text{nucl}(E_e)\rangle-\Box_n)~,
\end{equation}
where $\langle\dots\rangle$ denotes the average over the beta spectrum.  

To facilitate the numerical integration over the loop momentum, we first perform a Wick rotation to the box diagram integral. It leads to the following separation:
\begin{equation}
\Box=\Box_\text{Wick}+\Box_{\text{res},e}+\Box_{\text{res},T_3}~,
\end{equation}
where the first term is the Wick-rotated term, the second is the pole contribution from the electron propagator, and the third terms is the pole contribution from $T_3$. The combination of the first two terms is a regular function of $E_e$, which can be expanded as:
\begin{equation}
    \Box_\text{Wick}+\Box_{\text{res},e}=\boxminus_0+\boxminus_1 E_e+\mathcal{O}(E_e^2)~.
\end{equation}
In the single-nucleon sector, the smallest energy scale is $m_\pi$ and $E_e/m_\pi\sim 10^{-2}$, which renders the linear term $\boxminus_1 E_e$ irrelevant. Therefore we can express $\delta_\text{NS}$ as:
\begin{equation}
    \delta_\text{NS}\approx 2(\boxminus_0^\text{nucl}-\boxminus_0^n)+2\boxminus_1^\text{nucl}\langle E_e\rangle+\langle(\Box(E_e))_{\text{res},T_3}\rangle~.
\end{equation}

The first two expansion coefficients are the focus of \textit{ab initio} calculations:
\begin{eqnarray}
    \boxminus_0&=&\int\frac{d^4q_E}{(2\pi)^4}\frac{M_W^2}{M_W^2+Q^2}\frac{1}{(Q^2)^2}\frac{Q^2-\nu_E^2}{\nu_E}\frac{T_3(i\nu_E,Q^2)}{MM_F^{(0)}}\nonumber\\
    \boxminus_1&=&-\frac{8}{3}e^2\mathfrak{Re}\int\frac{d^4q_E}{(2\pi)^4}\frac{Q^2-\nu_E^2}{(Q^2)^3}\frac{iT_3(i\nu_E,Q^2)}{MM_F^{(0)}}~,
\end{eqnarray}
where $q_E=(\mathbf{q},\nu_E)$ is the Euclidean loop momentum, and $Q^2=\mathbf{q}^2+\nu_E^2$. An alternative representation of $\boxminus_{0,1}$ can be derived through the dispersion relation of $T_3$. It reads:
\begin{eqnarray}
    \boxminus_0&=&\frac{\alpha}{\pi}\int_0^\infty dQ^2\frac{M_W^2}{M_W^2+Q^2}\int_{\nu_0}^\infty \frac{d\nu}{\nu}\frac{\nu+2\sqrt{\nu^2+Q^2}}{(\nu+\sqrt{\nu^2+Q^2})^2}\frac{F_{3,-}(\nu,Q^2)}{MM_F^{(0)}}\nonumber\\
    \boxminus_1&=&\frac{2\alpha}{3\pi}\int_0^\infty dQ^2\int_{\nu_0}^\infty \frac{\nu+3\sqrt{\nu^2+Q^2}}{(\nu+\sqrt{\nu^2+Q^2})^3}\frac{F_{3,+}(\nu,Q^2)}{M M_F^{(0)}}~,
\end{eqnarray}
which is expressed in terms of the nuclear response functions:
\begin{equation}
    F_{3,\pm}(\nu,Q^2)=-\frac{iM\nu}{2|\vec{q}|}\sum_X\delta (E_X-M-\nu)\left\{\langle\phi_f|J_\text{em}^x(\vec{q})|X\rangle\langle X|(J_{W5}^{\dagger y}(-\vec{q})|\phi_i\rangle\mp \langle\phi_f|J_{W5}^{\dagger y}(-\vec{q})|X\rangle\langle X|(J_\text{em}^{x}(\vec{q})|\phi_i\rangle \right\}~.\label{eq:F3}
\end{equation}
Therefore, one can either choose to compute $T_3$ from Eq.\eqref{eq:T3GF} or $F_{3,\pm}$ from Eq.\eqref{eq:F3} with \textit{ab initio} methods. Both expressions involve the summation of all intermediate states $X$, which makes it computationally more demanding; however, with this approach one explicitly captures the full contribution to $\delta_\text{NS}$ from physics at all scales. 

The current algebra approach has been adopted to compute $\delta_\text{NS}$ for ${}^{10}\text{C}\rightarrow {}^{10}\text{B}$ using NCSM~\cite{Gennari:2024sbn}, which we will describe in Section~\ref{sec:deltaNSNCSM}. Follow-up studies on $^{14}\text{O}\rightarrow{}^{14}\text{N}$ and $^{18}\text{Ne}\rightarrow{}^{18}\text{F}$ are ongoing.  

\subsubsection{Effective field theory formalism}\label{sec:eft}

\begin{figure}
	\centering
   \includegraphics[width=0.7\columnwidth]{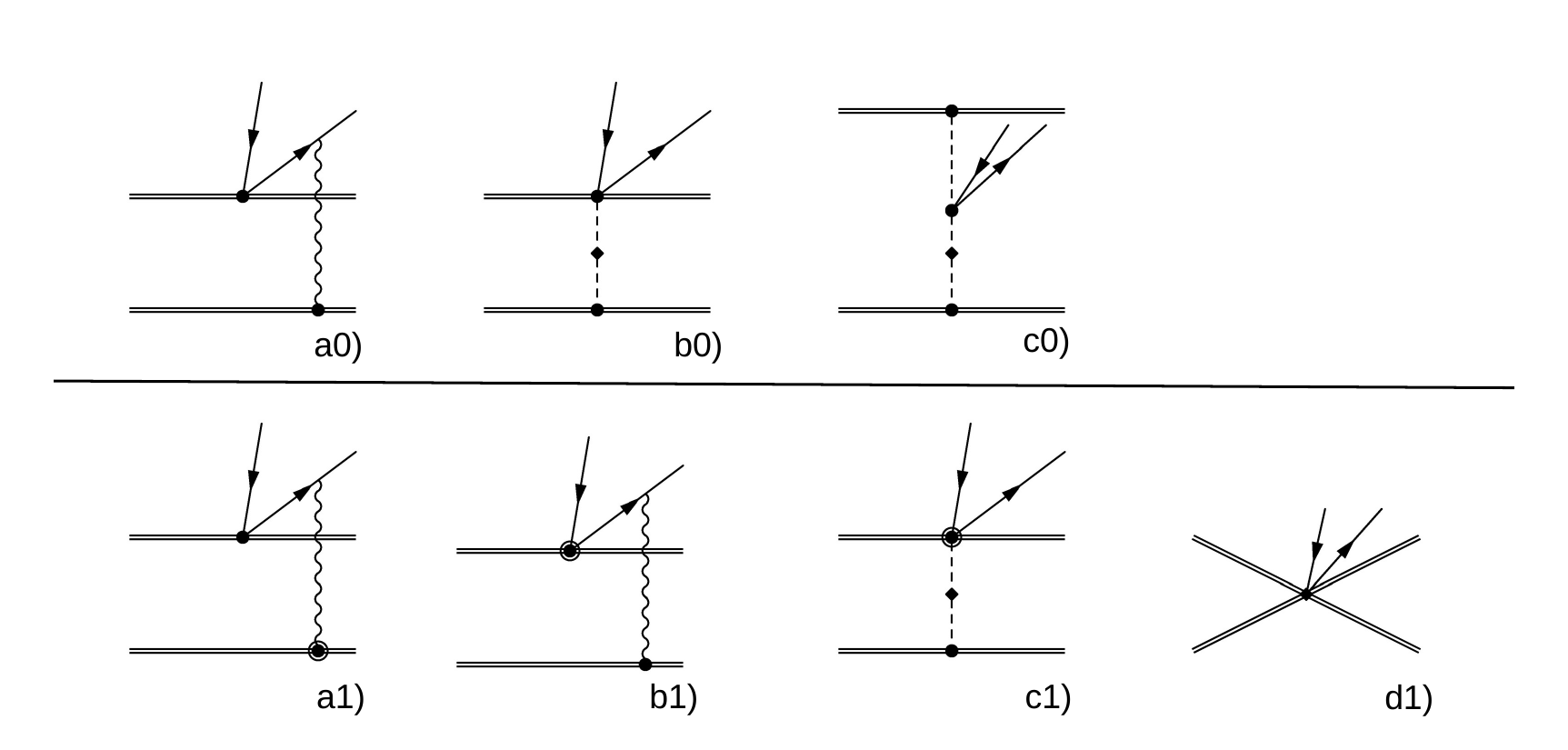}
	\caption{Lowest-order Feynman diagrams contributing to the effective two-body potentials. Figures courtesy of Wouter Dekens.\label{fig:potentials}} 
\end{figure} 

Refs. \cite{Cirigliano:2024msg,Cirigliano:2024rfk} took an alternative path, namely the effective field theory (EFT) approach to describe the RC. It consists of first splitting the loop integral into different regions of the loop momentum $q$.  Two regions are relevant to the structure-dependent RC: (1) The ``ultrasoft'' region where $q_0\sim \mathbf{q}\sim 10^0~\text{MeV}$, and (2) The ``potential'' region, where $q_0\sim 10^0~\text{MeV}\ll \mathbf{q}\sim 10^2~\text{MeV}$. The contribution from the ultrasoft region is sensitive to details of nuclear excited states and thus need to be evaluated explicitly. On the other hand, the contribution from the potential region turns out to be largely independent of the nuclear intermediate states. Therefore, in this region one may derive an ``effective potential'' from few-nucleon Feynman diagrams and simply evaluate its ground-state nuclear matrix element. 

In the EFT framework, the weak Hamiltonian responsible for nuclear beta decays reads:
\begin{equation}
    H_\text{EW}=\sqrt{2}G_F V_{ud}\bar{e}_L\gamma_\mu v_L\mathcal{J}^\mu_W~,
\end{equation}
where the effective weak current reads\footnote{\red{In ordinary circumstances, one may need to worry about ISB effects in the weak current, such as that in Ref.\cite{Krivoruchenko:2015rza}. However, in evaluating $\delta_\text{NS}$ which is itself already a $\mathcal{O}(\alpha/\pi)$ correction, the assumption of isospin symmetry is valid because an additional ISB correction on top of $\delta_\text{NS}$ is beyond the desired precision goal of $10^{-4}$.} }:
\begin{eqnarray}
    \mathcal{J}_W^\mu&=&\sum_{n=1}^A(g_V \delta^{\mu 0}-g_A\delta^{\mu i}\sigma^{(n)i})\tau^{(n)+}+(\mathcal{J}^{2b})^\mu +\dots\nonumber\\
    &&+\delta^{\mu 0}(\mathcal{V}^0+E_0\mathcal{V}_E^0)+\delta^{\mu i}\mathcal{V}_i+p_e^\mu\mathcal{V}_{m_e}+\dots~,\label{eq:VE}
\end{eqnarray}
with $E_0$ the electron end-point energy. 
We retain here only the one-body and two-body contributions. The one-body term in the first line consists of the vector and axial coupling $g_V$ and $g_A$ which are already renormalized by nucleus-independent RC~\cite{Cirigliano:2023fnz,Cirigliano:2024nfi}, and $(\mathcal{J}^{2b})^\mu$ is the pure-QCD two-body current. Meanwhile, the second line consists of two-body potentials that can be obtained diagrammatically from Fig. \ref{fig:potentials}. First, at $\mathcal{O}(\alpha)$, the lepton-energy-dependent potentials read:
\begin{eqnarray}
\mathcal{V}_E^0&=&\frac{1}{3}\left(\frac{1}{2}+\frac{4E_e}{E_0}\right)\mathcal{V}_E+\mathcal{V}_E^\pi\nonumber\\
\mathcal{V}_{m_e}&=&\frac{1}{2}\mathcal{V}_E+\mathcal{V}_{m_e}^\pi~,
\end{eqnarray}
where $\mathcal{V}_E$ is obtained from Fig. \ref{fig:potentials}(a0):
\begin{equation}
    \mathcal{V}_E(\mathbf{q})=g_V\sum_{j<k}e^2\frac{1}{\mathbf{q}^4}(\tau^{+(j)}P_p^{(k)}+P_p^{(j)}\tau^{+(k)})~,
\end{equation}
and $\mathcal{V}_E^\pi$ comes from the electromagnetically-induced pion mass splitting, i.e. Figs. \ref{fig:potentials}(b0),(c0), characterized by the coupling constant $Z_\pi$:
\begin{eqnarray}
    \mathcal{V}_E^\pi(\mathbf{q})&=&\frac{g_A^2Z_\pi e^2}{3}\sum_{j<k}(\tau^{+(j)}\tau_3^{(k)}+\tau_3^{(j)}\tau^{+(k)})\frac{1}{[\mathbf{q}^2+m_\pi^2]^2}\nonumber\\
    &&\times\left\{\mathbf{\sigma}^{(j)}\cdot\mathbf{\sigma}^{(k)}\left(1-\frac{1}{3}\frac{\mathbf{q}^2}{\mathbf{q}^2+m_\pi^2}-\frac{2}{3}\frac{\mathbf{q}^4}{(\mathbf{q}^2+m_\pi^2)^2}\right)+\frac{2}{3}S^{(jk)}\left(\frac{1}{2}\frac{\mathbf{q}^2}{\mathbf{q}^2+m_\pi^2}+\frac{\mathbf{q}^4}{(\mathbf{q}^2+m_\pi^2)^2}\right)\right\}\nonumber\\
    \mathcal{V}_{m_e}^\pi(\mathbf{q})&=&-\frac{g_A^2Z_\pi e^2}{3}\sum_{j<k}(\tau^{+(j)}\tau_3^{(k)}+\tau_3^{(j)}\tau^{+(k)})\frac{1}{[\mathbf{q}^2+m_\pi^2]^2}\nonumber\\
    &&\times\left\{\mathbf{\sigma}^{(j)}\cdot\mathbf{\sigma}^{(k)}\left(1-\frac{4}{3}\frac{\mathbf{q}^2}{\mathbf{q}^2+m_\pi^2}-\frac{2}{3}\frac{\mathbf{q}^4}{(\mathbf{q}^2+m_\pi^2)^2}\right)+\frac{2}{3}S^{(jk)}\left(2\frac{\mathbf{q}^2}{\mathbf{q}^2+m_\pi^2}+\frac{\mathbf{q}^4}{(\mathbf{q}^2+m_\pi^2)^2}\right)\right\}~.
\end{eqnarray}
Here we have defined operators in the spin and isospin space:
\begin{equation}
    P_{p,n}^{(j)}=\frac{\mathbbm{1}^{(j)}\pm \tau_3^{(j)}}{2}~,~S^{(jk)}=\mathbf{\sigma}^{(j)}\cdot\mathbf{\sigma}^{(k)}-\frac{3\mathbf{q}\cdot\mathbf{\sigma}^{(j)}\mathbf{q}\cdot\mathbf{\sigma}^{(k)}}{\mathbf{q}^2}~.
\end{equation}

More interesting are the lepton-energy-independent potential $\mathcal{V}^0$. At $\mathcal{O}(\alpha)$, it consists of three terms derived from the last four diagrams in Fig. \ref{fig:potentials}:
\begin{equation}
    \mathcal{V}^0=\mathcal{V}_0^\text{mag}+\mathcal{V}_0^\text{rec}+\mathcal{V}_0^\text{CT}~.\label{eq:V0}
\end{equation}
First, the long-distance ``magnetic'' and ``recoil'' potential obtained from Figs. \ref{fig:potentials}(a1)-(c1) read~\footnote{There was a typo in the expression of $\mathcal{V}_0^\text{rec}(\mathbf{q},\mathbf{P})$ in Ref.\cite{Cirigliano:2024msg}, which affected their final numerical result.}:
\begin{eqnarray}
    \mathcal{V}_0^\text{mag}(\mathbf{q})&=&\sum_{j<k}\frac{e^2}{3}\frac{g_A}{m_N}\frac{1}{\mathbf{q}^2}\left(\mathbf{\sigma}^{(j)}\cdot\mathbf{\sigma}^{(k)}+\frac{1}{2}S^{(jk)}\right)[(1+\kappa_p)\tau^{+(j)}P_p^{(k)}+\kappa_n\tau^{+(j)}P_n^{(k)}+(j\leftrightarrow k)]\nonumber\\
    \mathcal{V}_0^\text{rec}(\mathbf{q},\mathbf{P})&=&\sum_{j<k}\left[i\frac{e^2g_A}{2m_N}\frac{\tau^{+(j)}P_p^{(k)}}{\mathbf{q}^4}(\mathbf{P}_k\times\mathbf{q})\cdot\mathbf{\sigma}^{(j)}-\frac{Z_\pi e^2g_A^2}{m_N}\frac{\tau^{+(j)}\tau_3^{(k)}}{(\mathbf{q}^2+m_\pi^2)^2}\mathbf{\sigma}^{(j)}\cdot\mathbf{q}\:\mathbf{\sigma}^{(k)}\cdot\mathbf{P}_j+(j\leftrightarrow k)\right]~.\nonumber\\
\end{eqnarray}
For future purpose, we can split the ``recoil'' potential into two pieces:
\begin{equation}
    \mathcal{V}_0^\text{rec}=\mathcal{V}_0^\text{rec,1}+\mathcal{V}_0^\text{rec,2}
\end{equation}
which are linear and quadratic to $g_A$, respectively. There are also $\mathcal{O}(\alpha^2)$ terms in the potential which we do not display here. 

Renormalization analysis shows that the nuclear matrix element of $\mathcal{V}_0^\text{mag}$ depends logarithmically on the ultraviolet (UV) cutoff. This signals sensitivity to the UV physics which must be absorbed by an extra ``counterterm'' potential coming from Fig. \ref{fig:potentials}(d1):
\begin{equation}
    \mathcal{V}_0^\text{CT}=e^2(g_{V1}^{NN}O_1+g_{V2}^{NN}O_2)~,
\end{equation}
where 
\begin{equation}
    O_1=\sum_{j\neq k}\tau^{+(j)}\mathbbm{1}^{(k)}~,~O_2=\sum_{j<k}[\tau^{+(j)}\tau_3^{(k)}+(j\leftrightarrow k)]~.
\end{equation}
This potential introduces two new low-energy constants (LECs) $g_{V1}^{NN}$, $g_{V2}^{NN}$, which values are not fixed by chiral symmetry and must be determined by other means. This  is very similar to the EFT description of neutrinoless double beta decay~\cite{Cirigliano:2018hja,Cirigliano:2019vdj}, \red{except that the latter arises from the product of two charged weak currents and involves only one unknown LEC.}

The EFT approach has been adopted to compute $\delta_\text{NS}$ for $^{14}\text{O}\rightarrow{}^{14}\text{N}$~\cite{Cirigliano:2024rfk,Cirigliano:2024msg} and ${}^{10}\text{C}\rightarrow {}^{10}\text{B}$~\cite{King:2025fph} using quantum Monte Carlo, which we will describe in Section~\ref{sec:deltaNSQMC}. There is also an ongoing, parallel effort to compute $\delta_\text{NS}$ in superallowed decays of $^{10}$C, $^{14}$O, $^{26}$Al, $^{34}$Cl, $^{38}$K, $^{42}$Sc, $^{46}$V, $^{50}$Mn and $^{54}$Co using coupled cluster method \footnote{Private communication with Samuel Novario.}. It is highly desirable to have multiple \textit{ab initio} calculation of the same nuclei in order to better understand the method-dependence of the final result and thus better quantify the theory uncertainties.    

\subsubsection{\label{sec:matching}Connection between the two approaches}

{The EFT formalism in Refs.\cite{Cirigliano:2024msg,Cirigliano:2024rfk} attempts to cover \textit{all} EM effects that enter superallowed decays, including short-distance QED effects, Fermi function, shape factor, single-nucleon RC, nucleus-dependent RC, and even the ISB correction (which is also due to the EM interaction), which are all treated differently in the traditional approach, under a unified framework. As a consequence, there is a lot of reshuffling of different pieces of physics encoded in different terms at the right hand side of Eq.\eqref{eq:Vud}. The EFT version of that equation reads:
\begin{equation}
    |V_{ud}|^2_{0^+}=\frac{\pi^3\ln 2}{G_F^2m_e^5\overline{\mathcal{F}}(\mu)t[C_{\text{eff}}^{g_V}(\mu)]^2}\ ,\label{eq:VudEFT}
\end{equation}
with explanations as follows. First, in replacement of $1+\Delta_R^V$ in Eq.\eqref{eq:Vud}, the EFT master formula utilizes $[C_\text{eff}^{g_V}]^2$ that encodes the effect of hard and soft photons through the running and matching from the electroweak scale to an arbitrary, low renormalization scale $\mu$. The $\mu$-dependence is canceled by $\overline{\mathcal{F}}(\mu)$, an EFT version of $\mathcal{F}$:
\begin{equation}
\overline{\mathcal{F}}(\mu)=[1+\overline{\delta_R'}(\mu)](1+\overline{\delta_\text{NS}})(1-\overline{\delta_C})\bar{f}(\mu)\ ,    
\end{equation}
where $\bar{f}(\mu)$, $\bar{\delta}_R'(\mu)$, $\overline{\delta_\text{NS}}$ and $\overline{\delta_C}$ are the EFT versions of the statistical rate function, outer RC, nucleus-dependent RC and ISB correction, respectively. Terms that involve renormalization group running to sum up large logarithms will depend on the renormalization scale $\mu$. Since $V_{ud}$ is a scheme-independent quantity, the right hand side of Eq.\eqref{eq:Vud} and \eqref{eq:VudEFT} must be the same, although each of the individual components can be quite different. A qualitative comparison of different components in the two formalisms can be found in Table I of Ref. \cite{Cirigliano:2024msg}.

The main difference between the traditional and EFT master formula of $|V_{ud}|^2_{0^+}$ originates from the reshuffling of electroweak logarithms and higher-order ($\mathcal{O}(\alpha^n)$, $n\geq 2$) perturbative QED corrections from one term to another. Meanwhile, the positions of the non-perturbative terms at $\mathcal{O}(\alpha)$, which carry the largest theory uncertainties, remain untouched. This allows for a meaningful comparison between the two approaches.
For instance, the single-nucleon axial $\gamma W$ box diagram that enters $\Delta_R^V$ is encoded in $[C_\text{eff}^{g_V}(\mu)]^2$ in the EFT formalism. At the nuclear level, the traditional (current algebra) approach for $\delta_{NS}$ deals explicitly with the nuclear ``axial'' $\gamma W$-box diagram, while $\overline{\delta_\text{NS}}$ in the EFT approach includes contributions from the Feynman diagrams as well.} To connect the two approaches we simply identify components in the EFT approach that come from the axial $\gamma W$-box diagram. In long-distance two-body contributions, this is represented by the two-body potentials that are linear to $g_A$, namely $\mathcal{V}_0^\text{mag}$ and $\mathcal{V}_0^\text{rec,1}$. Additionally, the counterterm potential serves to cancel the UV divergence of the nuclear matrix element of $\mathcal{V}_0^\text{mag}$, so they have to appear together. As a result, the following matching relation can be obtained~\cite{Seng:2025hnz}:
\begin{equation}
    \delta_\text{NS}=\frac{2}{M_F^{(0)}}\langle \mathcal{V}_0^\text{mag}+\mathcal{V}_0^\text{rec,1}+\mathcal{V}_0^\text{CT}\rangle_{fi}~.\label{eq:deltaNSmatch}
\end{equation}

This relation is useful in the following sense. On the one hand, $\delta_\text{NS}$ can be computed explicitly with the current algebra approach, but it involves the summation over all nuclear intermediate states which can be computationally demanding. \red{In fact, the master formula of the current algebra formalism, Eq.\eqref{eq:T3GF}, is very similar to the amplitude of a two-neutrino double beta decay~\cite{Engel:2016xgb,Pirinen:2015sma}, both of which involve two nuclear matrix elements of electroweak currents in the numerator and an energy denominator, summing over all intermediate states. The main difference is that, in two-neutrino double beta decays the contributing intermediate states are restricted by kinematics, while in $\delta_\text{NS}$ such limitation does not exist since the loop momentum can take any value.} On the other hand, the ground-state nuclear matrix elements of effective two-body potentials are much easier to compute, but the theory precision is limited by the two unknown LECs, \red{similar to the case of neutrinoless double beta decay}. So, a useful way to make progress is to first choose a pair of superallowed transitions that can be easily handled in both methods. One then computes the LHS of Eq.\eqref{eq:deltaNSmatch} using the current algebra approach, and the RHS using the EFT approach. With that, we are able to extract the values of the two unknown LECs by the above matching relation. Once the LECs are known, one can proceed with the EFT approach for other transitions without being limited by their large theory uncertainties. Thus, we verify that the physics in the counterterms can indeed be obtained explicitly in the current algebra approach. 

\begin{figure}
	\centering
   \includegraphics[width=0.2\columnwidth]{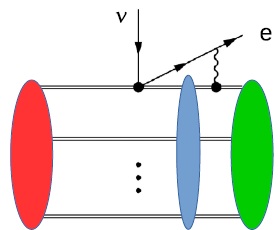}
	\caption{RC diagram with the weak and electromagnetic vertices acting on the same nucleon. Figure courtesy of Wouter Dekens.\label{fig:singleN}} 
\end{figure} 

Of course, for the matching to be useful we need to be sure that the direct \emph{ab initio} calculation of $\delta_\text{NS}$ through the summation of nuclear intermediate states will include the physics contained in the EFT counterterms, and indeed it does. This can be seen by investigating the type of diagrams depicted in Fig. \ref{fig:singleN}, where the weak and electromagnetic vertices act on a single nucleon. In a full theory valid in all energy scales, this diagram involves the integration of the single-nucleon form factors convoluted with the nucleon momentum distribution in a nucleus (the so-called ``Fermi smearing'', which gives rise to the quasi-elastic nucleon contribution to $\delta_\text{NS}$); it probes the loop momentum up to a few hundred MeV, and such a contribution is explicitly included in the calculation of $T_3$ through the summation of nuclear intermediate states. On the other hand, as explained in Ref. \cite{Cirigliano:2024msg}, the same diagram in the EFT language probes only physics at the ultrasoft region, i.e. $q\sim 10^0$~MeV. Therefore, the contribution from the quasielastic nucleons does not reside in Fig.\ref{fig:singleN}, but instead in the nuclear matrix element of the counterterm potentials. 

\subsection{\label{sec:deltaNSNCSM} Nuclear structure RC for ${}^{10}$C$\rightarrow {}^{10}$B with No-Core Shell Model}

Ref. \cite{Gennari:2024sbn} presents the first \textit{ab initio} calculation of $\delta_\text{NS}$ in the ${}^{10}$C$\rightarrow {}^{10}$B superallowed beta decay based on the current algebra approach using NCSM. Two strategies are adopted to facilitate the computation of the P-odd invariant amplitude $T_3$:
\begin{itemize}
    \item The Lanczos strength method~\cite{haydock1974inverse,Dagotto:1993ajt,Marchisio:2002jx} is used to enable an efficient summation over all nuclear intermediate states $X$.
    \item The electroweak current operators are expanded in terms of multipole operator with increasing angular quantum number $J$~\cite{Donnelly:1975ze,Donnelly:2017aaa}. Matrix elements of multipole operators with respect to harmonic oscillator states are well-documented, so one can make use of results in the literature to simplify the calculation. 
\end{itemize}

\begin{figure}
\begin{centering}
\includegraphics[scale=0.5]{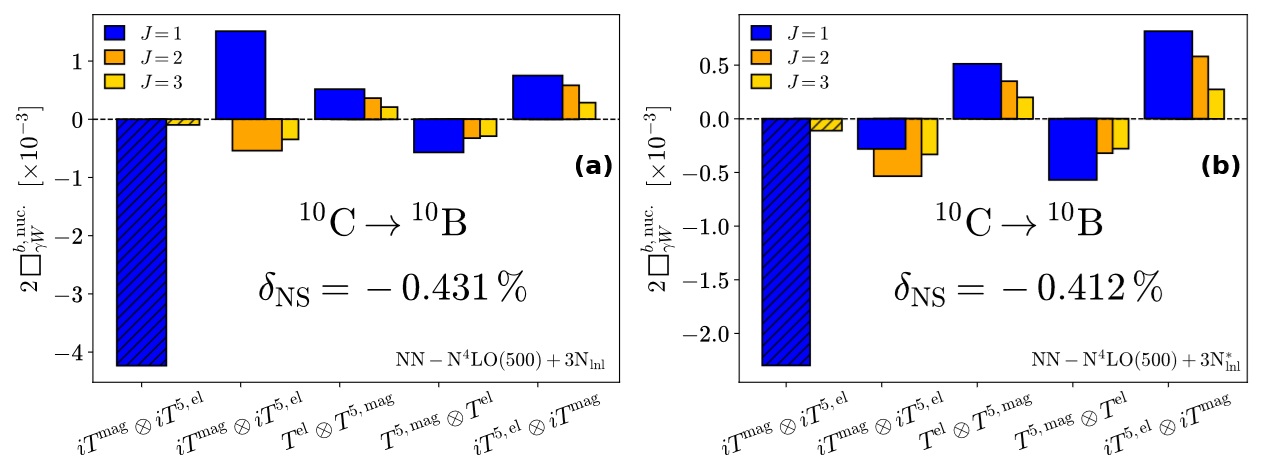}
\includegraphics[scale=0.65]{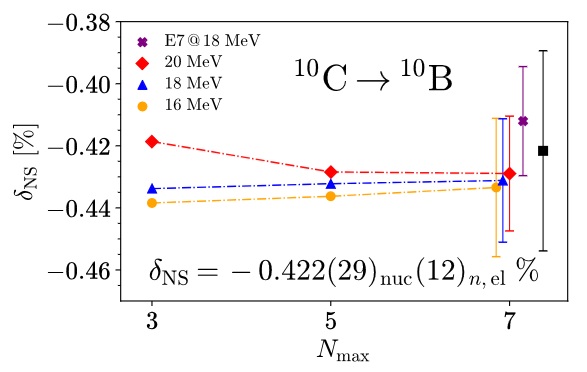}
\par\end{centering}
\caption{\label{fig:Box10C}Main results of the $\delta_\text{NS}$ calculation for $^{10}$C$\rightarrow$$^{10}$B in Ref.\cite{Gennari:2024sbn} (see the text for explanation).}
\end{figure}

The main results of Ref. \cite{Gennari:2024sbn} are summarized in Fig.\ref{fig:Box10C}. The first two plots show the contributions from different products of multipole operators to the nuclear box diagram, and one indeed finds that the contribution decreases rapidly with increasing $J$. These two diagrams are based on two different choices of three-nucleon forces~\cite{Gysbers:2019uyb,Kravvaris:2022eyf}, which has a mild effect to the central value of $\delta_\text{NS}$ that is taken as a measure of the uncertainty from the chiral interactions. The third diagram demonstrates the quick convergence of our result with increasing $N_\text{max}$, which becomes basically flat at $N_\text{max}=7$, much faster than most typical NCSM calculations. The variation of the result with respect to the oscillator frequency $\Omega$ is also shown in the same plot. 

Despite these promising features, nuclear \textit{ab initio} calculations cannot capture all nuclear physics that reside in $\delta_\text{NS}$. The reason is apparent: from the dispersive analysis, we know that $\delta_\text{NS}$ is induced by the difference between the single-nucleon absorption spectrum and the nuclear spectrum spectrum, both are contributed by nucleonic and non-nucleonic DOFs. Since \textit{ab initio} calculations are based on chiral Lagrangian that typically consists only of nucleon fields, it can only account for nuclear modifications to contribution from the nucleonic DOFs. In other words, it only covers contributions from intermediate states $X$ in Eq.\eqref{eq:T3GF} that are just made out of nucleons, while the contributions from all other intermediate states involving non-nucleonic particles are missed. The uncertainty due to this missing effects need to be included in the error budget. 

To get a more complete picture, we start from the single-nucleon box diagram, and split it into the ``elastic'' and ``inelastic'' pieces:
\begin{equation}
    \Box_n=\Box_n^\text{el}+\Box_n^\text{inel}~,
\end{equation}
where the elastic piece depends only on free nucleon form factors. This splitting is possible in the dispersive analysis~\cite{Seng:2018yzq,Shiells:2020fqp}. 
The nuclear modification to the first term, $\Box_n^\text{el}$, is accounted for in \textit{ab initio} calculation, which we denote as $\Box_\text{nucl}^{\text{ab-initio}}$. So, we can write the full nuclear box diagram as
\begin{equation}
    \Box_\text{nucl}=\Box_\text{nucl}^{\text{ab-initio}}+\left(\Box_n^\text{inel}+\delta(\Box_n^\text{inel})_\text{shad}\right)~,
\end{equation}
where $\delta(\Box_n^\text{inel})_\text{shad}$ denotes nuclear modification to $\Box_n^\text{inel}$ due primarily to the so-called ``shadowing'' effect at high energy~\cite{Armesto:2006ph,Kopeliovich:2012kw}. In Ref. \cite{Gennari:2024sbn}, this term is
included as an uncertainty. To estimate its size, one first splits the nucleon part into:
\begin{equation}
    \Box_n^\text{inel}=(\Box_n^\text{inel})^<+(\Box_n^\text{inel})^>~,
\end{equation}
where the superscripts ``<'' and ``>'' denote the contribution from the loop integral below and above $Q^2=2$~GeV$^2$. For $Q^2>2$~GeV$^2$, the integral is related to a sum rule of the parity-odd structure function $F_3$, similar to the well-know Gross-Llewellyn Smith (GLS) sum rule~\cite{Gross:1969jf}; the latter is experimentally demonstrated to hold even for heavy nuclei~\cite{Leung:1992yx,Kim:1970} after including higher-order perturbative QCD effects, so one may assume that this part is not modified. Meanwhile, the nuclear modification to $(\Box_n^\text{inel})^<$ can be inferred from the experimental knowledge of the ratio $R$ between the nucleon and nuclear structure function $F_2$. Depending on the value of the Bjorken variable $x_\text{B}$, $R$ may deviate positively or negatively from 1 due to different physical mechanisms, such as nuclear shadowing, anti-shadowing, EMC effect and Fermi motion. For $A\approx 10$, the maximum deviation of $R$ from 1 is about 20\%~\cite{Kopeliovich:2012kw}. So, one may simply multiply $(\Box_n^\text{inel})^<$ by 20\%, and take it as a conservative estimation of the absolute size of $\delta(\Box_n^\text{inel})_\text{shad}$ which will be included as an error. 

The full result of $\delta_\text{NS}$ for ${}^{10}$C$\rightarrow {}^{10}$B obtained in Ref. \cite{Gennari:2024sbn} reads\footnote{We do not include the uncertainty due to the subtraction of the single-nucleon box diagram which was displayed in Fig.\ref{fig:Box10C}, because it goes away upon combining $\delta_\text{NS}$ and $\Delta_R^V$.}:
\begin{equation}
    \delta_\text{NS}=-0.422(14)_\text{PME} (4)_\Omega(9)_\chi(24)_\text{sh}\%~, 
\end{equation}
where the sources of (nucleus-dependent) theoretical errors are: (i) PME: The ``partial model error'', due to many-body basis truncation, multipole expansion truncation, and single-nucleon dipole form factors; (ii) $\Omega$: The choice of oscillator frequency; (iii) $\chi$: The chiral expansion truncation; and (iv) sh: The nuclear shadowing effect. It is informative to track the ``evolution'' of this number over time:
\begin{eqnarray}
    \delta_\text{NS}(\%)&:&-0.345(35)~~,~~\text{Shell model}\nonumber\\
    &\rightarrow&-0.400(50)~~,~~\text{DR + Fermi gas model}\nonumber\\
    &\rightarrow &-0.422(29)~~,~~\text{\textit{Ab initio} NCSM}
\end{eqnarray}
The first calculation with shell model was based on the type-A/type-B picture. This was later discovered in the dispersive analysis to contain a large theory systematic, which was estimated using a crude Fermi gas model~\cite{Seng:2018qru,Gorchtein:2018fxl}; this resulted in a shift to $\delta_\text{NS}$ to the more negative side, accompanied by an inflated theory uncertainty. This result is then confirmed by the \textit{ab initio} calculation in Ref. \cite{Gennari:2024sbn} but with a substantially reduced uncertainty. 

\subsection{\label{sec:deltaNSQMC} Nuclear structure RC for $^{10}\text{C}\rightarrow{}^{10}\text{B}$ and $^{14}\text{O}\rightarrow{}^{14}\text{N}$ with Quantum Monte Carlo}

\begin{figure}
	\centering
   \includegraphics[width=0.5\columnwidth]{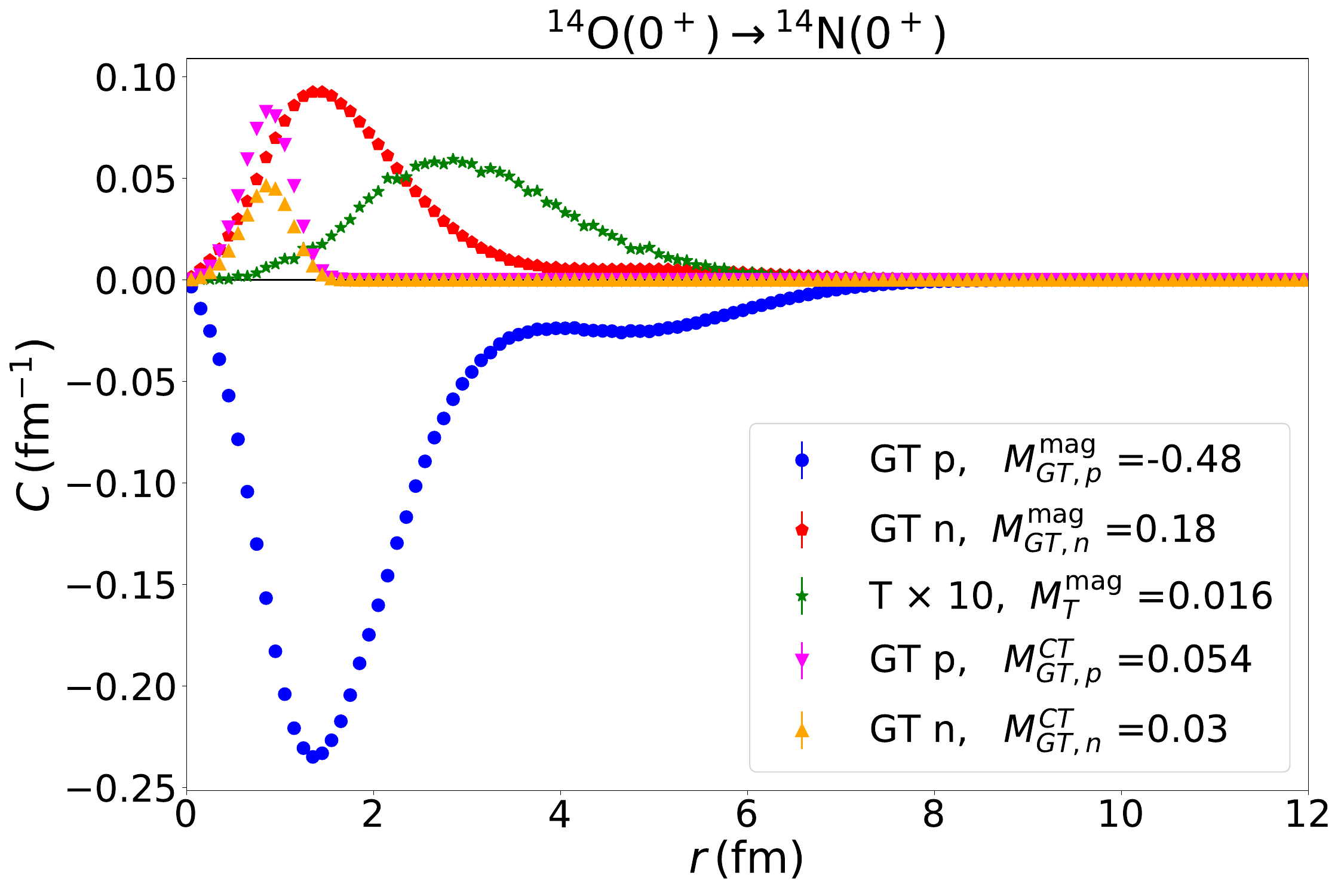}
	\caption{Nuclear matrix element densities of various potentials contributing to $\overline{\delta_{\rm NS}}$ in $^{14}$O$\rightarrow{}^{14}$N computed with VMC plotted as a function of interparticle spacing $r$. The integrated curves correspond to the matrix elements in the legend of the plot. \label{fig:MGT14}} 
\end{figure} 

\begin{table}
\begin{center}
\begin{tabular}{||c| c | c | c | c |  c | c | c || c||}
\hline
Model & Method & $\delta_{\rm NS}^{(0)}$  &  $\overline{\delta_{\rm NS}^E}$ \\ 
\hline 
NV2+3-Ia  & VMC & $ -\left( 4.03 + 0.25 \pm 0.69\right) \cdot 10^{-3} $  & $1.01 \cdot 10^{-3}$
\\
&GFMC & $-\left( 4.43+ 0.21 \pm 0.77\right) \cdot 10^{-3}$ &  $0.97 \cdot 10^{-3}$  \\
\hline 
NV2+3-Ia$*$ & VMC & $-\left( 4.18 + 0.31 \pm 0.62 \right) \cdot 10^{-3}$ & $1.06 \cdot 10^{-3}$ \\
&GFMC &  $-\left(4.25 + 0.31  \pm 0.66\right) \cdot 10^{-3}$  & $1.08 \cdot 10^{-3}$ \\
\hline
\hline 
AV18+UX & VMC & $-\left(4.48 + 0.27 \pm 0.67\right) \cdot 10^{-3}$ & $1.02 \cdot 10^{-3}$ \\
&GFMC & $-\left( 4.06 +0.40 \pm 0.48\right) \cdot 10^{-3}$ & $1.17 \cdot 10^{-3}$\\ 
\hline
AV18+IL7 & VMC & $-\left( 4.55 + 0.26 \pm 0.61 \right) \cdot 10^{-3}$ & $1.01 \cdot 10^{-3}$ \\
&GFMC &$-\left( 4.32 + 0.29 \pm 0.64\right) \cdot 10^{-3}$ & $1.06 \cdot 10^{-3}$\\ \hline
\end{tabular}
\end{center}
\caption{Summary of the results for $\bar\delta_{\rm NS}$.
For the energy-independent component, $\delta_{\rm NS}^{(0)}$,
the first term encodes the contribution of the magnetic and spin-orbit two-body operators. The second term is the $\mathcal O(\alpha^2)$ correction arising from $M_{\text{F}, p}^+$. The last contribution arises from the contact interactions, with LECs set to the {\it  arbitrary} values $g_{V1}^{NN} \pm g^{NN}_{V2} = \pm \frac{1}{m_N} \frac{1}{(2 F_\pi)^2}$. The last column shows the energy dependent part, $\overline{\delta_{\rm NS}^{\rm E}}$. 
}\label{Tab:deltaNS0}
\end{table}

Next, we report the pioneering studies of $\delta_\text{NS}$ based on the EFT approach, by computing the matrix elements for the potentials defined in Section~\ref{sec:eft} with many-body methods. Recently, such evaluations have been performed using quantum Monte Carlo methods; namely, VMC and GFMC for $^{10}$C~\cite{King:2025fph}, and VMC and AFDMC for $^{14}$O~\cite{Cirigliano:2024msg}. As an example, the VMC calculation of the nuclear matrix elements of various potentials is shown in Fig.\ref{fig:MGT14}.

To make connections with experimental data, for $\beta^+$ decay, one combines the matrix elements of the potentials to give both the energy-independent part (see Eq.\eqref{eq:V0}, including the $\mathcal{O}(\alpha^2)$ contribution not displayed),
\begin{equation}
    \delta^{(0)}_{\rm NS} = \frac{2}{g_V(\mu_{\pi})M_F^{(0)}}\sum_{N=n,p}\left[\alpha\left(M_{{\rm GT},N}^{\rm mag} + M_{{\rm T},N}^{\rm mag} + M_{{\rm GT},N}^{\rm CT} + M_{{\rm LS},N} \right) + \alpha^2 M^+_{{\rm F},N} \right] \, ,
\end{equation}
and the energy-dependent part (see Eq.\eqref{eq:VE}), 
\begin{equation}
    \overline{\delta^E_{\rm NS}} =  -\frac{2}{g_V(\mu_{\pi})M_F^{(0)}}\left[ M^0_E E_0 + \frac{m_e^2}{E_e}M_{m_e}\right] \, ,
\end{equation}
where the vector coupling constant $g_V$ is evaluated at a scale $\mu_{\pi} = m_{\pi}$ giving $g_V(m_{\pi})=1.01494(12)$~\cite{Cirigliano:2023fnz}, and the electron energy $E_e$ is averaged over phase-space. The matrix element $M^{X}_{i,N}$ denotes
\begin{equation}
    M^{X}_{i,N} = \mel{f}{V^X_{i,N}}{i}\, 
\end{equation}
for a given potential $V^X_{i,N}$ defined in Ref.~\cite{Cirigliano:2024msg}. These potentials are obtained by Fourier-transforming the momentum-space potentials $\mathcal{V}(\mathbf{q},\mathbf{P})$ in Section~\ref{sec:eft} into coordinate space, and scaling out appropriate powers of $\alpha$. The nuclear initial and final states are denoted as $\ket{i}$ and $\ket{f}$, respectively. The values of these contributions sum to give the overall correction,
\begin{equation}
\overline{\delta_{\rm NS}} = \delta_\text{NS}^{(0)}+\overline{\delta_\text{NS}^E}~.
\end{equation}

For ${}^{10}\text{C}$, the GFMC results were obtained for four different models of the nuclear interaction. Namely, two phenomenological models-- the Argonne $v_{18}$~\cite{Wiringa:1994wb} plus either the Illinois-7 (IL7)~\cite{Pieper:2001ap} or Urbana X (UX)~\cite{Wiringa:2013ala} three nucleon forces-- and two Norfolk chiral potential models~\cite{Piarulli:2016vel,Piarulli:2017dwd,Baroni:2018fdn}. Norfolk potential models are denoted as NV2+3-Ia and NV2+3-Ia* in the literature. The main results are summarized for VMC and GFMC calculations in $^{10}$C in Table~\ref{Tab:deltaNS0}. For VMC calculations of $^{14}$O using the interaction of Refs.~\cite{Gezerlis:2014zia,Lynn:2015jua}, the results are~\footnote{Results for $\delta_\text{NS}^{(0)}$ and $V_{ud}$ were corrected for the re-evaluation of $\mathcal{V}^\text{rec}_0$ matrix element. Private communication with Emanauele Mereghetti and Wouter Dekens.}:
\begin{equation}
\delta_\text{NS}^{(0)}=-2.84(88)\cdot 10^{-3}~~,~~\overline{\delta_\text{NS}^E}=2.06(41)\cdot 10^{-3}~.
\end{equation}

The comparison with traditional approach (i.e. Ref.\cite{Hardy:2020qwl}) is most conveniently done in terms of the full $V_{ud}$~\cite{Hardy:2014qxa}. The GFMC values of $|V_{ud}(^{10}C)|$ are compared with the traditional extraction below,
\begin{align}
 & \text{NV2+3-Ia}     &\left. V_{ud} \right|_{^{10} \rm C} &= 0.97355 (66)_{\rm exp} (12)_{g_V}  (17)_{\mu} (9)_{\delta_C}  (38)_{g_V^{\rm NN}}, \\
 & \text{NV2+3-Ia*}     &\left. V_{ud} \right|_{^{10} \rm C} &= 0.97345 (66)_{\rm exp} (12)_{g_V}  (17)_{\mu} (9)_{\delta_C}  (32)_{g_V^{\rm NN}},
\\
&\text{AV18+UX}  & \left. V_{ud} \right|_{^{10} \rm C} &= 0.97336 (66)_{\rm exp} (12)_{g_V}  (17)_{\mu} (9)_{\delta_C}  (23)_{g_V^{\rm NN}} , \\
&\text{AV18+IL7} & \left. V_{ud} \right|_{^{10} \rm C} &= 0.97349 (66)_{\rm exp} (12)_{g_V}  (17)_{\mu} (9)_{\delta_C}  (31)_{g_V^{\rm NN}}, \\
&\text{Traditional} & \left. V_{ud} \right|_{^{10} \rm C} &= 0.97318(66)_{\rm exp}(9)_{\Delta_R^V}(24)_{\delta_{\rm NS}}(9)_{\delta_C}\, .
\end{align}
For $^{14}$O, the comparison gives,
\begin{align}
 & \text{EFT}     &\left. V_{ud} \right|_{^{14} \rm O} &= 0.97411 (10)_{\rm exp} (12)_{g_V}  (22)_{\mu} (12)_{\delta_C}  (43)_{g_V^{\rm NN}}(20)_{\delta_\text{NS}^E}, \\
&\text{Traditional} & \left. V_{ud} \right|_{^{14} \rm O} &= 0.97405(13)_{\rm exp}(9)_{\Delta_R^V}(31)_{\delta_{\rm NS}}(12)_{\delta_C}\, .
\end{align}
The main theoretical uncertainty in the EFT approach is due to two undetermined LECs, namely, $g_{V1}^{NN}$ and $g_{V2}^{NN}$. The magnitudes of these two LECs can only be estimated using chiral power counting. Reducing this uncertainty in the future may involve lattice calculations of two-nucleon matrix elements, which can be quite challenging in practice. An alternative approach suggested in Ref. \cite{Cirigliano:2024msg} is to take these LECs as free fitting parameters, and perform a global fit, together with $V_{ud}$ to different superallowed transition rates. In this review we point out the possibility to pin down the LECs through the matching procedure outlined in Section~\ref{sec:many.body}.
Other uncertainties include variations in the matching scales of the EFT ($\mu$), the value of the coupling $g_V(\mu)$ at the matching scales, the uncertainty in the isospin breaking corrections taken from~\cite{Hardy:2014qxa} ($\delta_C$), and missing higher order energy-dependent terms in the chiral expansion denoted by $\delta^E_{\rm NS}$. In comparison, the traditional extraction has uncertainties from its approach to computing $\delta_{\rm NS}$, as well as for the transition independent radiative corrections $\Delta_R^V$; the latter is included in $g_V(\mu)$ in the EFT approach.

\subsection{\emph{Ab initio} efforts on constraining $\delta_C$}

Another important theory input in Eq.\eqref{eq:bigF} is the isospin-symmetry-breaking correction $\delta_C$, which is defined in terms of the deviation of the Fermi matrix element $M_F$ from its isospin-symmetric limit:
\begin{equation}
    M_F^2=(M_F^0)^2(1-\delta_C)~.
\end{equation}
It arises from the electromagnetic interactions between the nucleons in the nuclei (primarily from the static Coulomb interaction, thereby the subscript $C$), and according to the Behrends-Sirlin-Ademollo-Gatto theorem~\cite{Behrends:1960nf,Ademollo:1964sr}, it scales quadratically to the ISB potential $V$. The theoretical calculation of $\delta_C$ has been an unsettled problem for more than 6 decades~\cite{MacDonald:1958zz,Towner:2002rg,Towner:2007np,Hardy:2008gy,Xayavong:2017kim,Ormand:1989hm,Ormand:1995df,Satula:2011br,Satula:2016hbs,Liang:2009pf,Auerbach:2008ut} 

Earlier attempts to study ISB effects in nuclear beta decays with \textit{ab initio} methods included the computation of $\delta_C$ in ${}^{10}\text{C}\rightarrow {}^{10}\text{B}$ using NCSM~\cite{Caurier:2002hb}, where a very slow convergence of the outcome with respect to the increasing $N_\text{max}$ was observed. New attempts have been made using in-medium similarity renormalization group (IMSRG) \cite{hergert2016medium}, another \textit{ab initio} method that can cover a larger range of superallowed systems of interest, but no definite conclusion has been achieved~\cite{Martin:2021bud}. Another difficulty for IMSRG is that spurious ISB effects may arise even under the completely isospin-symmetric setup, during the process of operator truncation~\cite{Farren:2024spl}. Nevertheless, calculations of $\delta_C$ across several superallowed systems are underway using IMSRG and, for some of the nuclei, show better convergence with increasing model space than what was observed in earlier ${}^{10}\text{C}\rightarrow {}^{10}\text{B}$ calculations\footnote{Private communication with Jason Holt.}. Very recently, first QMC calculations of $\delta_C$ for the ${}^{10}\text{C}\rightarrow {}^{10}\text{B}$ decay have been performed \cite{piarulli2026quantum}. This work presented results consistent with existing evaluations of $\delta_C$, however, within continuum QMC methods, the uncertainty in $\delta_C$ remains the dominant theoretical limitation for the $^{10}$C system. Hence, \textit{ab initio} calculations of $\delta_C$ are still in the
preliminary stage, in contrast to the recent rapid progress of the $\delta_\text{NS}$ calculations. Yet, they are of high interest and predictions from various models will be vital for further constraining $V_{ud}$.

It is worth noting that all the aforementioned studies of $\delta_C$ were based on the computation of the full Fermi matrix element $M_F$ in the presence of ISB effects in the inter-nucleon interactions. This strategy naturally suffers from the challenge to isolate a small quantity $\delta_C$ from a large number $M_F$, which requires a high accuracy in the theory computation. An alternative approach to circumvent this difficulty is instead to start from a perturbative expression of $\delta_C$ (at the order $\mathcal{O}(V^2)$), and compute the involved nuclear matrix elements in the expression~\cite{Miller:2008my,Miller:2009cg,Seng:2023cvt}, which now requires only a moderate accuracy since the small parameter is explicitly factored out. An extra advantage is that this formalism points towards a connection between $\delta_C$ and the ISB effect in nuclear charge radii~\cite{Seng:2022epj}, and the latter can in principle be measured in experiment and thus serves as a benchmark for theory calculations~\cite{brown2025motivations}. The difficulty, on the other hand, is that this method requires a summation over all intermediate nuclear states just like the dispersive formalism in $\delta_\text{NS}$. It is desirable that future \textit{ab initio} studies of $\delta_C$ proceed in both ways to test their respective efficiency.

\newpage
\section{Recoil-order corrections to allowed beta decays}

\subsection{{Background}}

In-depth investigations of angular correlations in nuclear $\beta$ decay have historically been instrumental in uncovering the ``vector-minus-axial vector'' (V$-$A) nature of the charged current in the electroweak interaction~\cite{lee1956question,jackson1957possible}. Presently, research in nuclear $\beta$ decay continues to be at the forefront of probing for potential deviations from the Standard Model, particularly in the search for additional scalar (S) and tensor (T) Lorentz-invariant interactions that can naturally emerge in extensions beyond the Standard Model. Precise measurements of the $\beta$–$\nu$ angular correlation, the $\beta$ asymmetry, and the Fierz interference term serve as critical tests constraining possible new physics. Ongoing and upcoming experiments aim to enhance the precision of these measurements further \cite{gonzalez2019new, Brodeur2023nuclear}. 

Recoil-order terms are generally omitted in $\beta$-decay theory because they are proportional to $q/m_N$ or higher, where $q$ represents the momentum transfer---typically a few MeV/c---and $m_N$ is the nucleon mass \cite{Holstein1974}. Consequently, in most $\beta$ decays, recoil effects contribute less than one percent of the dominant Fermi and Gamow-Teller (GT) amplitudes. Nonetheless, when measurements reach very high precision, these recoil-order terms become significant and must be incorporated into the analyses of the experiments, particularly in cases where the primary contributions are suppressed or the recoil effects are unexpectedly large.

The study of recoil-order corrections in allowed $\beta$ decays provides a sensitive theoretical handle on sub-percent effects that can obscure or imitate signatures of BSM interactions. 
These corrections arise from higher-order terms in the nuclear weak current--such as weak magnetism, induced tensor, and higher-multipole axial components--which modify the $\beta$ spectrum shape and angular correlations at the $10^{-3}$ level. 
Quantifying these effects within \textit{ab initio} many-body approaches is therefore essential to distinguish genuine BSM contributions from Standard-Model nuclear structure effects. 
In this section, we review three complementary calculations that address this challenge: (i) the $^6$He$\!\to\!^6$Li decay computed with the NCSM, providing the first consistent evaluation of recoil and shape corrections with quantified uncertainties; (ii) the same transition studied using GFMC, incorporating explicit two-body weak currents and full spectral reconstruction; and (iii) the $^8$Li and  $^8$B $\beta$ decays analyzed with the SA-NCSM, where correlations between recoil-order form factors and nuclear deformation help reduce uncertainties on the \emph{ab initio} predictions. 

\subsection{{Theory framework}}

\red{The differential distribution for a $\beta^{-}$ ($\beta^{+}$) decay emitting an electron (positron) of energy $E_e$ and momentum $\vec{k_e}$,
together with an anti-neutrino (neutrino) of momentum $\vec{\nu}$, can be written as
\begin{equation}
  \label{eq:rate}
  \frac{d\Gamma}{dE_e d\Omega_{e} d\Omega_{\nu}} \red{= \frac{G_{\text{F}}^2V_{ud}^2}{2}} \left(E_0-E_e\right)^{2}k_e E_e F^{\red{\mp}}\!\left(Z_{f},E_e\right) C_{\text{corr}} \,
  \red{\frac{1}{2J_{\text{i}}+1}\, \Theta\left(q,\vec{\beta}\cdot\hat{\nu}\right)}\text{.}
\end{equation}
Here $E_0$ is the endpoint energy, $F^{\mp}(Z_f,E_e)$ is the Fermi function accounting for the Coulomb distortion of the charged-lepton wave function, and $C_{\text{corr}}$ denotes correction factors that are not sensitive to nuclear structure. The nuclear-structure dependence is contained in the shape factor $\Theta\left(q,\vec{\beta}\cdot\hat{\nu}\right)$, with $\vec{\beta}=\vec{k}_e/E_e$ and $\hat{\nu}=\vec{\nu}/|\vec{\nu}|$.}

For a pure GT transition, the \red{shape factor} can be written as
$\red{\Theta} \propto 1+a\, \vec{\beta}\cdot\hat{\nu} +b\,\frac{m_{e}}{E_e}$,
where \red{$m_e$ is the electron mass.}
The coefficients $a$ and $b$ encode the leading angular correlation and spectral observables and are particularly sensitive to BSM tensor interactions. Specifically, the $\beta$--$\nu$ angular-correlation coefficient $a$ is quadratically sensitive to the tensor coupling $\epsilon_T$ \cite{gonzalez2019new,sternberg2015limit,  araujo2019scalar}, while the Fierz interference term $b$ depends linearly on $\epsilon_T$~\cite{Falkowski2021,Holstein1977Fierz}.
In the Standard Model limit, one has $a=-1/3$ and $b=0$, and current experiments aim to detect per-mil-level deviations from these values as possible signatures of BSM physics.
However, at \red{such} level of precision, \red{small Standard-Model} nuclear-structure effects become \red{phenomenologically} relevant.
Recoil and shape corrections arise at next-to-leading order (NLO) in several small expansion parameters, namely the non-relativistic $p_\mathrm{F}/m_N\approx0.2$ \red{governing the small components of the Dirac bispinors of the nucleons}, the momentum transfer $qR\approx \red{0.02 A^{1/3}}$ \red{(for a typical endpoint of $2$MeV)}, and the nucleon recoil $q/m_N\approx \red{0.003}$ \red{(for $2$MeV), where $p_{\mathrm{F}}$ is a typical Fermi momentum and $R$ is the nuclear radius. These corrections shift} the effective values of \red{the angular-correlation and Fierz coefficients} and, if not properly accounted for, can \red{therefore imitate or obscure} the effects of BSM couplings.

\subsubsection{{Multipole operators decomposition}}

\red{A systematic way to account for all nuclear-structure effects is through a multipole decomposition of the weak nuclear current. Among the several conventions used in the literature, we focus here on the Donnelly--Walecka formulation. Introduced for precision applications in Ref.~\cite{glick2022formalism}, its advantages include the generality for arbitrary $\beta$ transition, the ability to extend the calculation to any required order, the possibility to quantify remaining theoretical uncertainties, and the use of operators that are well-established in the nuclear-structure community.}

\red{
Within this framework, the shape factor of a general $\beta^{\mp}$-decay takes the form~\cite{Walecka:1995}:
\begin{multline}
\Theta\left(q,\vec{\beta}\cdot\hat{\nu}\right)=
\sum_{J=1}^{\infty}\left[\left(1-\left(\hat{\nu}\cdot\hat{q}\right)\left(\vec{\beta}\cdot\hat{q}\right)\right)\left(|\braket{\Vert \hat{E}_{J}\Vert} |^{2}+|\braket{\Vert \hat{M}_{J}\Vert} |^{2}\right)
\pm\hat{q}\cdot\left(\hat{\nu}-\vec{\beta}\right)
2\mathfrak{Re}\left(\braket{\Vert \hat{E}_{J}\Vert} \braket{\Vert \hat{M}_{J}\Vert}^{*}\right)\vphantom{\left|\left\langle \left\Vert \hat{M}_{J}\right\Vert \right\rangle \right|^{2}}\right]\\
+\sum_{J=0}^{\infty}\left[\left(1-\vec{\beta}\cdot\hat{\nu}+2\left(\hat{\nu}\cdot\hat{q}\right)\left(\vec{\beta}\cdot\hat{q}\right)\right)|\braket{\Vert \hat{L}_{J}\Vert} |^{2}+\left(1+\vec{\beta}\cdot\hat{\nu}\right)|\braket{\Vert \hat{C}_{J}\Vert} |^{2}
-\hat{q}\cdot\left(\hat{\nu}+\vec{\beta}\right)
2\mathfrak{Re}\left(\braket{\Vert \hat{L}_{J}\Vert} \braket{\Vert \hat{C}_{J}\Vert}^{*}\right)\vphantom{\left|\left\langle \left\Vert \hat{M}_{J}\right\Vert \right\rangle \right|^{2}}\right]\text{,}\\
\label{eq:multipole_exp}
\end{multline}
where $\braket{\Vert \hat{O}_{J}\Vert}$ denotes a reduced matrix element of a rank $J$ spherical tensor operator $\hat{O}_{J}$, built from the nuclear Axial-vector or polar-Vector current between the initial- and final-nuclear state wave functions.
The multipole operators are the Coulomb, longitudinal, electric, and magnetic operators, defined by:
%
\label{eq:multipoles}
\begin{align}
\hat{C}_{J}\left(q\right) & \equiv  \int d^{3}r
\, M_{J}\left(q\vec{r}\right) 
\, J_{0}\left(\vec{r}\right), \\
\hat{L}_{J}\left(q\right) & \equiv  \frac{i}{q}\int d^{3}r\left[\vec{\nabla}
M_{J}\left(q\vec{r}\right) 
\right]\cdot\vec{J}\left(\vec{r}\right), \\
\hat{E}_{J}\left(q\right) & \equiv \frac{1}{q}\int d^{3}r\left[\vec{\nabla}\times
\vec{M}_{JJ1}\left(q\vec{r}\right) 
\right]\cdot\vec{J}\left(\vec{r}\right), \\
\hat{M}_{J}\left(q\right) & \equiv  \int d^{3}r 
\, \vec{M}_{JJ1}\left(q\vec{r}\right) 
\cdot\vec{J}\left(\vec{r}\right)\mbox{,}
\end{align}
%
with
\begin{align}
M_{J}\left(q\vec{r}\right) & \equiv  j_{J}\left(qr\right)Y_{J}\left(\hat{r}\right), \\
\vec{M}_{JL1}\left(q\vec{r}\right) & \equiv  j_{L}\left(qr\right)\vec{Y}_{JL1}\left(\hat{r}\right)\mbox{.}
\end{align}
Here $\vec{J}\left(\vec{r}\right)$ and $J_0\left(\vec{r}\right)$ are the nuclear current and nuclear charge, respectively, $j_{J}$ are spherical Bessel functions, and $Y_{JM}$ ($\vec{Y}_{Jl1}^{M}$) are spherical harmonics (vector spherical harmonics).}

Following Ref.~\cite{glick2022formalism}, the \red{full shape factor of a pure GT transition} including recoil and shape corrections can be \red{simplified to}
\begin{equation}
  \label{eq:GT}
  \red{\Theta\left(q,\vec{\beta}\cdot\hat{\nu}\right) = \,\,}
  3 \left(1+\delta_{1}\right)
  \left[1 -\frac{1}{3} \left(1+\tilde{\delta}_{a}\right) \vec{\beta}\cdot\hat{\nu} +\delta_b \frac{m_{e}}{E_e}\right]
  \vert\braket{\Vert \hat{L}_{1}^{A}\Vert} \vert^{2},
\end{equation}
\red{where} $\braket{\Vert \hat{L}_{1}^{A}\Vert}$ corresponds to the 
reduced matrix element of the leading GT operator,
while $\delta_1$, $\tilde{\delta}_a$, and $\delta_b$ parameterize recoil and shape corrections arising from subleading multipole contributions.
These corrections depend on reduced matrix elements of higher multipole operators of the weak current and can be expressed as:

\begin{equation}
  \label{eq:corrections}
  \begin{split}
    \delta_{1} &\equiv\frac{2}{3}\mathfrak{Re}\left[
      -E_0
      \frac{\braket{\Vert \hat{C}_{1}^{A}/q\Vert} }{\braket{\Vert\hat{L}_{1}^{A}\Vert} }
      \red{\pm}\sqrt{2}\left(E_0-2E_e\right)
      \frac{\braket{\Vert \hat{M}_{1}^{V}/q\Vert} }{\braket{\Vert \hat{L}_{1}^{A}\Vert} }
      \red{+\sqrt{2}\frac{\braket{\Vert \hat{E}_{1}^{A '}\Vert}}{\braket{\Vert \hat{L}_{1}^{A}\Vert} }}\right]
    , \\
    \tilde{\delta}_{a} &\equiv\frac{4}{3}\mathfrak{Re}\left[ 2E_0\frac{\braket{\Vert \hat{C}_{1}^{A}/q\Vert} }{ \braket{\Vert \hat{L}_{1}^{A}\Vert} }
    \red{\pm}\sqrt{2}\left(E_0-2E_e\right)
      \frac{\braket{\Vert \hat{M}_{1}^{V}/q\Vert} }{ \braket{\Vert \hat{L}_{1}^{A}\Vert} }
    \red{+\sqrt{2}\frac{\braket{\Vert \hat{E}_{1}^{A '}\Vert}}{\braket{\Vert \hat{L}_{1}^{A}\Vert}}}\right]
    , \\
    \delta_{b} &\equiv\frac{2}{3}m_{e}\mathfrak{Re}\left[ \frac{\braket{ \Vert \hat{C}_{1}^{A}/q\Vert} }{\braket{ \Vert \hat{L}_{1}^{A}\Vert} }
      \red{\pm}\sqrt{2}
      \frac{\braket{ \Vert \hat{M}_{1}^{V}/q\Vert} }{\braket{ \Vert \hat{L}_{1}^{A}\Vert} } \right].
  \end{split}
\end{equation}
\red{Here the superscript $A(V)$ denotes that the multipole operator was calculated using the axial-(polar-)vector contribution to the weak nuclear current or charge.
An extra simplification was obtained by utilizing the relation
\begin{equation}
    \hat{E}_{J}=\sqrt{\frac{J+1}{J}}\hat{L}_{J}-i\sqrt{\frac{2J+1}{J}}\int d^{3}rj_{J+1}\left(qr\right)\vec{Y}_{J,J+1,1}\left(\hat{r}\right)\cdot\vec{J}\left(\vec{r}\right) \equiv \sqrt{\frac{J+1}{J}}\hat{L}_{J} + \hat{E}_{J}^{'},
\end{equation}
that leaves a small residual correction $\hat{E}_{J}^{'}$, of $\hat{E}_{J}$ regarding $\hat{L}_{J}$.}

\red{For simplicity, here we introduced only the corrections arising from the rank-1 operators, as those are the only ones contributing to the decay of $^6$He presented here using this formalism (see Sec. \ref{sec:6he.ncsm} and \ref{sec:6he.qmc}).
A general $\beta$-decay transition with a $J_{\text{i}}\rightarrow J_{\text{f}}$ angular momentum change, will contain corrections from all the integer total angular momenta $J$ which uphold the selection rule $\left|J_{\text{i}}-J_{\text{f}}\right|\leq J\leq J_{\text{i}}+J_{\text{f}}$. An example of such a case is seen in Sec.~\ref{sec:8ncsm}.}

\red{For the applications discussed below, it is useful to express the multipoles in the impulse approximation, using one-body axial and vector operators constructed as in Ref.~\cite{Donnelly:1979ezn}:
\begin{equation}
  \label{eq:multipoles1b}
  \begin{split}
    \hat{C}_{J}^{A}\left(q\right) &= \frac{i}{m_N}
    \sum_{j=1}^{A} \tau^{\pm}_j
    \left[g_A M_{J}\left(q\vec{r}_{j}\right) \vec{\sigma}_j \cdot \vec{\nabla}
    +\frac{1}{2} \left(g_A - \frac{E_0 + \Delta E_c}{2m_N}\tilde{g}_P
     \right)
       \vec{\nabla} M_{J}(q\vec{r}_j) \cdot \vec{\sigma}_j \right],\\
    \hat{L}_{J}^{A}\left(q\right) &=\frac{i}{q} \left(g_A+\frac{q^2}{4m_N^2}\tilde{g}_P
    \right)
    \sum_{j=1}^{A} \tau^{\pm}_j
    \vec{\nabla} M_{J}(q
      \vec{r}_j) \cdot \vec{\sigma}_j
    , \\
      \hat{E}_{J}^{A}\left(q\right) & =\frac{1}{q} \left(g_A+\frac{q^2}{4m_N^2}\tilde{g}_P
    \right)
\sum_{j=1}^{A}  \tau_{j}^{\pm} \left[ \vec{\nabla}\times\vec{M}_{JJ1}\left(q\vec{r}_{j}\right)\right]\cdot\vec{\sigma}_j, \\
\hat{M}_{J}^{A}\left(q\right) & =\left(g_A+\frac{q^2}{4m_N^2}\tilde{g}_P
    \right)
\sum_{j=1}^{A}  \tau_{j}^{\pm} \vec{M}_{JJ1}\left(q\vec{r}_{j}\right)\cdot\vec{\sigma}_j,\\
\hat{M}_{J}^{V}\left(q\right) &= -\frac{i}{m_N}
    \sum_{j=1}^{A} \tau^{\pm}_j
    \left\{ g_V \vec{M}_{JJ1}\left(q\vec{r}_{j}\right)\cdot \vec{\nabla}
    +\frac{i}{2}
      \mu \left[ 
      \vec{\nabla} \times \vec{M}_{JJ1}(q
      \vec{r}_j) \right] \cdot \vec{\sigma}_j \right\},\\
  \end{split}
\end{equation}
with additional electromagnetic contributions from Coulomb energy shift $\Delta E_c$ following Refs.~\cite{behrens1982electron, glick2022formalism}. All the sums run up to the nuclear mass number $A$ and $\sigma_j/2$ ($\tau_j/2$) is the spin (isospin) of the $j^\mathrm{th}$ particle.
$g_A$, $g_V$, $\tilde{g}_P(q^2) \simeq -\frac{(2m_N)^2}{m_{\pi}^2-q^2}g_A(0)$,
and $\mu$ are the axial, vector, induced pseudoscalar form factors, and magnetic moment, respectively. All are functions of $q^2$ which is not written explicitly for simplicity.}

\red{An alternative formalism leverages the fact that the charge and current operators in momentum space can be written in terms of the reduced multipoles. For nuclear states with initial and final nuclear angular momenta $J_i$ and $J_f$, respectively, having projection $M$ via the relations~\cite{Carlson:1997qn}
\begin{eqnarray} \label{eq:charge}
\mel{J_fM}{\rho^{\dagger}_{\pm}(\bfq)}{J_i M} &=& (-1)^{J_i-M} \sum_J \sqrt{4\pi} (-i)^J P_J(\cos\theta) c^{M}_{J_fJ_iJ} \reduce{}{C_J}{} \\ 
\label{eq:longit} \mel{J_fM}{\bfqh\cdot\bfj^{\dagger}_{\pm}(q)}{J_i M} &=& (-1)^{J_i-M} \sum_L \sqrt{4\pi} (-i)^J P_J(\cos\theta) c^{M}_{J_fJ_iJ} \reduce{}{L_J}{}\\ \nonumber
\mel{J_fM}{\bfeh_1\cdot\bfj^{\dagger}_{\pm}(\bfq)}{J_i M} &=& (-1)^{J_i-M+1} \sum_{J\geq 1} \sqrt{4\pi} (-i)^J c^{M}_{J_fJ_iJ} \frac{P^1_J(\cos\theta)}{\sqrt{J(J+1)}} \times \\
&& [\reduce{}{E_J}{}\cos\phi - i\reduce{}{M_J}{} \sin\phi] \label{eq:elec} \\ \nonumber
\mel{J_fM}{\bfeh_2\cdot\bfj^{\dagger}_{\pm}(\bfq)}{J_i M} &=& (-1)^{J_i-M+1} \sum_{J\geq 1} \sqrt{4\pi} (-i)^J c^{M}_{J_fJ_iJ} \frac{P^1_J(\cos\theta)}{\sqrt{J(J+1)}} \times \\
&& [\reduce{}{E_J}{} \sin\phi + i\reduce{}{M_J}{}\cos\phi]\, , \label{eq:mag}
\end{eqnarray}
where $\theta$ and $\phi$ are, respectively, the polar and azimuthal angle that $\bfq$ makes with respect to the spin-quantization axis, $P_J(\cos\theta)$ is a Legendre polynomial, $P^1_J(\cos\theta)$ is an associated Legendre Polynomial, the Clebsch-Gordan coefficient is defined as,
\begin{equation}
    c^M_{J_fJ_iJ} = \inner{J_f,M;J_i,-M}{J,0}\, ,
\end{equation}
and the unit vectors are given by $\bfeh_2 = \bfzh \times \bfqh$ and $\bfeh_1 = \bfeh_2 \times \bfqh$. In the $\beta$ decay process, we are dealing with the charge changing electroweak charge and current operators. Additionally, $\bfj_{\pm}^{\dagger} = \bfj_x^{\dagger} \pm i\bfj_y^{\dagger}$-- with the subscripts $x$ and $y$ indicating isospin components-- and $\rho_{\pm}^{\dagger}$ is similarly defined.}

\subsubsection{{Holstein’s form factors}}

\red{
Holstein's form-factor formalism offers a complementary organization of the same physics for allowed $\beta$ decays. 
Widely used for experimental analyzes, it parametrizes the leading long-wavelength description of the nuclear structure effects in terms of a small number of reduced form factors.
The leading Gamow--Teller response is denoted by the form factor $c$,
while the dominant recoil-order Standard-Model corrections are described by the induced-tensor form factor $d^I$,
the weak-magnetism form factor $b_{\mathrm{WM}}$,
and the higher-rank axial form factors $j_K$.
With the phase convention used here\footnote{Note that in Ref. \cite{Holstein1974} the convention of angular momentum couplings is such that the final state's angular momentum is coupled to the spherical tensor to yield the initial state's angular momentum. Here, we couple the initial state to the spherical tensor to yield the final state. Hence, the expressions in Eq. (\ref{eq:Holstein}) differ from Ref. \cite{Holstein1974} by a factor of $(-1)^{(J_f-J_i)}$.}, these form factors are defined as:
\begin{eqnarray}
c &=&  \frac{(-)^{(J_f-J_i)}}{\sqrt{2J_i+1}} g_A\braket{J_f \Vert \sum_{i=1}^A \tau^{\pm}_i  \sigma_i  \Vert J_i} = \frac{(-)^{(J_f-J_i)}}{\sqrt{2J_i+1}}g_A M_{GT},\nonumber\\
d^I &=& A\frac{(-)^{(J_f-J_i)}}{\sqrt{2J_i+1}}  g_A \braket{J_f \Vert  \sum_{i=1}^A \tau^{\pm}_i  \sqrt{2}[L_i \times \sigma_i ]^1 \Vert J_i}, \nonumber \\
b_{\mathrm{WM}} &=& A\frac{(-)^{(J_f-J_i)} }{\sqrt{2J_i+1}} \Big[g_V \braket{J_f \Vert  \sum_{i=1}^A \tau^{\pm}_i L_i \Vert J_i} +  g_M\braket{J_f \Vert  \sum_{i=1}^A \tau^{\pm}_i  \sigma_i  \Vert J_i}  \Big], \nonumber \\
j_K &=& - \frac{2}{3}
A^2 m_N^2 
\frac{(-)^{(J_f-J_i)}}{\sqrt{2J_i+1}} g_A \braket{J_f \Vert \sum_{i=1}^A \tau^{\pm}_i  [Q_i \times \sigma_i]^K \Vert J_i}, 
{\rm with\,} K=1,2,3. 
\label{eq:Holstein}
\end{eqnarray}
The operators $L_i$ and $Q_i=\sqrt{16\pi/5}r_i^2Y_{2\mu}(\hat r_i)$ denote the single-particle orbital angular momentum and quadrupole tensor, respectively.
Thus, $c$ is proportional to the allowed GT matrix element, $d^I$ probes the axial spin-orbit recoil operator,
$b_{\mathrm{WM}}$ probes the vector magnetic response, and $j_K$ contain quadrupole-spin matrix elements.
The latter contribute only when the angular momentum selection rules allow rank-$K$ transition operators.
}

\subsubsection{\label{sec:matchingGH}{Connection between the approaches}}

\red{
The Donnelly--Walecka and Holstein descriptions are not independent formalisms. 
The multipole expansion provides the full nuclear-current response, while Holstein's form factors correspond to its leading long-wavelength limits.
For $q/m_N \ll1$ and $qR\ll1$, the spherical Bessel functions entering the multipole operators can be expanded in powers of $q/m_N$ and $qR$, and the leading terms reduce to the simple one-body structures appearing in Eq.~\eqref{eq:Holstein}.}

\red{Taking the leading $qR$ order of the one-body impulse-approximation operators of Eq.~\eqref{eq:multipoles1b}, the rank-one multipole matrix elements then reduce to the Holstein form factors as
\begin{equation}
\label{eq:mapping}
\begin{split}
\braket{J_f\Vert \hat{L}_1^A \Vert J_i}
&\simeq
i\,
\frac{\sqrt{2J_i+1}}{(-)^{(J_f-J_i)}}
\frac{c}{2\sqrt{3\pi}},
\\
\braket{J_f\Vert \hat{C}_1^A/q \Vert J_i}
&\simeq
i\,
\frac{\sqrt{2J_i+1}}{(-)^{(J_f-J_i)}}
\frac{d^I}{2\sqrt{3\pi}A m_N},
\\
\braket{J_f\Vert \hat{M}_1^V/q \Vert J_i}
&\simeq
i\,\,
\frac{\sqrt{2J_i+1}}{(-)^{(J_f-J_i)}}
\,\frac{b_{\mathrm{WM}}}{2\sqrt{6\pi}A m_N}.
\end{split}
\end{equation}
In writing Eq.~\eqref{eq:mapping}, the complex couplings appearing in the multipole operators at Eq.~\eqref{eq:multipoles1b} have been reduced to $g_A(q^2)$ in the strict long-wavelength limit.
This is a adequate for
$\hat L_1^A$, $\hat E_1^A$ and $\hat{M}_1^V$, 
where the induced-pseudoscalar term is suppressed by $q^2/(2m_N)^2$.
The axial Coulomb multipole $\hat C_1^A$ requires more care because it contains the combination
$(E_0+\Delta E_c)/(2m_N)$, including the Coulomb displacement energy.
This contribution is not part of the Holstein parametrization and can
be numerically important in some cases, see, for example, Ref.~\cite{glick2026ab}.
}

\red{
The same long-wavelength expansion relates higher-rank axial multipoles to the quadrupole--spin operators entering Holstein's form factors $j_K$. In the conventions of Eq.~\eqref{eq:Holstein}, one obtains:
\begin{equation}
\begin{split}
\braket{J_f\Vert \hat{M}_2^A(q)/q^2 \Vert J_i}
&\simeq
-
\frac{\sqrt{2J_i+1}}{(-)^{J_f-J_i}}
\frac{j_2}{40 \sqrt{5\pi} A^2 m_N^2}
,
\\[1em]
\braket{J_f\Vert \hat{L}_3^A/q^2 \Vert J_i}
&\simeq
-i
\frac{\sqrt{2J_i+1}}{(-)^{J_f-J_i}}
\frac{3}{8}
\frac{j_3}{\sqrt{105\pi} A^2 m_N^2}.
\end{split}
\end{equation}
%
%
These two relations contribute only when the initial and final nuclear
angular momenta allow rank-2 or rank-3 reduced matrix elements. They
therefore do not enter the $^6$He ground-state decay, but become relevant for
the $A=8$ transitions discussed below.
}

\red{The leading correspondence between the two descriptions can therefore be
summarized as
\begin{equation}
\boxed{
\hat{L}_{1}^{A} \rightarrow c,\qquad
\hat{C}_{1}^{A} \rightarrow d^{I},\qquad
\hat{M}_{1}^{V} \rightarrow b_{\mathrm{WM}},\qquad
\hat{M}_{2}^{A} \rightarrow j_{2},\qquad
\hat{L}_{3}^{A} \rightarrow j_{3}.
}
\end{equation}
This notation should be read as a long-wavelength reduction of the multipole matrix elements, not as an equality between the full operators and Holstein's form factors. 
Corrections beyond this leading correspondence arise from subleading terms in the multipole expansion. 
}

\subsection{$^6$He$\to^6$Li beta decay corrections with NCSM \label{sec:6he.ncsm}} 

The $\beta$-decay of $^6$He offers a particularly clean probe of BSM tensor currents thanks to its pure GT character, and its light nuclear mass which makes it computationally tractable within \textit{ab initio} many-body frameworks, allowing high-precision theoretical predictions.
Several experimental programs are currently pursuing precision studies of this decay. At the Laboratoire de Physique Corpusculaire de CAEN~\cite{cirigliano2019precision},
and the University of Washington CENPA through the He6-CRES Collaboration~\cite{asner2015single}, efforts focus on measuring the $\beta$-electron energy spectrum, while at SARAF, SNRC~\cite{ohayon2018weak} the emphasis is on the angular correlation between the emitted $\beta$ particle and neutrino. As these measurements strive for per-mil level accuracy, precise Standard Model calculations are essential to fully exploit their potential for constraining new physics.

Ref.~\cite{glick2022nuclear} carried out the first fully consistent \textit{ab initio} calculation of recoil and shape corrections
including uncertainty quantification essential for precision searches.
Nuclear wave functions were obtained using the NCSM with $\chi$EFT interactions~\cite{epelbaum2009,machleidt2011,PhysRevLett.84.5728,BARRETT2013131}.
Two Hamiltonians were employed: NNLO$_\mathrm{opt}$ (two-nucleon forces only) and NNLO$_\mathrm{sat}$ (including three-nucleon forces)~\cite{N2LOopt,NNLOsat2015}. 
Matrix elements were computed in the impulse approximation, with one-body axial and vector operators constructed as in \red{Eq. \eqref{eq:multipoles1b},
%
with} electromagnetic contributions from Coulomb energy shift $\Delta E_c \approx 0.85$MeV 
evaluated using Refs.~\cite{PhysRevC.56.1720,ANTONY19971}, as well as \red{additional} Coulomb relevant order corrections from Refs.~\cite{RevModPhys.90.015008, hayen2020consistent}.
Translational invariance was ensured by using one-body densities in Jacobi coordinates~\cite{Navratil2004}, thus eliminating spurious center-of-mass motion.
The NCSM calculations were performed using model spaces up to $N_{\max}=12$ and with varying HO parameter $\hbar\Omega=16$–24 MeV to assess convergence.
Method and parametric uncertainties from the many-body solver were quantified by varying model spaces and comparing the two chiral interactions.
Uncertainties due to the omission of two-body weak currents were estimated from EFT power counting ($\epsilon_{\mathrm{EFT}}\lesssim0.15$) and from comparisons with experimental observables such as the $^{6}$Li magnetic moment and $B(M1)$ transition, and the $^6\text{He}$ half-life~\cite{PhysRevC.87.035503,PhysRevLett.126.102501,PhysRevC.79.065501}.
The subleading multipoles and higher-order terms omitted from the nonrelativistic expansion were estimated following the controlled formalism of Ref.~\cite{glick2022formalism}\red{, with $p_\mathrm{F}/m_N\approx0.2$, $qR\approx0.05$, $\alpha Z_f \approx 0.02$, and $q/m_N\approx0.004$ to match the specific decay (here, $\alpha$ is the fine structure constant)}.

\begin{figure}[htb]
    \centering
    \includegraphics[width = 0.6\textwidth]{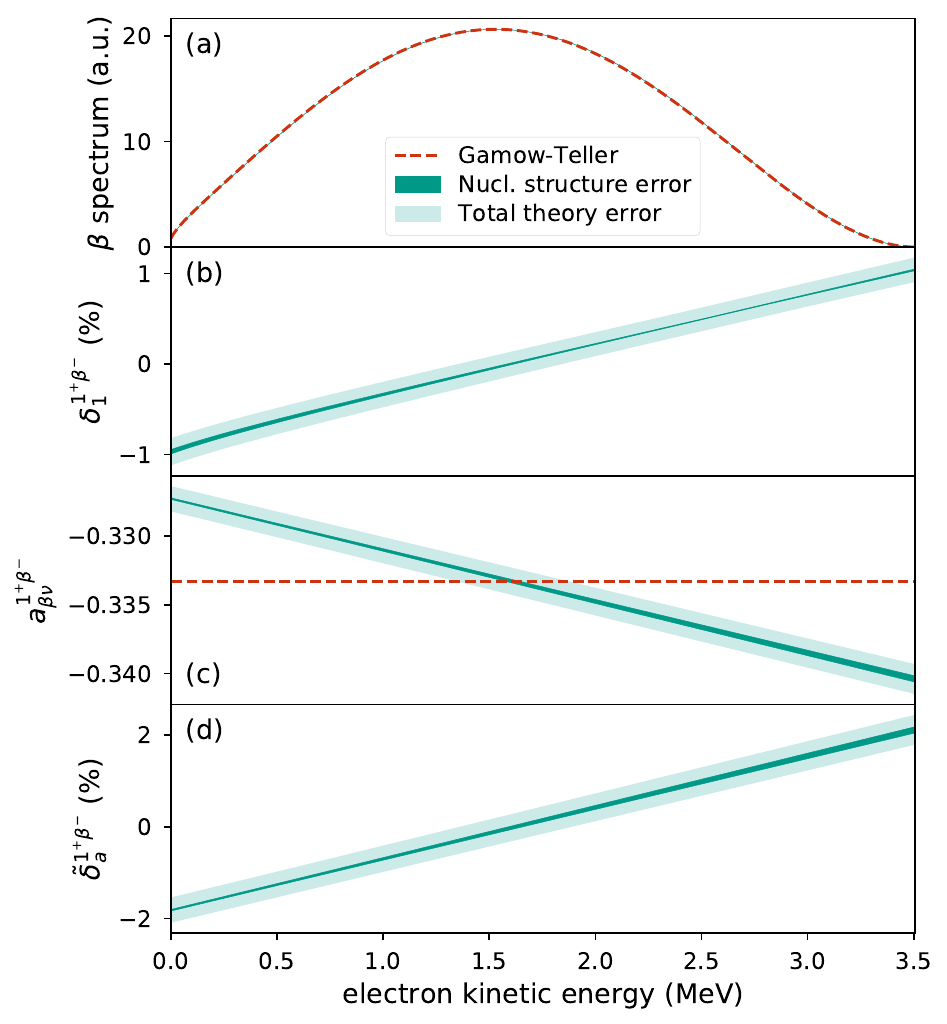}
    \caption{(a) Calculated energy dependence of the spectrum of $^{6}$He $\beta$-decay, in arbitrary units. Dashed line is the pure GT spectrum, while the filled bands include nuclear-structure dependent corrections. (b) The residual nuclear structure correction $\delta_{1}$ compared to the pure GT spectrum (Eq.~\eqref{eq:corrections}).
    (c) Energy dependence of the angular correlation $a_{\beta\nu}$. Dashed line corresponds to the SM value, $a_{\beta\nu}=-1/3$. (d) Relative size of the
    $\tilde{\delta}_{a}$ correction from~\eqref{eq:corrections}.
    The width of the dark filled bands shows the variation with the employed nuclear Hamiltonian and NCSM model space parameters for HO frequency $\hbar\Omega = 16, 20, 24$~MeV, $N_\mathrm{max}=8, 10, 12$ ($10, 12, 14$) using the NNLO${}_{\mathrm{sat}}$ (NNLO${}{_\mathrm{opt}}$) interaction, using translationally-invariant one-body densities. The width of the light filled band shows the total estimated theory error.
    Reprinted figure from~\cite{glick2022nuclear}.
    \label{fig:spect}}
\end{figure}

\begin{table}[h!]
\centering
\caption{Calculated nuclear-structure corrections for the $^{6}$He$\!\rightarrow\!^{6}$Li $\beta$ decay obtained in Ref.~\cite{glick2022nuclear} using the \textit{ab initio} NCSM with chiral EFT interactions\red{, and in Ref.~\cite{King:2022zkz} using GFMC.}
Angle brackets $\langle\cdot\rangle$ denote averaging over the $\beta$-energy spectrum.
\red{NCSM results include contributions from Coulomb energy shift $\Delta E_c$ and other Coulomb corrections not included in the GFMC calculations, whereas GFMC results include contributions from two-body currents not included explicitly in the NCSM calculations, though their effects are accounted for in the quoted NCSM uncertainties.
The} quoted NCSM uncertainties include methodological and parametric uncertainties of the many-body solver, as well as contributions from subleading multipoles, higher-order\red{s in the non-relativistic expansion}, and omitted two-body currents (EFT truncation) \red{which dominate the total uncertainties.
The quoted GFMC uncertainties represent the statistical uncertainty from fitting the multipoles with model uncertainties included on nuclear matrix elements for the energy range used to fit the two-nucleon interaction and for different versions of the three-nucleon force.}}
\begin{tabular}{lccccl}
\hline\hline
Observable & Symbol & NCSM~\cite{glick2022nuclear} & \red{GFMC~\cite{King:2022zkz} }\\
\hline
Fierz-like term & $\delta_b$ & $-1.52(18) \times10^{-3}$ & \red{$-1.47(3) \times10^{-3}$} \\
Angular-correlation correction & $\langle\tilde{\delta}_a\rangle$ & $-2.54(68) \times10^{-3}$ & \red{$-1.43(20)\times 10^{-3}$} \\
Total extracted angular-correlation & $a$ & $-0.3331 (32)$ & \red{$-0.3327(30)$} \\
\hline\hline
Multipole Operator & Symbol & NCSM~\cite{glick2022nuclear} & \red{GFMC~\cite{King:2022zkz} } \\
\hline
Longitudinal & $\langle \hat{L}_{1}^{A} / i \rangle$ & 0.464(16) 
& \red{$-0.447(7)$} \\
Electric & $\langle \hat{E}_{1}^{A} / i \rangle$ & 0.657(23) & \red{$-0.633(10)$}  &  \\
Magnetic \red{$\left[1/\text{MeV}\right]$} 
& $\langle \hat{M}_{1}^{V} / i q \rangle$ & $-1.36(15)\times10^{-3}$ & \red{$1.36(2)\times 10^{-3}$} &  \\
Coulomb \red{$\left[1/\text{MeV}\right]$} 
& $\langle \hat{C}_{1}^{A} / i q \rangle$ & $-1.39 (18)\times10^{-4}$ & \red{$1.10(5)\times 10^{-4}$}  &  \\
\hline\hline
Holstein's Form Factor & Symbol &  NCSM~\cite{glick2022nuclear}  &\red{GFMC~\cite{King:2022zkz}} & Shell Model~\cite{PhysRevC.12.2016} &\red{Exp.} \cite{PhysRevC.12.2016} \\
\hline
Gamow--Teller & $|c|$ &  $2.85(14)$ & \red{$2.74(4)$} & &$2.75(3)$\\
Weak magnetism & $|b_{\mathrm{WM}}|$ & $66.7(8.3)$ & \red{66.4(1.1)} &  &$69.0(1.0)$\\
Induced tensor & $|d^{I}|$ &  $7.73 (59)$ & \red{$5.37(23)$} & $2.4$ &\\
Induced tensor to GT ratio & $|d^{I}/(A c)|$  & $0.45(4)$ &\red{$0.33(1)$} & $0.12$ &$2.0(1.5)$ \\
\hline\hline
\end{tabular}
\label{tab:6He_results}
\end{table}

The resulted calculations, presented in Fig.~\ref{fig:spect} and summarized in Table~\ref{tab:6He_results}, revealed that nuclear-structure corrections produce measurable modifications to the GT observables.
After averaging over the $\beta$-energy spectrum, the correction to the angular correlation was found to be $\langle \tilde{\delta}_a \rangle = -2.54(68) \cdot 10^{-3}$,
which corresponds to a 0.7\% modification of the GT angular correlation $a=-1/3$.
The extracted angular-correlation coefficient from available experimental data~\cite{PhysRev.132.1149}, after including radiative, atomic, and the newly calculated structure effects, becomes
$a=-0.3331(32)$,
slightly shifted from the naive GT value but consistent with the standard model.
Additionally, a nonzero Fierz-like contribution $\delta_b = -1.52(18)\cdot 10^{-3}$ arises entirely due to nuclear structure effects.
These value implies an energy-dependent distortion of the spectrum at the $10^{-3}$ level, comparable to the target precision of current experiments.

The results were also benchmarked against the 1975 calculation by Calaprice~\cite{PhysRevC.12.2016}, which expressed the transition in terms of the Gamow–Teller ($c$), weak magnetism ($b_{\mathrm{WM}}$), and induced tensor ($d$) form factors. 
The NCSM values for the Gamow–Teller and weak-magnetism form factors, $c = 2.85(14)$ and $b_{\mathrm{WM}} = 66.7(8.3)$, agree well with Calaprice’s values ($c = 2.75(3)$, $b_{\mathrm{WM}} = 69.0(1.0)$) extracted from measurements of half-life and M1 transition, respectively.
For the induced tensor component, however, the NCSM result $d^{I} \approx 7.73(59)$ is significantly larger than the value obtained in the shell-model calculation reported by Calaprice ($d^{I} \approx 2.4$). 
Nonetheless, when expressed as a ratio $d^{I}/(A c) \approx 0.45(4)$, the NCSM prediction lies closer to the value $2.0(1.5)$ inferred from the angular-correlation measurements analysis, which assumes all other recoil corrections and other effects are well known (which, of course, is not the case).

\begin{figure}[htb]
    \centering
    \includegraphics[width=0.6\textwidth]{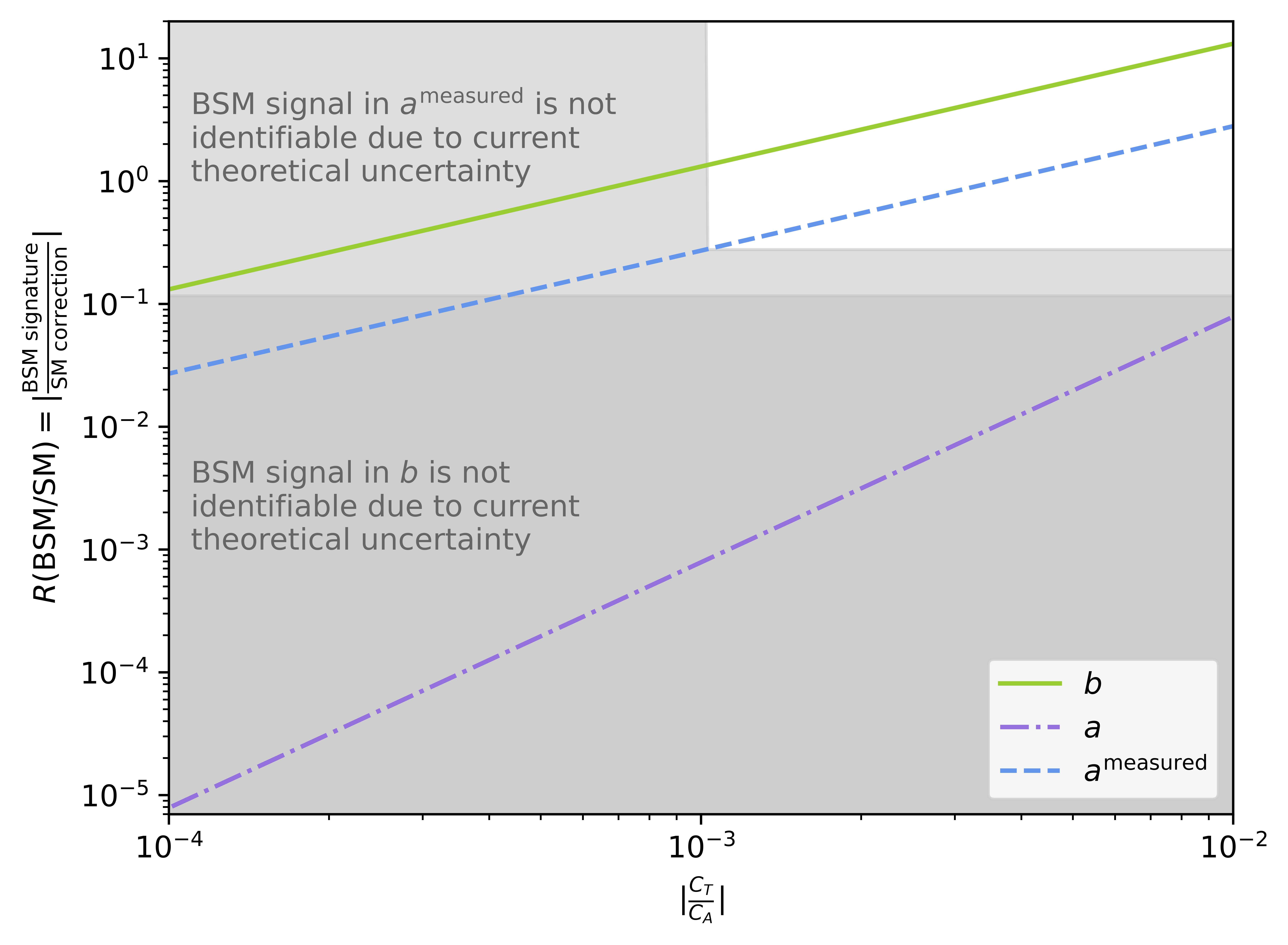}
    \caption{
    Ratio of possible BSM-induced signals to SM nuclear-structure corrections for the $^{6}$He $\beta$ decay~\cite{glick2022nuclear}. 
    The solid green, dash-dotted purple, and dashed blue lines correspond to the Fierz term $b$, angular correlation $a$, and experimentally measured $a^{\text{measured}}$, respectively. 
    If the ratio is of the order of $1$, the corrections should be calculated explicitly. The limit of theoretical uncertainty consequently occurs when the ratio is about the size of the theoretical uncertainty in calculating the SM corrections.
    The white region indicates where the theoretical uncertainty of the SM corrections (below $10^{-4}$) allows clean separation between SM and BSM effects. In the other domains, separation is limited by theoretical uncertainties.
    Reprinted figure with permission from~\cite{glick2023multipole}. Copyright 2023 by the American Physical Society.
    \label{fig:BSM_signatures}}
\end{figure}

A key outcome of these calculations is their quantified theoretical uncertainties, achieved for the first time. The resulting accuracy of order $10^{-4}$ places future measurements within the discovery window for BSM physics, as illustrated in Fig.~\ref{fig:BSM_signatures}. 
Experiments observing additional per-mil deviations from these SM predictions would imply nonzero BSM tensor couplings. 
Furthermore, the indirect dependence of the measured angular correlation on the Fierz term $b$ transforms its naive quadratic sensitivity to $\epsilon_T$ into an effective linear one, significantly enhancing experimental reach for tensor interactions within angular correlation measurements. 
Overall, these NCSM results establish a high-precision Standard Model benchmark for the $^{6}$He$ \rightarrow {}^{6}$Li decay. 

\subsection{$^{6}$He$ \rightarrow {}^{6}$Li beta decay spectrum with QMC \label{sec:6he.qmc}}

\begin{figure}
    \centering
    \includegraphics[width=0.5\linewidth]{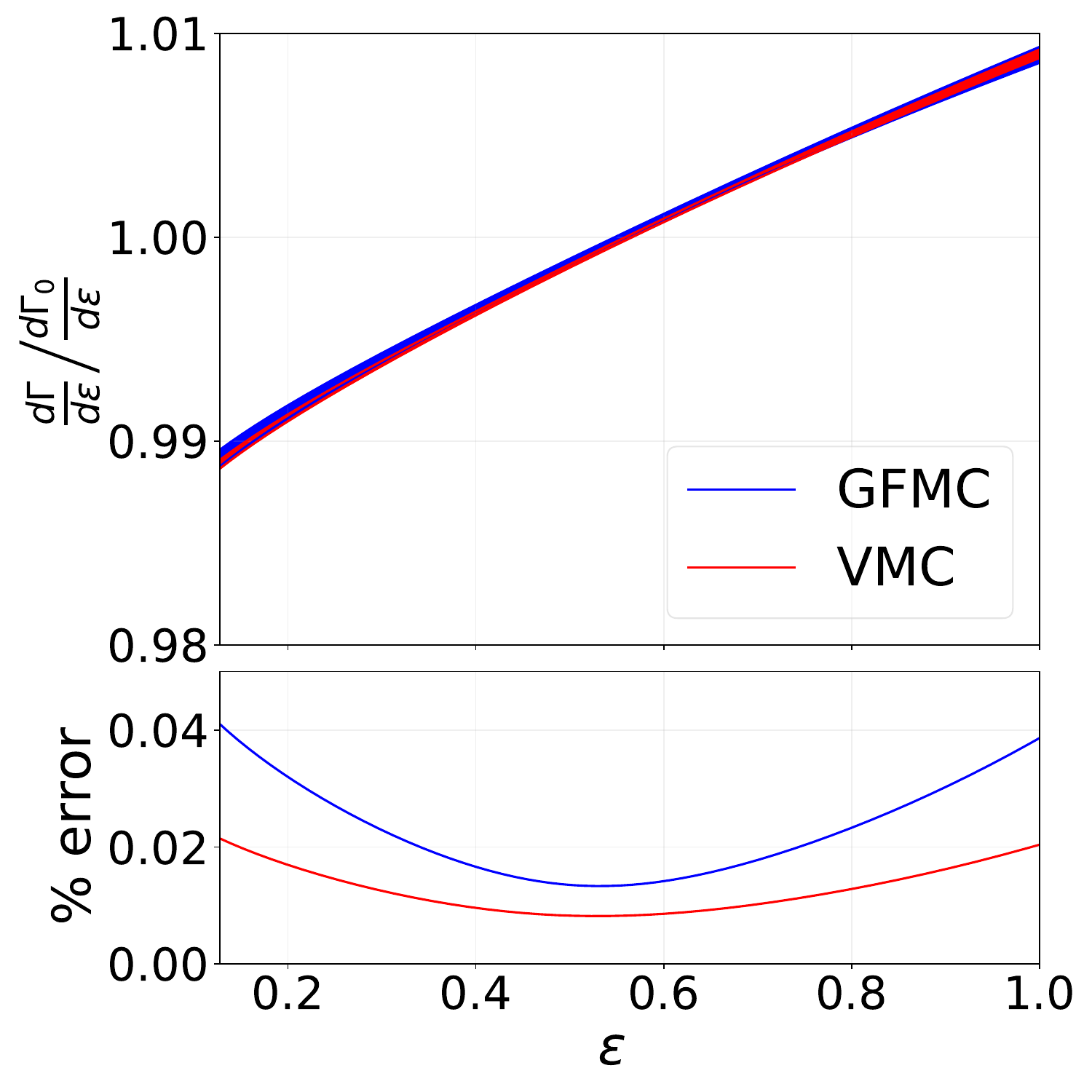}
    \caption{The Standard Model (SM) spectrum for $^6$He $\beta$ decay relative to the zero recoil approximation computed with GFMC (blue band), compared to results including BSM tensor (red band) and pseudoscalar (green band) contributions. The width of the band represents variations in the model Hamiltonian. Figure reprinted with permission from Ref.~\cite{King:2022zkz}.}
    \label{fig:gfmc.spectrum}
\end{figure}

\begin{figure}
    \centering
    \includegraphics[width=0.5\linewidth]{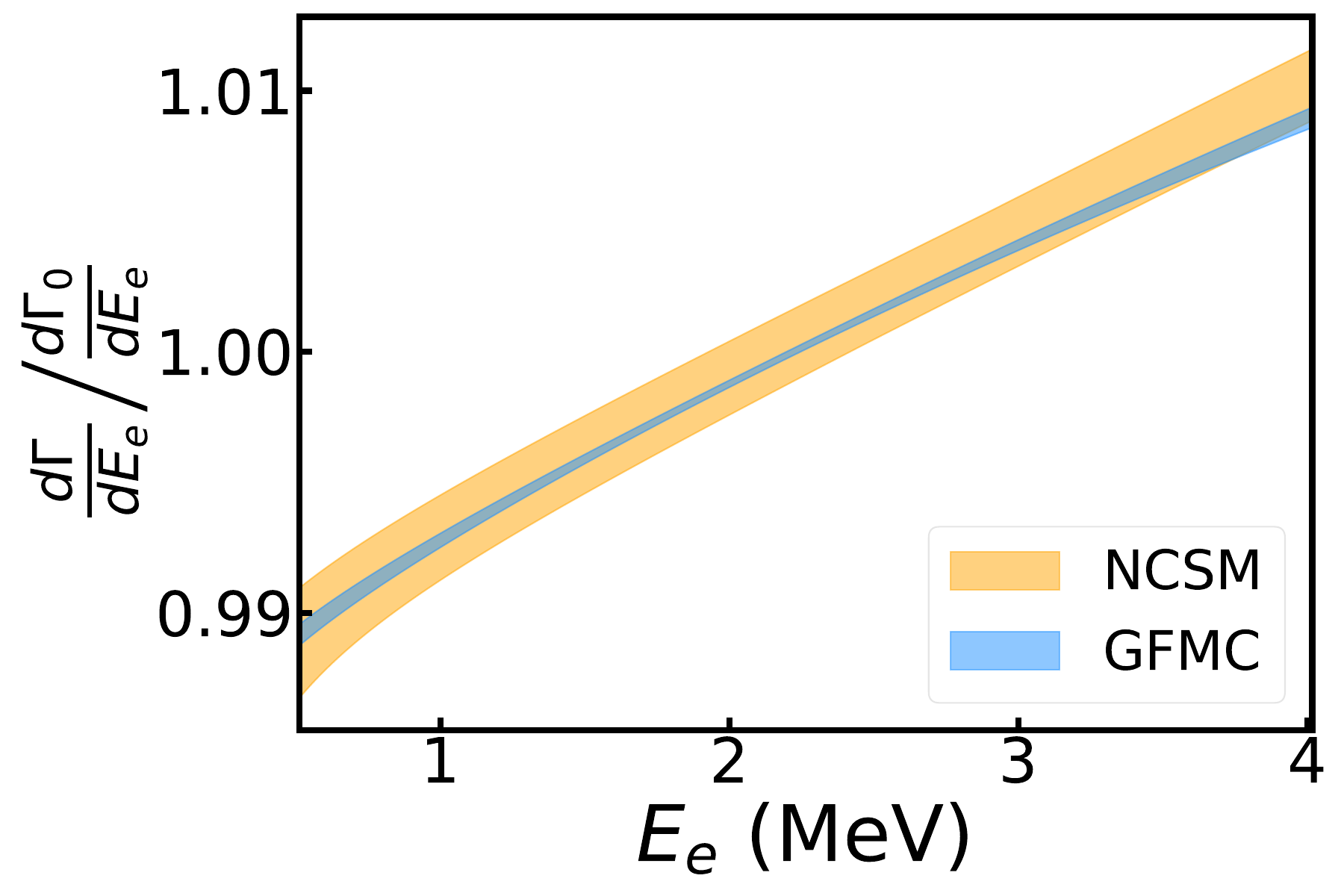}
    \caption{\red{Comparison of the distortion in the shapes of the spectra computed with NCSM (orange) and GFMC (blue) as a function of $E_e$. The width of the band indicates the uncertainty on each calculation. The NCSM result includes uncertainties from methodological and parametric uncertainties for the many-body solver, as well as uncertainties from subleading multipoles, higher-order terms in the non-relativistic expansion, and omitted two-body current contributions. The GFMC result includes uncertainties from the nuclear Hamiltonian on the multipoles that enter into the spectrum shape.}}
    \label{fig:he6.comp}
\end{figure}

\red{In Ref.~\cite{King:2022zkz}, VMC and GFMC calculations of the $^6$He $\beta$-decay spectrum were carried out and included the effects of two-body electroweak currents. In order to include the two-body effects, the authors adopted the expressions in Eq.~\ref{eq:charge} to Eq.~\ref{eq:elec} for the case $J_i=0$ and $J_f=1$ with $M=0$. This restricts the multipoles to $J=1$ and either purely axial or vector according to selection rules. The four multipoles entering the rate can be recast as matrix elements which one can compute using QMC techniques~\cite{Carlson:1997qn},
\begin{eqnarray}
\reduce{}{C_1^A}{} &=&\frac{i}{\sqrt{4\pi}}\langle {}^6{\rm Li},10\rvert \rho_{A,+}^\dagger(q\hat{\bf z})\rvert {}^6{\rm He},00\rangle \label{eq:c1} \\
\reduce{}{L_1^A}{} &=&\frac{i}{\sqrt{4\pi}}\langle {}^6{\rm Li},10\rvert \hat{\bf z}\cdot{\bf j}_{A,+}^\dagger(q\hat{\bf z})\rvert {}^6{\rm He},00\rangle\, \label{eq:l1} \\
\reduce{}{M_1^V}{} &=&-\frac{1}{\sqrt{2\pi}}\langle {}^6{\rm Li},10\rvert \hat{\bf y}\cdot{\bf j}_{V,+}^\dagger(q\hat{\bf x};V)\rvert {}^6{\rm He},00\rangle\, , \label{eq:e1} \\
\reduce{}{E_1^A}{} &=&-\frac{i}{\sqrt{2\pi}}\langle {}^6{\rm Li},10\rvert \hat{\bf z}\cdot{\bf j}_{A,+}^\dagger(q\hat{\bf x};A)\rvert {}^6{\rm He},00\rangle\, \label{eq:e1}  
\end{eqnarray}
where the letters $A$ and $V$ indicate either the axial or vector component of the current. The bras and kets denote the initial and final state nuclei, as well as their $JM$ quantum numbers. Because the $Q$-value of the decay, $Q\approx3.5$ MeV, limits the momentum transfer to the nucleus $q$ such that $q/m_{\pi} = qr_{\pi} \lesssim 0.03$, it was noted in Ref.~\cite{King:2022zkz} that one can perform an expansion of the multipoles of Ref.~\cite{Walecka:1995} in terms of a small parameter $qr_{\pi}$. Noting also that the multipoles are either purely even or odd polynomials in $q$, one obtains the following expressions,
\begin{eqnarray}
\reduce{}{C_1^A(q)}{} &=& - i \frac{q r_\pi }{3} \left(C^{(1)}_1  - \frac{(q r_\pi)^2}{10} C^{(3)}_{1}  +{\cal O}\left[(q r_\pi)^4\right] \right) \, , \label{Fit1}\\
\reduce{}{L_1^A(q)}{} &=& -\frac{i}{3} \left( L^{(0)}_1 - \frac{(q r_\pi)^2}{10} L^{(2)}_1+{\cal O}\left[(q r_\pi)^4\right] \right)\, , \label{Fit2}\\
\reduce{}{M_1^V(q)}{} &=& -i \frac{q r_\pi}{3} \left( M_1^{(1)}  - \frac{(q r_\pi)^2}{10} M^{(3)}_{1}  +{\cal O}\left[(q r_\pi)^4\right] \right) \, ,\label{Fit3}\\
\reduce{}{E_1^A(q)}{} &=&-\frac{i}{3} \left( E_1^{(0)}-\frac{(q r_\pi)^2}{10} E_1^{(2)}+{\cal O}\left[(q r_\pi)^4\right]\right)\, , \label{Fit4}
\end{eqnarray}
where we have included higher order terms in long-wavelength expansion and denoted the explicit $q$ dependence of the reduced multipoles. The leading coefficients in the notation of Ref.~\cite{King:2022zkz} may be related to those of Holstein's formalism, as this expansion is based on the Walecka approach. Namely, for the case of $^6$He $\beta$ decay, we have,
\begin{eqnarray}
L_1^{(0)} &=& \sqrt{\frac{3}{4\pi}} c\, , \\
M_1^{(1)} &=& \frac{\sqrt{3}}{2\sqrt{2\pi}r_{\pi}Am_N} b_{\rm WM}\, ,\\
C_1^{(1)} &=& \frac{\sqrt{3}}{2\sqrt{\pi}r_{\pi}Am_N} d^I\, .
\end{eqnarray}
The values extracted from the GFMC calculation are presented in Table~\ref{tab:6He_results}. Similarly to the NCSM results, good agreement with the Gamow-Teller ($c$) form factor (${\lesssim } 0.01\%$) is achieved when comparing to the value from Calaprice~\cite{PhysRevC.12.2016}. When comparing with the weak magnetism form factor $b_{\rm WM}$ from Calaprice, there is ${\sim} 1.7\sigma$ disagreement; however, the GFMC and NCSM results are consistent and agree within error bar. It is worth noting that the $M1$ transition data from Refs.~\cite{Tilley:2002vg,Bergstrom:1975ax} would give a value of $b_{\rm WM}=68.0(7)$, agreeing within ${\sim} 1\sigma$ with the GFMC results. Like the NCSM, the GFMC predicts a larger induced tensor form factor $d^I$ (=5.37) than the shell-model result in Calaprice (=2.4). The GFMC and NCSM values are also discrepant, though notably the formalism used in the NCSM adopts a Coulomb correction $\Delta E_c$ to the energy transfer $E_0$ in the pseudoscalar form factor (see Eq.~\ref{eq:multipoles1b}). When rescaling the results of the NCSM formalism for $C_1^A$ by $E_0/(E_0+\Delta E_C)$, there is agreement between the two approaches. In the GFMC, the Coulomb corrections were treated with the formalism of Ref.~\cite{Hayen:2018} and thus the authors opted to use the energy transfer explicitly appearing in the charge operator.}

\red{Expanding the multipoles in this way has the advantage that, when combined with electroweak charge and current operators that admit an expansion in $m_{\pi}/\Lambda_{\chi}$, means that one can obtain a spectrum retaining terms up to the desired precision $q/\Lambda_{\chi}$. In the case of $^6{\rm He}$ $\beta$ decay, one needs only up to the quadratic term in $q$ to achieve a below $0.1\%$ accuracy for a given model. Working to this order, the expression of the spectrum in terms of the multipole coefficients is,
\begin{eqnarray}
\frac{d \Gamma}{d E_e} &=& \frac{G_F^2  E_0^4 V_{ud}^2}{2\pi^3} \sqrt{1 - \frac{\mu_e^2}{\varepsilon^2}}\,  \varepsilon^2 (1 - \varepsilon)^2F^{-}(Z, E_e)C_{\rm corr} \nonumber \\
& &\frac{4 \pi}{3} \Bigg\{ 
 \left| L_1^{(0)} \right|^2  \left[1 + \alpha ZE_0R\left(\frac{2}{35}-\frac{233}{630}\frac{\alpha Z}{E_0R}-\frac{1}{70}\frac{\mu_e^2}{\varepsilon}-\frac{4}{7}\varepsilon\right)\right]  \nonumber \\
& &+ 2 E_0 r_\pi \Bigg[\left(1 - 2 \varepsilon + \frac{\mu_e^2}{\varepsilon }\right)  {\rm Re} (E_1^{(0)} M_1^{(1) *})- 
\left(1 - \frac{\mu_e^2}{ \varepsilon}\right)
  {\rm Re} (L_1^{(0)} C_1^{(1) *})   \Bigg] \nonumber \\
& & + \frac{(E_0 r_\pi)^2}{3} \Bigg[
\left(
3 - 4 \varepsilon (1 - \varepsilon) - \mu^2_e \frac{2 + \varepsilon}{\varepsilon}
\right) |C_1^{(1)}|^2 
- \frac{3}{5}\left( 1 - \frac{\mu_e^2}{\varepsilon} (2 - \varepsilon)\right)  \textrm{Re} \left( L_1^{(0)} L_1^{(2) *} \right) \nonumber \\ & & +  \left(3 - 10 \varepsilon (1 - \varepsilon) + \mu_e^2 \frac{4 - 7 \varepsilon}{\varepsilon}
\right) \left( \left|M_1^{(1)}\right|^2 - \frac{1}{5} \textrm{Re}\left( E^{(0)}_{1} E^{(2)}_{1}\right) \right) \Bigg] \nonumber \\
& & - \frac{4}{7}  \frac{\alpha Z E_0 r_\pi^2}{R}  (1- \varepsilon) \left(\frac{E_1^{(0)}E_1^{(2)}}{2}-L_1^{(0)}L_1^{(2)}\right)\Bigg\}, \label{rate}
\end{eqnarray}
where the scaled variables $E_e = E_0 \varepsilon$ and $m_e = E_0 \mu_e$ have been introduced. $M_i$ and $M_f$ are the masses of the initial and final state nuclei. The effects of nuclear recoil are retained to leading order in $E_0/M_f$, and nuclear recoil makes it so that the electron endpoint energy shifts from $E_e = E_0$ to $E_e = W_0 - \frac{W_0^2 - m_e^2}{2 M_f}$. The above spectrum has also incorporated Coulomb corrections, which depend on the nuclear charge $Z$, the fine structure constant $\alpha$, and the nuclear radius $R$. There are additional corrections from the distortion of the outgoing electron wave function, $F^{-}(Z,E_e)$ is the Fermi function in Eq.~\ref{eq:rate}, and higher order electromagnetic, atomic shielding, and recoil kinematic corrections make up $C_{\rm corr}$ and the specifics are detailed Ref.~\cite{King:2022zkz} following Ref.~\cite{Hayen:2018}. Note that dropping terms dependent on powers of $E_0r_{\pi}$ and $\alpha$, one recovers the GT spectrum in the zero-recoil limit, $d\Gamma_0/dE_e$. Note also that if one retains only terms up to $E_0r_{\pi}$, again dropping order $\alpha$ corrections, one can recover $\delta_1$ and $\delta_b$ from Eq.~\ref{eq:corrections}. }

While the resultant spectrum would naively give the required precision to connect with experiment, one has not taken into account the effects of using different models on the uncertainty. The goal of Ref.~\cite{King:2022zkz} was to perform the calculation with various models of the nuclear interaction in order to estimate the model dependence. In particular, four models the Norfolk two- and three-nucleon interactions~\cite{Piarulli:2016vel,Piarulli:2017dwd,Baroni:2018fdn} were employed in the computation of the nuclear matrix elements. The upper panel of Fig.~\ref{fig:gfmc.spectrum} shows the distortion of the $\beta$ decay spectrum relative to $d\Gamma_0/d\varepsilon$ as a function of the scaled variable $\varepsilon$ for VMC (red) and GFMC (blue). The spectrum receives distortions of up to $\approx 1\%$, with the dominant correction coming from the $M_1$ multipole. Because it was found that the isospin symmetry breaking between $M_1$ obtained from $^6{\rm He}$ $\beta$ decay and the electromagnetic decay of $^6{\rm Li}(1^+;0)$ to the ground state was smaller than the experimental uncertainty on the latter, the experimental value was adopted to further constrain any potential model dependence. The dominant source of uncertainty on the spectrum comes from the model dependence on the theoretical values of $L_1^{(2)}$ and $E_1^{(2)}$, and the overall spectrum uncertainty seen in the lower panel of Fig.~\ref{fig:gfmc.spectrum} is below the $0.1\%$ goal needed to compare with experiment. Finally, it was found that the SM Fierz term induced by recoil is $\delta_b=-1.47(3)\times10^{-3}$ in the GFMC calculation, which agreed with the previous {\it ab initio} evaluation discussed in Section~\ref{sec:6he.ncsm}, while also greatly reducing the uncertainty because of the explicit inclusion of two-body currents. \red{One can also obtain the correction to the angular correlation $\avg{\tilde{\delta}_a}$ from the multipoles. The dominant contributions are linear in $M_1^{(1)}$ and $C_1^{(1)}$, and the overall correction is $-1.43(20)\times 10^{-3}$, which is smaller than the NCSM prediction; however, after accounting for radiative and atomic effects following the same procedure as Ref.~\cite{glick2022nuclear}, $a=-0.3327(30)$. Thus, the values of $a$ are consistent within the error, which is dominated by experimental uncertainty at present. Figure~\ref{fig:he6.comp} shows the distortion of the spectrum obtained when one takes the corrections from NCSM (orange) and compares it to the GFMC result (blue). Within their uncertainties, the two calculations show agreement for the distortion of the spectrum. The GFMC uncertainty band is narrower primarily due to the explicit inclusion of two-body current effects.}

Because of the level of precision obtained in the theoretical calculation, it was possible to incorporate BSM physics effects to see the effects on the spectrum. Including leading-order tensor and psuedoscalar currents indicated that experiments would have sensitivity to tensor contributions not currently excluded by other analyses. Further, it was shown that in a minimal extension to the SM with one heaver sterile neutrino with mass 1 MeV, a characteristic kink would be present in the spectrum. Thus, the GFMC prediction can help to place further constraints on BSM physics scenarios. 

\subsection{$^{8}$Li$ \rightarrow {}^{8}$Be and $^{8}$B$ \rightarrow {}^{8}$Be  beta decays corrections with SA-NCSM \label{sec:8ncsm}}

In recent years, a series of high-precision angular correlations measurements have been performed using $^8$Li and $^8$B beta decays to set stringent constraints on the BSM tensor currents in the weak interactions \cite{sternberg2015limit, BurkeySGS2022, GallantSSC2023, LongfellowGSB2024}. In these experiments, the largest source of the systematic errors on the $\beta$-$\nu$ correlation parameter $a$ stemmed from the uncertainty in the recoil-order terms. In particular, four recoil-order terms have been determined by the simulations to contribute most to the systematic error budget, namely, the second-forbidden axial vectors ($j_{2}$ and $j_{3}$), induced tensor ($d^I$), and weak magnetism ($b_{\mathrm{WM}}$), with $j_{2}$ and $j_{3}$  being especially large. 
The operators of these recoil-order terms  in the impulse approximation (IA) are given in Eq. \ref{eq:Holstein}.

The matrix elements in Eq. (\ref{eq:Holstein}) have been computed translationally invariant in the SA-NCSM. 
These recoil-order form factors are usually reported as the
ratios $j_{2,3}/A^2c$, $d/Ac$, and $b_{\red{\mathrm{WM}}}/Ac$ that enter into the expression of the $\beta$-decay rate for nuclei undergoing delayed $\alpha$-particle emission \cite{Holstein1974, sternberg2015limit, Burkey:2019ant, BurkeySGS2022}. 
For the calculations of these recoil-order terms, authors of Ref. \cite{Sargsyan_A8} adopted various chiral nucleon-nucleon (NN) potentials without renormalization in nuclear medium, namely, N3LO-EM \cite{EntemM03}, NNLOopt \cite{Ekstrom13}, and NNLOsat \cite{NNLOsat2015}, and in addition, the soft JISP16 phase-equivalent NN interaction \cite{Shirokov2010nn}.

\begin{figure}[ht]
    \centering
    \includegraphics[width=0.49\textwidth]{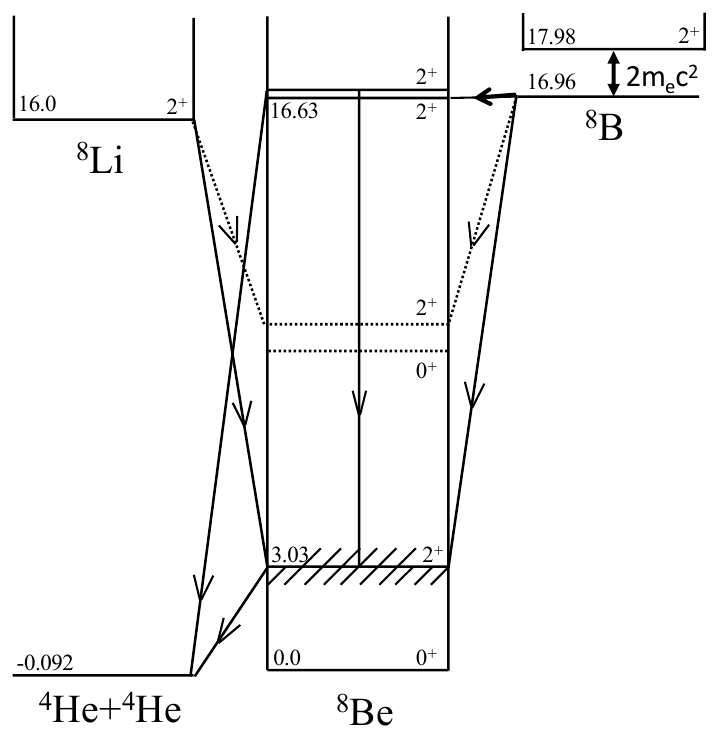}
    \caption{Level diagram of the $^8$Li and $^8$B $\beta$ decays to $^8$Be. All of $^8$Be states are above the $\alpha+\alpha$ separation threshold. The dotted levels correspond to the  states that have not been directly observed experimentally, but calculated in the SA-NCSM and proposed in earlier studies (see text for details).  
    }
        \label{fig:A8_levels}
\end{figure}

\begin{figure}[ht]
   \includegraphics[width=0.99\textwidth]{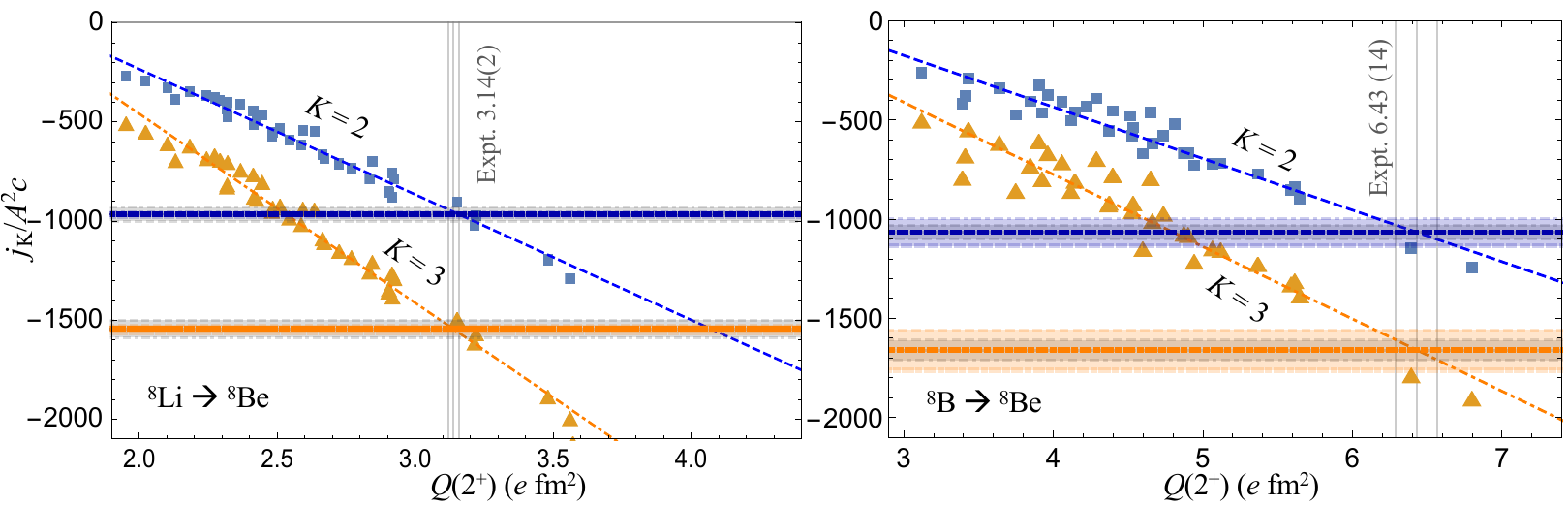}
    \caption{
     Calculated $j_2/A^2c$ and $j_3/A^2c$ from SA-NCSM (squares and triangles, respectively) and their predicted values (upper and lower horizontal 
    lines, respectively) for the $^8\mathrm{Li}\; \beta$ decay (left) and $^8\mathrm{B}\; \beta$ decay (right) to  $2_1^+$ in \Be{} vs. the calculated quadrupole moments [$Q(2^+)$] of the initial nuclei. Calculations use the NNLO$_\mathrm{opt}$, NNLO$_\mathrm{sat}$ and ${\rm N}^{3}{\rm LO}$ chiral potentials, and the JISP16 NN, in \Nmax=6-12 model spaces. The vertical gray lines show the experimental values of \Li~and  $^8$B $Q(2^+)$ with uncertainties \cite{BorremansBBG2005Q2,SumikamaNOI2006}. The darker horizontal bands are uncertainties solely from the $Q(2^+)$  experimental uncertainty while the lighter bands also include the linear regression uncertainty. Figures adapted from Refs. \cite{Sargsyan_A8, LongfellowGSB2024} with permissions.
    }
    \label{fig:jk_vs_Q}
\end{figure}

\Li~and $^8$B $\beta$ decays predominantly proceed via a nearly pure
\cite{WiringaPPM2013} Gamow-Teller transition from the $J_{\red{\text{i}}}^{\pi} = 2^{+}$, isospin
$T=1$ $^{8}\mathrm{B}$ ground state to the broad $J_{\red{\text{f}}}^{\pi} = 2^{+}$,
$T=0$ resonance in $^{8}\mathrm{Be}$ at 3 MeV which immediately breaks into
two $\alpha$ particles (Fig. \ref{fig:A8_levels}). Since this transition to the lowest 2$^+$ state in \Be~accounts for most of the statistics in the experiment, it is vital to obtain the corresponding recoil-order terms with high precision. Remarkably, there appears to be a strong correlation between $j_{2,3}/A^2c$ and the ground state quadrupole moments of \Li~and $^8$B based on calculations across several NN interactions, \Nmax{} and \hw~parameters. 
In Fig. \ref{fig:jk_vs_Q}, each marker corresponds to one calculation with a specific model space and harmonic oscillator spacing \hw~using one of the NN potentials mentioned above. Using this correlation, the values of $j_{2}/A^2c$ and $j_{3}/A^2c$ for the transition to the \Be~$2_1^+$ state are predicted by performing a linear regression and using the \Li~ and $^8$B experimentally measured ground state quadrupole moments from Refs. \cite{BorremansBBG2005Q2,SumikamaNOI2006}. To accommodate large model spaces, clustering, and collectivity, Ref. \cite{Sargsyan_A8} uses the symmetry-adapted basis of SA-NCSM to select the model spaces of about one-third of the points in Fig. \ref{fig:jk_vs_Q} following the prescription of Ref. \cite{LauneyDSBD2020}. The total uncertainties on the predictions from the correlation plots arise from two parts: the \Li~and $^8$B ground state $Q(2^+)$ experimental uncertainty intersecting with the linear regression slope and the regression uncertainty from the Student's t-distribution at 95\% confidence level (depicted as dark and light horizontal bands, respectively, in Fig. \ref{fig:jk_vs_Q}). The linear dependence between $j_{2,3}/A^2c$ and $Q(2^+_{g.s.})$ persists regardless of errors that arise from many-body truncation (including SA model space choices), use of various NN interactions that have different regulators and momentum cutoffs, and two-body currents in operators for $Q(2^+_{g.s.})$ \cite{FilinMBEKR2021} and axial-vector beta transitions \cite{Gysbers:2019uyb,MarisEFG2021,King:2020wmp}. Therefore, thanks to the diverse range of inputs, the regression uncertainty contains the effects of the many-body truncation and these additional corrections making the predictions parameter and model (interaction) independent. Similar correlations between different observables in \emph{ab initio} calculations are also used in other works to reduce uncertainties on predictions for slowly converging observables (see, \eg, Refs \cite{CalciR2016, CaprioMF2025I, CaprioFM2025II}).

\begin{figure}[ht]
   \includegraphics[width=0.99\textwidth]{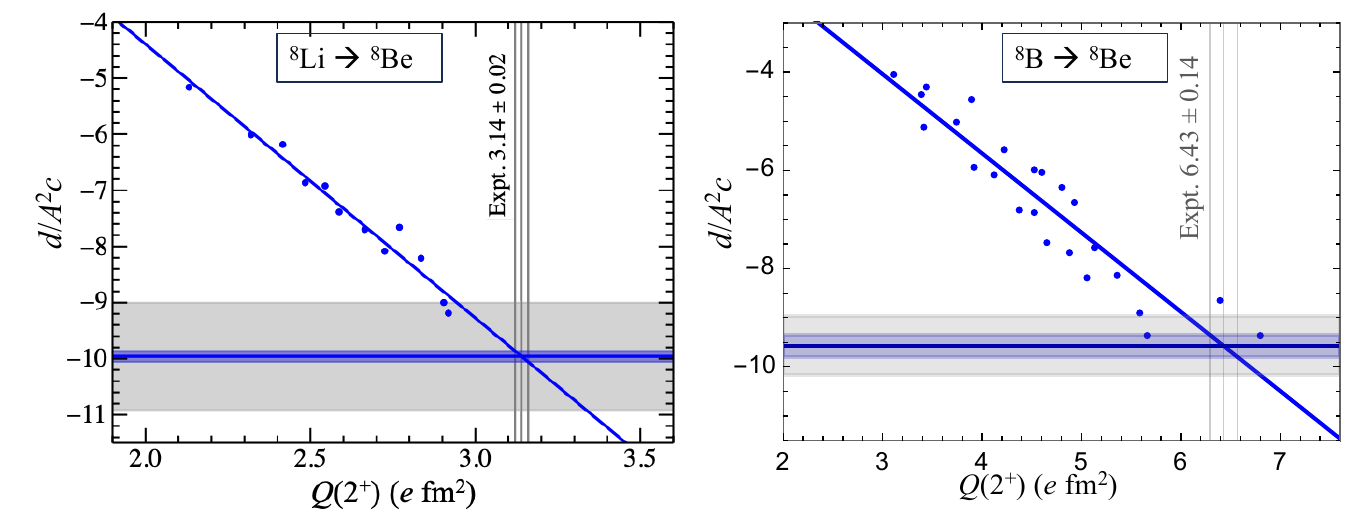}
    \caption{
     Calculated $d/A^2c$ from SA-NCSM (blue dots) and their predicted values for the $^8\mathrm{Li}\; \beta$ decay (left) and $^8\mathrm{B}\; \beta$ decay (right) to  $2_1^+$ in \Be{} vs. the calculated quadrupole moments [$Q(2^+)$] of the initial nuclei. The vertical gray lines show the experimental values of \Li~and  $^8$B $Q(2^+)$ with uncertainties \cite{BorremansBBG2005Q2,SumikamaNOI2006}. The  horizontal blue bands are uncertainties solely from the $Q(2^+)$  experimental uncertainty while the gray bands also include the linear regression uncertainty. 
    }
    \label{fig:d_vs_Q}
\end{figure}

 The predictions for $j_{2}/A^2c$ and $j_{3}/A^2c$ recoil-order terms, as well as the weak magnetism $b_{\mathrm{WM}}/Ac$ and induced tensor $d^I/Ac$ terms, for the lowest four  $2^+$ states in \Be~calculated from SA-NCSM, are summarized in Table \ref{tb:recoil}.
The $d/Ac$ prediction for $2_1^+$  is based on a  correlation similar to the one for $j_{2,3}/A^2c$
(Fig. \ref{fig:d_vs_Q}). For $b_{\mathrm{WM}}/Ac$  to all states in \Be~and for the other recoil-order terms to higher-lying $2^+$
states, the values are calculated using \NNLOopt~and
JISP16 interactions with uncertainties from varying \hw~from
20 MeV by 5 MeV and model-space sizes from
\Nmax=6–12. In addition to being necessary for measurements of BSM terms in the weak interaction, $b_{\mathrm{WM}}/Ac$ predictions are of interest to experiments that test the CVC hypothesis, while $d^I/Ac$ is of importance to determining the existence of second-class currents \cite{DeBraeckeleer1995}.   

\begin{table}[h!]
\caption{The recoil-order terms from SA-NCSM. Results for the 2$^+_1$ $j_{2,3}/A^2c$  and $d/Ac$ are based on the correlation to $Q(2_{\rm g.s.}^+)$; all other calculations use \NNLOopt{} and JISP16 and have error bars from variations in \hw{} by 5 MeV and in model-space sizes up to \Nmax=12. Table adapted from Ref. \cite{LongfellowGSB2024} with permission.
}
\label{tb:recoil}
\centering
\begin{tabular}{llllll}
\hline \hline
\multicolumn{1}{c}{} & \multicolumn{1}{c}{} & \multicolumn{1}{c}{$j_2/A^2c$} & \multicolumn{1}{c}{$j_3/A^2c$} & \multicolumn{1}{c}{$d^I/Ac$} & \multicolumn{1}{c}{$b_{\mathrm{WM}}/Ac$} \\ \hline
\Li $\rightarrow$ \Be & $2_1^+$             &  $-966     \pm 36$              & $-1546 \pm 44$                  & $10.0 \pm 1.0$                                     & $6.0 \pm 0.4 $                       \\
 & $2^+_2 \mathrm{(intruder)}$           & $-10  \pm 10$                 & $-80  \pm 30$                & $-0.5  \pm 0.5$                    & $3.7  \pm 0.4  $                      \\
 & $2^+_3 \rm{(doublet\, 1)}$            & $12  \pm 5$                  & $-60  \pm 15  $                & $0.3  \pm 0.2  $                   & $3.8  \pm 0.2$                         \\
 & $2^+_4 \rm{(doublet\, 2)}$       & $11  \pm 3$                   & $-65 \pm 11$                 & $0.2 \pm 0.2$                 & $3.8 \pm 0.2$     \\
 $^8$B $\rightarrow$ \Be & $2_1^+$             &  $-1067     \pm 68$              & $-1660 \pm 102$                  & $9.6 \pm 0.6$                                     & $6.1 \pm 0.5 $                       \\
 & $2^+_2 \mathrm{(intruder)}$           & $10  \pm 45$                 & $-41  \pm 75$                & $-0.5  \pm 0.8$                    & $3.7  \pm 0.4  $                      \\
 & $2^+_3 \rm{(doublet\, 1)}$            & $8  \pm 4$                  & $-53  \pm 20  $                & $0.1  \pm 0.1  $                   & $3.8  \pm 0.2$                         \\
 & $2^+_4 \rm{(doublet\, 2)}$       & $7  \pm 5$                   & $-70 \pm 13$                 & $0.2 \pm 0.1$                 & $3.8 \pm 0.2$  \\
\hline \hline
\end{tabular}
\end{table}

Earlier experimentally deduced values for these recoil-order terms were reported in Ref. \cite{sumikama2011test} by Sumikama \etal~These values, namely $j_2/A^2c = -490 \pm 70$, $j_3/A^2c = -980 \pm 280$, $d^I/Ac = 5.5 \pm 1.7$, and $b_{\mathrm{WM}}/Ac = 7.5 \pm 0.2$, differ from the theoretical predictions, and were derived through a comprehensive fit to $\beta$-spin alignment data \cite{sumikama2011test} and $\beta$-$\alpha$ angular correlation measurements \cite{McKeownGG1980} from \Li~and $^8$B beta decays. Due to the limited magnitude of higher-order effects and the relatively large statistical uncertainties, the recoil-order terms $j_{2}/A^2c$, $j_{3}/A^2c$ and $d^I/Ac$ were assumed in Ref. \cite{sumikama2011test} to be independent of the \Be~excitation energy. Consequently, the reported values were averaged over the entire beta decay spectrum. 

Conversely, the SA-NCSM wavefunctions are computed for specific states; therefore, the predictions in Table \ref{tb:recoil} correspond solely to the each of the 2$^+$ final states, in the \Be~spectrum. The SA-NCSM calculations indicate significant differences in the recoil-order terms between the lowest 2$_1^+$ state and higher-lying states, notably for the $j_K/A^2c$ terms, where the values differ by nearly two orders of magnitude (see Table \ref{tb:recoil}). Averaging of the calculated recoil-order terms over the entire \bt spectrum would likely yield values similar to the ones obtained by Sumikama \etal~Accordingly, the latest angular-correlation experiment described in Refs. \cite{BurkeySGS2022} and  \cite{LongfellowGSB2024} minimize sensitivity to higher-lying states by restricting the analysis to decays associated primarily with the lowest $2_1^+$ state in \Be~for which the calculated recoil-order terms are well constrained.

\begin{figure}[ht]
    \centering
   \includegraphics[width=0.53
\textwidth]{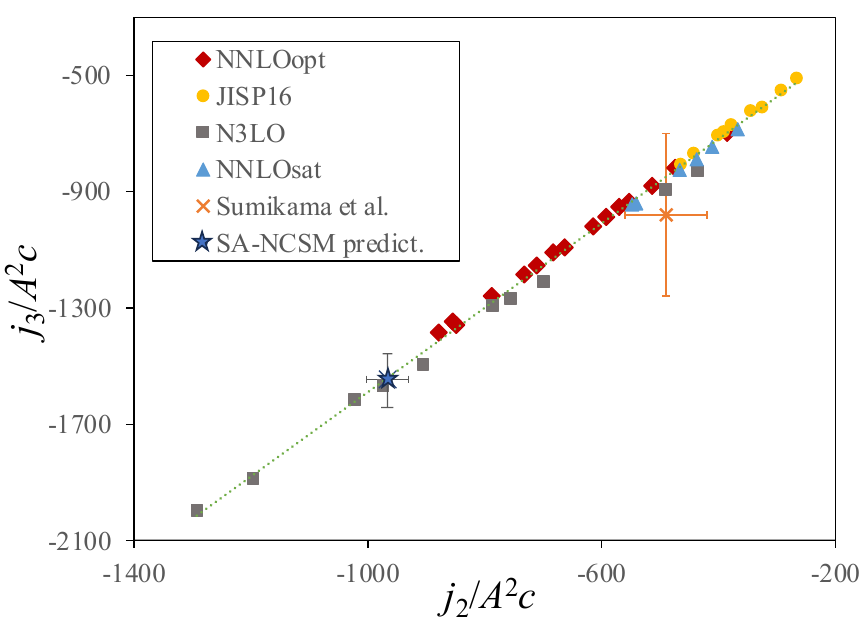}
    \caption{Calculated \Li~beta decay $j_3/A^2c$ vs. $j_2/A^2c$  values to the lowest $2^+$ state in \Be~with \NNLOopt, NNLO$_\mathrm{sat}$, N3LO-EM and JISP16 interactions for the same model spaces as in Fig. \ref{fig:jk_vs_Q} $^8$Li$\rightarrow{}^8$Be plot, along with the values from Sumikama \etal~ \cite{sumikama2011test} (orange cross) and the SA-NCSM prediction (blue star) from Table \ref{tb:recoil}. The linear fit (dotted green line) is added to guide the eye. Figure adapted from Ref. \cite{Sargsyan_A8} Supplemental Material with permissions. 
}
    \label{fig:j2_vs_j3}
\end{figure}

Since both $j_2/A^2c$ and $j_3/A^2c$ for the \Be~lowest $2_1^+$ state are strongly correlated with the \Li~ and $^8$B g.s. quadrupole moments (Fig. \ref{fig:jk_vs_Q}), it is expected that there is a strong correlation between $j_2$ and $j_3$. Indeed, the calculated $j_2/A^2c$ vs. $j_3/A^2c$ for all four interactions under consideration follow an almost perfect linear trendline (Fig. \ref{fig:j2_vs_j3}). Notably, the experimentally deduced values of $j_2/A^2c$ vs. $j_3/A^2c$ from Sumikama \etal~\cite{sumikama2011test}, which have been determined  over the whole range of the \bt energy spectrum, align remarkably well with the calculated correlation. In other words, the ratio $j_3/j_2$ from experiment agrees with the ratio from all the calculations. 

 The comprehensive analysis of recoil-order terms in the \bt decays of \Li~and $^8$B, supported by SA-NCSM calculations across various nuclear interactions, reveals robust correlations that are largely independent of the chosen models and parameters. These correlations between the recoil-order form factors and the ground state quadrupole moments enhance the precision of theoretical predictions essential for constraining BSM weak tensor currents. The consistency between calculated and experimental ratios of $j_2/j_3$ recoil-order terms underscores the reliability of the current theoretical framework and its potential to refine the interpretation of \bt-decay measurements in the search for new physics.


    \section{Summary and outlook}\label{sec:sum}
Recent advances in \emph{ab initio} nuclear many-body theories have significantly enhanced our understanding of nuclear $\beta$ decays, especially in the context of probing BSM physics. State-of-the-art many-body methods, such as NCSM, SA-NCSM and QMC, now enable precise calculations of radiative and recoil-order corrections with quantified uncertainties. These developments are vital for interpreting high-precision experimental data and for extracting fundamental parameters like $V_{ud}$ with unparalleled accuracy.

The integration of EFT frameworks with \emph{ab initio} calculations has provided a robust theoretical foundation for evaluating structure-dependent radiative corrections. \emph{Ab initio} calculations of radiative corrections, such as the $\delta_{NS}$ term in superallowed $\beta$ decays of $^{10}$C $\rightarrow{}^{10}$B and $^{14}$O $\rightarrow{}^{14}$C have achieved unprecedented precision. These results have significantly reduced theoretical uncertainties to the order of $10^{-4}$. This level of accuracy enables current and future experiments to probe BSM physics with enhanced sensitivity, as even small deviations could indicate new interactions or particles. Experiments observing deviations beyond these refined predictions could potentially reveal new physics at energy scales up to 10 TeV. 

Similarly, advances in \emph{ab initio} many-body calculations of recoil-order terms have played a crucial role in refining the limits on BSM weak interaction currents. High-precision predictions of recoil corrections for  $^6$He, $^8$Li and $^8$B $\beta$ decays have helped improve the experimental bounds on possible non-SM contributions. These calculations  quantify uncertainties arising from many-body models and underlying NN interactions leading to more reliable interpretations of experimental data. As a result, they have tightened the constraints on exotic tensor couplings in the weak interaction, further restricting the level of new physics discovery. 

Looking ahead, several lines of theoretical work will be particularly impactful. First, increasing many‑body model space sizes and applying multiple complementary \emph{ab initio} methods will further reduce truncation and method‑dependent uncertainties. If possible, correlating decay observables with other measured nuclear properties can also constrain theory systematics. Second, continued improvement of nuclear interactions is needed. To this end, sensitivity studies to identify which interaction terms affect $\beta$-decay observables most would be useful to prioritize and guide the development of these improved interactions. Furthermore,
\red{while it is clear that there are number of many-body methods with different strengths, it is presently hard to standardize results from different many-body methods because of the variation in nuclear Hamiltonians. In order to better compare across methods, it would be beneficial to take advantage of the possibility to benchmark approaches with the softer-core local chiral interactions that are presently available. Further, developments like those in Ref.~\cite{Curry:2023mkm} to implement non-local terms into QMC calculations will also make benchmarks with a larger suite of interactions possible and should continue to be pursued by the community.} 

Even further, extending \emph{ab initio} calculations to a broader set of nuclei that are experimentally accessible (beyond the very light systems) will broaden the range of precision $\beta$-decay tests. Several experimental campaigns target heavier emitters such as $^{22}$Na, $^{23}$Ne, and $^{32}$Ar. Providing reliable theoretical predictions with quantified uncertainties for these and similar systems will be crucial to disentangling nuclear‑structure effects from genuine BSM signals. Moreover, unique forbidden $\beta$ decays and other non‑allowed transitions offer complementary sensitivity to exotic weak currents (including right‑handed components and low‑mass new states that modify spectral shapes). A clear theoretical priority is to extend \emph{ab initio} treatments to forbidden spectra with consistent recoil, radiative, and Coulomb corrections and robust error estimates. NCSM calculations of the unique first forbidden decay of $^{16}$N, including recoil-order and Coulomb corrections, have recently been completed~\cite{glick2026ab}, which will provide a benchmark in this direction and a template for similar studies in heavier systems. 

Finally, close synergy with experimental programs remains essential. As experimental uncertainties approach the sub‑per‑mil level, matched theoretical precision and transparent uncertainty budgets will be required to fully exploit those measurements \cite{Brodeur2023nuclear, King2026witepaper}. Integrating low‑energy $\beta$‑decay constraints with results from collider and other high‑energy searches will also strengthen overall BSM discovery potential by providing complementary information across energy scales.

 The input that \emph{ab initio} nuclear theory has provided to high-precision experiments is rapidly advancing our capacity to test the Standard Model and explore potential new physics. Continued efforts in theoretical developments, coupled with innovative experimental techniques, promise to keep nuclear \bt decays at the forefront of fundamental physics research in the coming decades.
	
	\newpage
	\section*{Acknowledgments}  
\red{We thank Emanuele Mereghetti for useful discussions when preparing the comparisons of the two formalisms for the $^6$He $\beta$ decay spectrum. We also thank Kristina Launey and Nicholas Scielzo for useful discussions.}
The work of CYS is supported in part by the U.S. Department of Energy (DOE), Office of Science, Office of Nuclear Physics, under award DE-FG02-97ER41014, by the FRIB Theory
Alliance award DE-SC0013617, by the DOE Topical Collaboration ``Nuclear Theory for New
Physics'', award No. DE-SC0023663, and by University of Tennessee, Knoxville.
The work of AGM is supported in part by the Hebrew University of Jerusalem through the Dalia and Dan Maydan Post-Doctoral Fellowship. The work of GBK is supported by Los Alamos National Laboratory's Laboratory Directed Research and Development program under project 20240742PRD1 (G.~B.~K.).
The work of GHS is supported by the U.S. Department of Energy, Office of Science, Office of Nuclear Physics, under the FRIB Theory Alliance award DE-SC0013617.
This work benefited from high performance computational resources provided by the National Energy Research Scientific Computing Center (NERSC), a U.S. Department of Energy Office of Science User Facility at Lawrence Berkeley National Laboratory operated under Contract No. DE-AC02-05CH11231, as well as the Frontera computing project at the Texas Advanced Computing Center. Frontera is made possible by National Science Foundation award OAC-1818253.


   \bibliography{beta_ppnp}
	


	
	\newpage
	\appendix
	\renewcommand*{\thesection}{\Alph{section}}
	

\end{document}